\documentclass[11pt,a4paper]{article}
\usepackage{mathtools}
\usepackage{mathrsfs}
\usepackage{amsmath}
\usepackage{tikz-feynman}
\usepackage[utf8x]{inputenc}
\usepackage{amssymb}
\usepackage{slashed}
\usepackage{bbm}
\usepackage{bm}
\usepackage{color}
\usepackage{tikz}
\usepackage{cite}
\usepackage{float}
\usepackage[]{todonotes}
\usepackage{jheppub}
\usepackage{dsfont}
\usepackage{hyperref}

\hypersetup{
colorlinks=true,         % false: boxed links; true: colored links, false is default
linkcolor=blue,          % color of internal links, red is default
citecolor=red,        % color of links to bibliography
urlcolor=red            % color of external links, cyan is default
}

\renewcommand{\[}{\begin{equation}\begin{aligned}}
\renewcommand{\]}{\end{aligned}\end{equation}}

\newcommand{\hel}{\eta} % Helicity label
\renewcommand{\d}{\mathrm{d}}
\newcommand{\dd}{\hat{\mathrm{d}}}
\newcommand{\del}{\hat{\delta}}
\newcommand{\dPhi}{\d \Phi}
\newcommand{\Del}{\hat{\delta}_\Phi}
\newcommand{\ket}[1]{| #1 \rangle}
\newcommand{\bra}[1]{\langle #1 |}
\newcommand{\amp}{\mathcal{A}}
\newcommand{\waveshape}[1]{\alpha^{(#1)}}
\newcommand{\barwaveshape}[1]{\bar{\alpha}^{(#1)}}
\def\Lexp{\biggl\langle\!\!\!\biggl\langle}
\def\Rexp{\biggr\rangle\!\!\!\biggr\rangle}
\newcommand{\KMOCav}[1]{\Lexp #1 \Rexp}

\renewcommand{\Re}{\operatorname{Re}}
\renewcommand{\Im}{\operatorname{Im}}

\newcommand{\Fop}{\mathbbm{F}}
\newcommand{\Aop}{\mathbbm{A}}
\newcommand{\Rop}{\mathbbm{R}}
\newcommand{\hop}{\mathbbm{h}}
\newcommand{\Pop}{\mathbbm{P}}
\newcommand{\Wop}{\mathbb{W}}

\newcommand{\Fcl}{F}
\newcommand{\Rcl}{R}

\def\polvh#1{\varepsilon^{(#1)}}
\def\polvhconj#1{\varepsilon^{(#1)*}}
\def\polvprimehconj#1{\varepsilon'^{(#1)*}}

\renewcommand{\v}[1]{\boldsymbol{{#1}}}

\newcommand{\kb}{\bar k}
\newcommand{\qb}{\bar q}
\newcommand{\hc}{\textrm{h.c.}}

\newcommand{\bark}{\bar k}
\newcommand{\barq}{\bar q}

\def\Lexp{\biggl\langle\!\!\!\biggl\langle}
\def\Rexp{\biggr\rangle\!\!\!\biggr\rangle}

\def\vmu{{\vphantom{\mu}}}
\definecolor{allOrderBlue}{rgb}{0.4,0.5,1}

\newcommand{\sd}{\mathrm{d}}
\newcommand{\pd}{\partial}

\newcommand{\cl}[1]{\mathcal{#1}}
\newcommand{\braket}[1]{\langle #1 \rangle}

\usepackage{tcolorbox}
\definecolor{airforceblue}{rgb}{0.36, 0.54, 0.66}
\definecolor{azure}{rgb}{0.0, 0.5, 1.0}
\newtcolorbox{tdbox}{colback=airforceblue!40!white,colframe=azure!90!black} 
\newcommand{\td}[1]{
	\if\notesOn1
	\begin{tdbox}
		#1
	\end{tdbox}
	\fi
}
\def\notesOn{1}
\title{The Uncertainty Principle and Classical Amplitudes}
\author[1]{Andrea Cristofoli,}
\author[2]{Riccardo Gonzo,}
\author[2,3]{Nathan Moynihan,}
\author[3]{Donal O'Connell,}
\author[3]{Alasdair Ross,}
\author[3]{Matteo Sergola,}
\author[4]{Chris D. White}

\affiliation[1]{School of Mathematics and Maxwell Institute for Mathematical Sciences, University of Edinburgh, EH9 3FD, Scotland}
\affiliation[2]{School of Mathematics \& Hamilton Mathematics Institute, Trinity College Dublin, College Green, Dublin 2, Ireland}
\affiliation[3]{Higgs Centre for Theoretical Physics, School of Physics and Astronomy, The University of Edinburgh, EH9 3FD, Scotland}
\affiliation[4]{Centre for Theoretical Physics, School of Physical and Chemical Sciences, Queen Mary University of London, 327 Mile End Road, London E1 4NS, United Kingdom}

\preprint{SAGEX-21-31-E, QMUL-PH-21-56}

\abstract{
We study the variance in the measurement of observables during scattering events, as computed using amplitudes. The classical
regime, characterised by negligible uncertainty, emerges as a consequence of an infinite set of relationships among multileg, multiloop amplitudes in a momentum-transfer expansion. We discuss two non-trivial examples in detail: the six-point tree and the five-point one-loop amplitudes in scalar QED.
We interpret these relationships in terms of a coherent exponentiation of radiative effects in the classical limit which generalises the eikonal formula, and show how to recover the impulse, including radiation reaction, from this generalised eikonal.
Finally, we incorporate the physics of spin into our framework.
}

\begin{document}
\maketitle

\section{Introduction}
\label{sec:intro}

Scattering amplitudes are quantum mechanical objects. They are matrix elements of the time evolution operator 
$S = U(\infty, -\infty)$ from the far past to the far future; on their distributional support, amplitudes evaluate to a complex number whose
square can be interpreted as a probability. Yet recent research has revealed that scattering amplitudes also have a useful role to play in 
classical physics.

Amplitudes have found this new role for three basic reasons. One is the double copy~\cite{Kawai:1985xq,Bern:2008qj,Bern:2010ue,Bern:2010yg}, which
simplifies the computation of amplitudes in general relativity allowing us to go to higher loops and legs, and therefore to higher theoretical
precision~\cite{Bern:2018jmv,Bern:2019crd,Carrasco:2021otn}. The second reason is the detection~\cite{LIGOScientific:2016aoc} of gravitational waves, leading to a  vigorous programme of research into their 
phenomenology. This programme demands a high precision understanding of gravitational wave physics. 
Finally, we have learned how to extract classical physics from the quantum mechanical amplitudes~\cite{Donoghue:1993eb,Donoghue:1994dn,Bjerrum-Bohr:2002fji,Holstein:2004dn,Neill:2013wsa,Bjerrum-Bohr:2013bxa,Cachazo:2017jef,Guevara:2017csg,Cheung:2018wkq,KMOC,Maybee:2019jus}, allowing the double copy to
be applied in astrophysically relevant situations~\cite{Bern:2019nnu,Antonelli:2019ytb}.

Of course there are many reasons to be interested in scattering amplitudes besides gravitational wave physics. Amplitudes are
particularly important in collider physics. They are also interesting in their own right because of their internal structure.
In many theories, including general relativity, amplitudes have uniqueness properties which allow us (in principle) to determine the entire $S$ matrix
from basic knowledge of the helicities of the interacting particles~\cite{Benincasa:2007xk,Arkani-Hamed:2017jhn}. In this way amplitudes provide a
way to \emph{define} a quantum field theory. Since amplitudes are intrinsically quantum-mechanical objects this definition hard-wires key aspects
of quantum mechanics, such as the uncertainty principle, into the physics.

Taking the viewpoint that scattering amplitudes define the physics, an old question reappears. How do we understand the classical limit?
It is more common to define quantum field theories in terms of path integrals: then (thanks to Feynman) it is clear how the classical
limit arises --- it comes from stationary phase in the path integral. But amplitudes, as a quantum-first definition of a quantum field theory,
do not have as clear a link to classical physics.

Our interest is in this question. How is the classical limit encoded in the quantum-first definition of field theory through scattering amplitudes?
To answer this question we take another look at the link between amplitudes and classical physics.

In fact by now there are several methods available for converting amplitudes into classical quantities. It has long been understood that 
four-point amplitudes are closely connected to classical potentials~~\cite{Donoghue:1993eb,Donoghue:1994dn,Bjerrum-Bohr:2002fji,Holstein:2004dn,Bjerrum-Bohr:2013bxa}, and it is possible to deduce interaction potentials from four-point multiloop amplitudes. More generally, effective Lagrangians are often used to compute amplitudes --- and this procedure can be reversed,
allowing Wilson coefficients in effective Lagrangians to be extracted from amplitudes~\cite{Neill:2013wsa,Cheung:2018wkq,Bern:2019crd,Bern:2019nnu}. The versatility of effective Lagrangians allows information and the potential, spin effects~\cite{Bern:2020buy,delaCruz:2021gjp}, etc to be readily 
extracted from amplitudes and applied to bound gravitational systems. 

In this article we will be particularly interested in two other links between amplitudes and classical physics. The first of these was 
introduced by Kosower, Maybee and one of the authors (KMOC)~\cite{KMOC,Maybee:2019jus}. This method is based on computing observables from scattering amplitudes. The basic idea is that if the observable is well-defined in both the classical and quantum-mechanical cases, then in 
the correspondence regime the result of a quantum calculation must agree with the classical calculation.
But it is worth emphasising that the methods of KMOC apply equally well to quantum observables which do not make sense
in the classical theory (for example, the number of photons radiated during a scattering event). In this sense KMOC is a quantum-first
method: it takes the quantum field theory as the basic starting point, making contact with classical physics only at the end of the
computation. 

The eikonal approximation is the second link between amplitudes and classical physics of particular interest to us. Eikonal physics
has a long history, going back to nineteenth century work on the relation between geometric and wave optics. The ``eikonal analogy''
was a guide in the early development of quantum mechanics. In its long history the precise meaning of the word ``eikonal'' seems to 
have undergone some drift. We are interested in the eikonal approximation to (especially four-point) massive amplitudes in quantum 
field theory~\cite{PhysRev.143.1194,Abarbanel:1969ek,Cheng:1969eh,PhysRev.186.1611,WALLACE1973190,PhysRev.186.1656,Wallace:1977ae, Cheng1981ConsequencesOT}. In this approximation, the incoming $s$-channel energy is large compared to momentum transfers. The key fact in this
limit is that the amplitude can be separated into two factors. One of these factors is the exponent of an eikonal function which we
will denote as $\chi$, divided by $\hbar$. The $\chi$ is analogous to a classical action, and this part of the amplitude describes the classical
physics. The remaining terms in the amplitude, which do not exponentiate, are quantum mechanical.

The quintessential difference between classical and quantum physics is uncertainty. Quantum observables naturally have a variance
which is absent in classical physics. We begin our study by computing the variance in important classical observables using scattering
amplitudes, following the methods of KMOC. In the classical limit the variance must be negligible. We find that this condition of negligible
variance becomes an infinite set of relations which involve different amplitudes. More specifically, in the correspondence regime we can
expand our amplitudes in powers of momentum transfer over centre-of-mass energy. This is a semi-classical expansion
because the momentum transfers in KMOC are order $\hbar$\footnote{Alternatively one can think of the expansion as a large-mass expansion because the centre-of-mass energy is dominated by the particle masses.}. The zero-variance conditions relate certain terms in
this ``transfer'' expansion between amplitudes with different numbers of loops and legs. For example, five-point amplitudes in 
the transfer expansion are related to lower loop five-point and four-point amplitudes.

At four points, the zero-variance conditions are very familiar: they are the relations required for eikonal exponentiation. Our 
work therefore shows that there is a generalisation of the eikonal formula beyond four points. We outline the structure of the generalisation,
which involves a coherent radiative state entangled with the four-point dynamics.

Throughout our work, we consider semiclassical scattering events involving two point-like particles. The particles may interact electromagnetically, gravitationally, or via classical Yang-Mills forces.
We begin our work
in section~\ref{sec:uncertain} with a discussion of a basic requirement for a successful quantum description of such a classical event: 
negligible variance in a measurement of the field strength. As we will see, this requirement becomes a non-trivial constraint on scattering amplitudes. The leading obstruction is given by the six-point tree amplitude; this amplitude must be suppressed relative to the corresponding six-point one-loop amplitude. As a check we compute the tree amplitude, demonstrating explicitly that it has the required suppression.
We build on this observation in section~\ref{sec:infinity_relations} to find an infinite series of constraints on multi-loop, multi-leg scattering amplitudes. 
The first non-trivial constraint on the five-point one-loop amplitude is verified in section~\ref{sec:one-loop} in the particular case of electrodynamics.
In section~\ref{sec:eikonal-review} we interpret these zero-variance conditions in eikonal terms, arguing that radiation
exponentiates in a manner analogous to the conservative terms. We propose a formula for the $S$ matrix acting on our state which
is a product of a coherent radiative state entangled with the more traditional eikonal (conservative) dynamics.
Section~\ref{sec:Schwinger} provides a partial derivation of our final state proposal in the soft case using path integral methods.
We turn to particles with spin in section~\ref{sec:spin} where we show that a simple generalisation of our final state neatly encodes the dynamics of spinning particles. 
The main body of our article concludes in section~\ref{sec:discussion} with a discussion.

\subsection*{Note added}

As our work developed, we became aware of independent progress in a similar direction by Paolo Di Vecchia, Carlo Heissenberg, Rodolfo Russo and Gabriele Veneziano. This group's work was presented most recently as a series of seminars~\cite{carlosTalk, rodolfosTalk, gabrielesTalk} at Saclay, and will soon appear in print. The themes most in
common between this work and that of Di Vecchia et al concern the factorisation of five-point one-loop amplitudes, which we discuss in section~\ref{sec:one-loop}, and the extension of the eikonal to the radiative case in our section~\ref{sec:Rad_State}.

\section{Negligible Uncertainty}
\label{sec:uncertain}

In classical electrodynamics, a key role is played by the field strength $\Fcl_{\mu\nu}(x)$. This object is a complete gauge-invariant characterisation of the field; once it is known, quantities such as the energy-momentum radiated to infinity and the field angular momentum are easily determined. In a quantum description, the field becomes an operator $\Fop_{\mu\nu}(x)$. In a semiclassical situation, the expectation value of this operator on a state $\ket{\psi}$ should equal the classical field, up to negligible quantum corrections:
\[
\langle \psi | \Fop_{\mu\nu}(x) | \psi \rangle = F_{\mu\nu}(x) + \mathcal{O}(\hbar) \,.
\]
Note that we have schematically indicated the presence of small, order $\hbar$, quantum corrections. More precisely, these corrections must be suppressed by dimensionless ratios involving Planck's constant; the precise ratios depend on the actual physical context.

Since in the quantum theory a single-valued field is replaced by the expectation value of an operator, we must address the quintessentially quantum mechanical issue of uncertainty. The uncertainty can be characterised by the variance
\[
\bra{\psi} \Fop_{\mu\nu}(x) \Fop_{\rho \sigma}(y) \ket{\psi} - \bra {\psi} \Fop_{\mu\nu}(x) \ket{\psi} \bra{\psi} \Fop_{\rho \sigma}(y) \ket{\psi} \,.
\label{eq:variance}
\]
In the domain of validity of the classical approximation, this variance must be negligible.

Precisely the same remarks hold in a quantum mechanical approach to GR. The curvature tensor $\Rcl_{\mu\nu\rho\sigma}(x)$ in the classical theory is replaced by the expectation value of the curvature operator $\Rop_{\mu\nu\rho\sigma}(x)$. The variance
\[
\bra{\psi} \Rop_{\mu\nu\rho\sigma}(x) \Rop_{\alpha\beta\gamma\delta}(y) \ket{\psi} - \bra {\psi} \Rop_{\mu\nu\rho\sigma}(x) \ket{\psi} \bra{\psi} \Rop_{\alpha\beta\gamma\delta}(y) \ket{\psi} 
\]
must be negligible.

This section is devoted to an investigation of this condition of negligible uncertainty. Working in the KMOC formalism at lowest order in perturbation theory, we will see that the expectations $\bra{\psi} S^\dagger \Fop_{\mu\nu}(x) \Fop_{\rho \sigma}(y) S \ket{\psi}$ and $\bra{\psi} S^\dagger \Rop_{\mu\nu\rho\sigma}(x) \Rop_{\alpha\beta\gamma\delta}(y) S \ket{\psi}$ are determined by tree-level \emph{six}-point amplitudes while $\bra {\psi} S^\dagger \Fop_{\mu\nu}(x) S \ket{\psi}$ and $\bra {\psi} S^\dagger \Rop_{\mu\nu\rho\sigma}(x) S\ket{\psi}$ are determined by \emph{five}-point tree amplitudes.
(We have included factors of the $S$ matrix here as we are mainly interested in variances of operators in the asymptotic future, given an initial state $\ket{\psi}$ in the very far past.)
We must then face the question of how it can be that the variance is negligible.

\subsection{Field strength expectations}

We begin by reviewing the evaluation of single field-strength observables in KMOC. Although this material appears elsewhere in the literature, the calculation will give us an opportunity to introduce notation which we will use in the remainder of our paper. We will discuss the electromagnetic case in some detail. The gravitational case is completely analogous to the electromagnetic case, so we only quote key results.

The electromagnetic field operator is
\[
\Aop_\mu(x) = \frac1{\sqrt{\hbar}} \sum_{\hel=\pm} \int \dPhi(k) \left[ a_\hel(k) \, \polvhconj{\hel}_\mu(k) \, e^{-i \kb \cdot x} + \hc \right] \,.
\label{eq:AopDef}
\]
This is a sum over helicities $\hel$; the notation $\hc$ indicates that the hermitian conjugate of the explicit term should be included. The measure is the Lorentz-invariant phase space measure, namely
\[
\d\Phi(k) \equiv \dd^4 k \, \del(k^2) \, \theta(k^0) \,,
\]
where the carets indicate factors of $2\pi$:
\[
\del(k) \equiv (2\pi) \delta(k), \quad \dd k \equiv \frac{\d k}{2\pi} \,.
\]
Furthermore, we follow reference~\cite{KMOC} by introducing the notation $\kb$ for the wavevector associated with the momentum $k$. These quantities differ only by a factor of $\hbar$:
\[
k = \hbar \kb \,.
\]
As we will soon see, it can at times be useful to keep factors of $\hbar$ explicit --- and we will do so throughout this article. Notice that the polarisation vectors $\polvh{\eta}(k)$ are homogeneous in the momentum, so they can equally be taken to be functions of the momentum or of the wavenumber. We normalise creation and annihilation operators so that
\[
[a_\hel(k), a^\dagger_{\hel'}(k')] = \delta_{\eta \eta'} \Del(k-k') \,,
\]
where $\Del(k)$ is
\[
\Del(k) \equiv 2 k^0 \, \del^3(\v{k} - \v{k}') \,.
\]

Although we could choose to discuss the variance of the field itself, we focus on the variance in the field strength operator. The explicit expression for this operator follows from eq.~\eqref{eq:AopDef} and can be written as
\[
\Fop_{\mu\nu}(x) = \frac{1}{\sqrt{\hbar}} \sum_{\hel=\pm} \int \dPhi(k) \left[-i a_\hel(k) \, \kb_{[\mu}^\vmu \polvhconj{\hel}_{\nu]}(k) \, e^{-i \kb \cdot x} + \hc \right] \,.
\label{eq:explicitF}
\]
Because we frequently encounter antisymmetric operators like this, we have defined the antisymmetrisation symbol
\[
a_{[\mu} b_{\nu]}  \equiv a_{\mu} b_{\nu} -  a_{\nu} b_{\mu}
\]
without any factor of $1/2$.

In classical GR, we define the spacetime metric to be
\[
g_{\mu\nu} = \eta_{\mu\nu} + \kappa h_{\mu\nu} \,.
\]
The constant $\kappa$ is given in terms of Newton's constant $G$ by\footnote{We retain this definition even in units where $\hbar \neq 1$. In that case the dimensions of the fields $A_\mu$ and $h_{\mu\nu}$ are both  $\sqrt{\textrm{mass}/\textrm{length}}$.}
\[
\kappa \equiv \sqrt{32\pi G} \,.
\]
In a quantum description, the corresponding field operator is
\[
\hop_{\mu\nu}(x) = \frac1{\sqrt{\hbar}} \sum_{\hel=\pm} \int \dPhi(k) \left[ a_\hel(k) \, \polvhconj{\hel}_\mu(k) \polvhconj{\hel}_\nu(k) \, e^{-i \kb \cdot x} + \hc \right] \,.
\]
Notice that we have written the polarisation tensor that conventionally appears in the graviton operator as an explicit outer product of two polarisation vectors; this is always possible. We are using the same symbol (namely, $a_\hel$) for annihilation operators in both electromagnetism and gravity; we hope context will make clear which operator is relevant. The linearised Riemann tensor operator follows from conventional definitions and is given by
\[
\Rop_{\mu\nu\rho\sigma}(x) &= \frac{\kappa}{2} \left( \partial_\sigma \partial_{\,[\mu} \hop_{\nu] \,\rho} - \partial_\rho \partial_{\,[\mu} \hop_{\nu] \,\sigma} \right) \\
&= -\frac{\kappa}{2} \frac1{\sqrt{\hbar}} \sum_{\hel=\pm} \int \dPhi(k) \left[ a_\hel(k) \, \kb_{[\mu}^\vmu \polvhconj{\hel}_{\nu]}(k) \, \kb_{[\sigma}^\vmu \polvhconj{\hel}_{\rho]}(k) \, e^{-i \kb \cdot x} + \hc \right] \,.
\]

The basic idea of KMOC~\cite{KMOC} is to take two stable incoming particles, described by a quantum state, which are very distant in the far past. We scatter these particles off one another gently, so that they do not approach too closely (and so that we can describe the scattering, if we choose, using classical field theory). The particles nevertheless radiate. We assume that the particles remain unbound throughout the process. 

We are therefore interested in classical point-like particles which should be located at some position and should have some velocity. In a quantum description, this can be achieved by placing the particles in appropriate wavefunctions which localise the particles both in position and in momentum space. We may write the initial state as
\[
\ket{\psi} = \int \dPhi(p_1) \dPhi(p_2) \, \phi_1 (p_1) \phi_2(p_2) e^{i b \cdot p_1 / \hbar}\, \ket{p_1, p_2} \,.
\label{eq:psi}
\]
This is a two-particle state; the impact parameter $b$ is introduced by translating particle one with respect to the other. The Lorentz invariant phase-space for a massive particle is
\[
\dPhi(p_1) = \dd^4 p_1 \, \del(p_1^2 - m_1^2) \, \theta(p_1^0) \,.
\]
The wavefunctions $\phi_i(p_i)$ for each particle have the property that they localise the particle's position with an uncertainty $\ell_w$ which is characteristic of the wavepacket\footnote{We could choose different uncertainties $\ell_w^{(i)}$ for each particle at the expense of a slightly more complicated notation. There is no reason for us to exploit this freedom in our article, so we choose the simpler case of one common $\ell_w$.}. At the same time, the wavepackets must localise the momenta of the particles to uncertainty $\Delta p$ of order $\hbar / \ell_w$. In the classical limit, we require that these uncertainties are negligible compared to the distance, of order $b$, between the particles
\[
\ell_w \ll b \,,
\]
and also require that the momentum-space uncertainty is negligible compared to the masses of the particles:
\[
\frac{\hbar}{\ell_w} \ll m \Rightarrow l_c \ll \ell_w\,,
\]
where $l_c \equiv \hbar / m$ is a measure of the Compton wavelength of the particle. Thus we have a set of inequalities on the initial parameters:
\[
l_c \ll \ell_w \ll b \,.
\]
More generally, if $\ell_s$ is the distance of closest approach of the particles during scattering, we require that
\[
l_c \ll \ell_w \ll \ell_s \,.
\]
Small-angle scattering has the property that $\ell_s \simeq b$.

Since we will encounter these wavepackets rather frequently, it is sometimes convenient to write the two-particle momentum-space wavefunction as 
\[
\phi_b(p_1, p_2) \equiv \phi(p_1,p_2) e^{i b \cdot p_1 / \hbar} \quad , \quad \phi(p_1,p_2) \equiv \phi_1 (p_1) \phi_2(p_2)\,.
\label{eq:twophi}
\]
We will further use the short-hand notation
\[
\dPhi(p_1, p_2, p_3 , \ldots) \equiv \dPhi(p_1) \dPhi(p_2) \dPhi(p_3) \cdots \,.
\label{eq:multiPhase}
\]
for integrals over the phase space of various particles.

Now that we have an incoming state, we may write the outgoing state in terms of the time evolution operator $U(+\infty, - \infty)$, which is equal to the $S$ matrix:
\[
\ket{\psi}_\textrm{out} &= S \ket{\psi} \\
&=  \int \dPhi (p_1, p_2) \, \phi_b(p_1, p_2)\, S\ket{p_1, p_2} \,.
\]
Since the $S$ matrix can be written in terms of scattering amplitudes, the outgoing state itself can be written in terms of integrals over amplitudes and the incoming state. This simple fact will be a recurring theme in this paper.

Let us now consider the leading-order field strength in our situation. In the far past, we have
\[
\langle \psi | \Fop_{\mu\nu}(x) | \psi \rangle &= \frac{1}{\sqrt{\hbar}} \sum_{\hel} \int \dPhi(k) \left[ \bra{\psi} -i a_\hel(k) \ket{\psi} \, \kb_{[\mu}^\vmu \polvhconj{\eta}_{\nu]}(k) \, e^{-i \kb\cdot x} + \hc \right] \\
&= 0 \,.
\]
The expectation value vanishes because there are no photons in the initial state: $a_\hel(k) \ket{\psi} = 0$. Classically, the interpretation is that the initial state contains no incoming radiation. Notice that the expectation value is not sensitive to the Coulomb fields of the incoming particles; instead, we are computing the asymptotic value of the field at infinity, namely the coefficient of the $1/\textrm{distance}$ piece of the field strength.

In the far future, the expectation value is
\[
\langle \psi | S^\dagger \Fop_{\mu\nu}(x) &S | \psi \rangle \\
&= \frac{1}{\sqrt{\hbar}} \sum_{\hel} \int \dPhi(k) \left[ -i \bra{\psi} S^\dagger a_\hel(k) S \ket{\psi} \, \kb_{[\mu}^\vmu \polvhconj{\eta}_{\nu]}(k) \, e^{-i \kb\cdot x} + \hc \right] \,.
\]
This no longer vanishes. 
We may evaluate it at lowest perturbative order by writing the $S$ matrix in terms of the transition matrix $T$ as $S = 1 + i T$. The matrix elements of $T$ on momentum eigenstates are the scattering amplitudes, which we may organise (as usual in perturbation theory) in terms of the coupling $e$ or $\kappa$ depending on whether we are interested in electrodynamics or gravity. Generically we will denote the perturbative coupling as $g$. Thus we have
\[
\bra{\psi} S^\dagger a_\hel(k) S \ket{\psi} &= i \bra{\psi} (a_\hel(k) T - T^\dagger a_\hel(k)) \ket{\psi} + \bra{\psi} T^\dagger a_\hel(k) T \ket{\psi}  \\
&=  i \bra{\psi} a_\hel(k) T  \ket{\psi}  + \bra{\psi} T^\dagger a_\hel(k) T \ket{\psi}  \\
&\simeq   i \bra{\psi} a_\hel(k) T \ket{\psi} \,.
\]
In the middle line above, we used the fact that $a_\hel(k) \ket{\psi} = 0$; in the last line we neglected the term involving two $T$ matrices which does not contribute at lowest order by counting powers of $g$. 

Further expanding the state using eq.~\eqref{eq:psi}, and taking advantage of the short-hand notation of eqs.~\eqref{eq:twophi} and~\eqref{eq:multiPhase}, we may write
\[
\langle \psi | S^\dagger \Fop_{\mu\nu}(x) S| \psi \rangle &= 2 \Re \frac1{\sqrt{\hbar}} \sum_{\hel} \int \dPhi(p_1', p_2', p_1, p_2, k) \phi_b^*(p_1', p_2') \phi_b(p_1, p_2) \times \\
& \hspace{120pt}\times \bra{k^\eta, p_1', p_2' } T \ket{p_1, p_2} \, \kb^\vmu_{[\mu} \polvhconj{\eta}_{\nu]} \, e^{-i \kb \cdot x} \,.
\label{eq:F_expectation}
\]
The matrix element $\bra{k^\eta, p_1', p_2' } T \ket{p_1, p_2}$ is, at lowest order, a five-point tree amplitude so it is proportional to $g^3$. This is consistent with a classical analysis of the outgoing radiation field. 

In GR, the equivalent expression is
\[
\langle \psi | S^\dagger \Rop_{\mu\nu\rho\sigma}(x) S| \psi \rangle &= 2 \Re \frac{-i}{\sqrt{\hbar}} \frac{\kappa}2 \sum_{\hel} \int \dPhi(p_1', p_2', p_1, p_2, k) \phi_b^*(p_1', p_2') \phi_b(p_1, p_2) \times \\
& \hspace{80pt}\times \bra{k^\eta, p_1', p_2' } T \ket{p_1, p_2} \, \kb^\vmu_{[\mu} \polvhconj{\eta}_{\nu]} \, \kb^\vmu_{[\sigma} \polvhconj{\eta}_{\rho]} \, e^{-i \kb \cdot x} \,.
\]

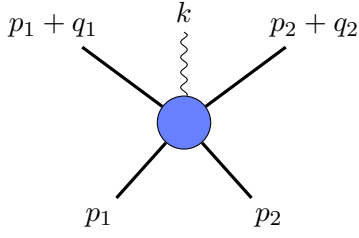
\begin{figure}
\centering
\begin{tikzpicture}
\begin{feynman}
	\vertex (v);
	\node [below left=1 and 0.8 of v] (i1) {$p_1$};
	\node [below right=1 and 0.8 of v] (i2) {$p_2$};
	\node [above left=1 and 1 of v] (o1) {$p_1+q_1$};
	\node [above right=1 and 1 of v] (o2) {$p_2+q_2$};
	\vertex [above = 1.2 of v] (g) {$k$};
	\draw [very thick] (i1) -- (v) -- (o1);
	\draw [very thick] (i2) -- (v) -- (o2);
	\diagram*{(v) -- [photon] (g)};
	\filldraw [color=white] (v) circle [radius=10pt];
	\filldraw [fill=allOrderBlue] (v) circle [radius=10pt];
\end{feynman}	
\end{tikzpicture}
\caption{The kinematic configuration we choose for the five-point amplitude which determines the leading-order radiation field.}
\label{fig:fivepoints}
\end{figure}

It will be useful for us to simplify these expressions further, following again the discussion of reference~\cite{KMOC} at leading order.
The matrix element $\bra{k^\eta, p_1', p_2' } T \ket{p_1, p_2}$ is the amplitude times a momentum-conserving delta function; our expectation value instructs us to integrate over all momenta in the amplitude. We may relabel these external momenta as shown in fig.~\ref{fig:fivepoints}. The measure can then be written as
\[
\dPhi(p_1', p_2', p_1, p_2, k) = \dPhi(p_1, p_2, k) \dd^4 q_1 \dd^4 q_2 \del(2 p_1 \cdot q_1 + q_1^2) \del(2 p_2 \cdot q_2 + q_2^2) \,.
\]
In this form, the overall momentum-conserving delta function reads $\del^4(q_1 + q_2 +k)$. Now, in the classical regime the photon momentum is of order $\hbar$, as are the momentum mismatches $q_1$ and $q_2$. The $q_i^2$ terms in the delta functions above are therefore small shifts compared to the width (or order $1/\ell_w$) of the wavefunctions in the expectation values. As the observable has the structure of a convolution of the sharply-peaked wavefunctions multiplied by delta functions and otherwise smooth functions, we may neglect the $q_i^2$ shifts\footnote{As we will see later in this article, care must be taken with these delta functions at higher order. Here we work at leading order, so the situation is simple.}. Similarly, the wavefunctions themselves are
\[
\phi_b^*(p_1+q_1, p_2+q_2) \phi_b(p_1 , p_2) = \phi^*(p_1+q_1, p_2+q_2) \phi(p_1, p_2) e^{- i  q_1 \cdot b/\hbar} \,. 
\] 
We may neglect the $q_i$ shifts in the wavefunctions $\phi_i^*(p_i + q_i)$ because the momenta $q_i$ are negligible compared to the width of the wavefunctions. Thus the field strength becomes
\[
\langle \psi | S^\dagger \Fop_{\mu\nu}(x) S| \psi \rangle &= 2 \Re \frac1{\sqrt{\hbar}} \sum_{\hel} \int \dPhi(p_1, p_2, k) |\phi(p_1, p_2)|^2  \, \dd^4 q_1 \dd^4 q_2 \, \del_\sigma(2 p_1 \cdot q_1) \del_\sigma(2 p_2 \cdot q_2) \,   \\
& \times\, \mathcal{A}_{5,0}(p_1 p_2 \rightarrow p_1+q_1, p_2+q_2, k^\eta )\,  \del^4(k +q_1 +q_2)  \, \kb^\vmu_{[\mu} \polvhconj{\eta}_{\nu]} \, e^{-i (\kb \cdot x + \qb_1 \cdot b)} \,.
\label{eq:fivepointstep}
\]
The notation $\del_\sigma(x)$ indicates that the delta functions have been ``broadened'' so that their width is of order $\sigma \sim \hbar / \ell_w$. In our point-particle treatment, we neglect this scale in the rest of this paper. The amplitude $ \mathcal{A}_{5,0}(p_1 p_2 \rightarrow p_1+q_1, p_2+q_2, k^\eta )$ is a five-point, tree amplitude. More generally, we will denote $n$ point $L$ loop amplitudes as $\mathcal{A}_{n,L}$ (in gauge theory) or $\mathcal{M}_{n,L}$ (in gravity).

The integral in eq.~\eqref{eq:fivepointstep} still depends on the wavefunctions of the particles. But now the role of the wavefunctions is transparent: they are steeply-peaked functions of the momenta $p_1$ and $p_2$ which allows us to simply replace these variables of integration with the incoming classical momenta $m_1 v_1$ and $m_2 v_2$. We thus write the field expectation value as
\[
\langle \psi | S^\dagger & \Fop_{\mu\nu}(x) S| \psi \rangle = 2 \Re \hbar^{7/2} \sum_{\hel} \Lexp \int \dPhi(\kb) \, \dd^4 \qb_1 \dd^4 \qb_2 \, \del(2 p_1 \cdot \qb_1) \del(2 p_2 \cdot \qb_2) \,   \\
& \times \, \mathcal{A}_{5,0}(p_1 p_2 \rightarrow p_1+q_1, p_2+q_2, k^\eta )\,  \del^4(\kb + \qb_1 + \qb_2)  \, \kb^\vmu_{[\mu} \polvhconj{\eta}_{\nu]} \, e^{-i (\kb \cdot x +\qb_1 \cdot b)} \Rexp \,.
\label{eq:F_expectation2}
\]
The double-angle brackets are shorthand notation that instructs us to evaluate the momenta at their classical values, and remind us to take care of $q^2$ shifts in the delta functions. Recalling that  $k$, $q_1$ and $q_2$ are all of order $\hbar$, we have scaled out all the $\hbar$ dependence except that of the scattering amplitude itself.
Similarly, in gravity, one finds
\[
\hspace{-5pt}
&\langle \psi | S^\dagger \Rop_{\mu\nu\rho\sigma}(x) S| \psi \rangle = -2 \Re \hbar^{7/2}  \frac{i\kappa}2 \sum_{\hel}\Lexp \int \dPhi(\kb) \, \dd^4 \qb_1 \dd^4 \qb_2 \, \del(2 p_1 \cdot \qb_1) \del(2 p_2 \cdot \qb_2)\,   \\
&  \times \mathcal{M}_{5,0}(p_1 p_2 \rightarrow p_1+q_1, p_2+q_2, k^\eta )  \del^4(\kb + \qb_1 + \qb_2)   \, \kb^\vmu_{[\mu} \polvhconj{\eta}_{\nu]} \, \kb^\vmu_{[\sigma} \polvhconj{\eta}_{\rho]} \, e^{-i (\kb \cdot x +\qb_1 \cdot b)} \Rexp \,.
\]
For these expressions to make sense classically, it better be that the overall $\hbar$ dependence of the amplitudes cancels that of the observable. Indeed we may write
\[
\mathcal{A}_{5,0}(p_1 p_2 \rightarrow p_1+q_1, p_2+q_2, k^\eta ) &= \hbar^{-7/2} \mathcal{A}_{5,0}^{(0)}(p_1 p_2 \rightarrow p_1+q_1, p_2+q_2, k^\eta ) + \mathcal{O}(\hbar)\,\\
\mathcal{M}_{5,0}(p_1 p_2 \rightarrow p_1+q_1, p_2+q_2, k^\eta ) &= \hbar^{-7/2} \mathcal{M}_{5,0}^{(0)}(p_1 p_2 \rightarrow p_1+q_1, p_2+q_2, k^\eta ) + \mathcal{O}(\hbar) \,,
\label{eq:seriesTaster}
\]
where the quantities $\mathcal{A}_{5,0}^{(0)}$ and $\mathcal{M}_{5,0}^{(0)}$ are independent of $\hbar$, as was noticed in~\cite{KMOC}. 
We will return to this structure below.

The physical interpretation of these expectation values is that they compute the radiative part of the field at large distances. To see this explicitly, the $\kb$ integral needs to be performed taking advantage of the large distance between the point of measurement $x$ and the particles. The integration can be performed using textbook methods and was recently reviewed in detail in ref.~\cite{Cristofoli:2021vyo}. The question of central interest to us in this section, however, is to compute the uncertainty in the field strength; to do so, we turn to computing the expectation of two field strengths.

\subsection{Expectation of two field strengths}
\label{sec:expectation_of_two_F}
\begin{figure}
\centering
\begin{tikzpicture}
\begin{feynman}
	\vertex (v);
	\node [below left=1 and 0.8 of v] (i1) {$p_1$};
	\node [below right=1 and 0.8 of v] (i2) {$p_2$};
	\node [above left=1 and 1 of v] (o1) {$p_1 + q_1$};
	\node [above right=1 and 1 of v] (o2) {$p_2 +q_2$};
	\vertex [above right= 1.2 and 0.3 of v] (g1) {$k$};
	\vertex [above left= 1.2 and 0.3 of v] (g2) {$k'$};
	\draw [very thick] (i1) -- (v) -- (o1);
	\draw [very thick] (i2) -- (v) -- (o2);
	\diagram*{(v) -- [photon] (g1)};
	\diagram*{(v) -- [photon] (g2)};
	\filldraw [color=white] (v) circle [radius=10pt];
	\filldraw [fill=allOrderBlue] (v) circle [radius=10pt];
\end{feynman}	
\end{tikzpicture}
\caption{The kinematic configuration we choose for the six-point amplitude appearing at leading-order expectation of a pair of field strength operators.}
\label{fig:sixpoints}
\end{figure}
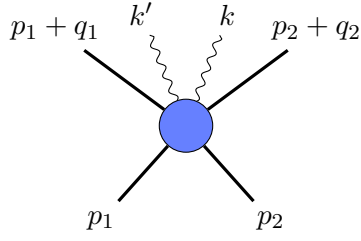

It will be quite straightforward for us to compute expectations of products of operators using precisely the methods of the previous subsection. In electrodynamics, we need to compute
\[
\langle \psi | S^\dagger \Fop_{\mu\nu}(x)  \Fop_{\rho \sigma}(y) S| \psi \rangle = \frac{1}{\hbar} \sum_{\hel, \hel'} \int& \dPhi(k',k) \bra{\psi} S^\dagger  \left[-i a_\eta(k) \, \kb^\vmu_{[\mu} \polvhconj{\eta}_{\nu]} \, e^{-i \kb \cdot x} + \hc \right] \\
& \hspace{20pt} \times \left[-i a_{\eta'}(k') \, \kb'_{[\rho} \polvprimehconj{\eta'}_{\sigma]} \, e^{-i \kb' \cdot y} + \hc \right] S \ket{\psi} \,.
\]
Working at lowest order, and taking advantage of the fact that $a_\eta(k) \ket{\psi} = 0$, the expectation simplifies to
\[
\langle \psi | S^\dagger \Fop_{\mu\nu}(x)  \Fop_{\rho \sigma}(y) S| \psi \rangle =- \frac{2}{\hbar} \Re \sum_{\hel, \hel'} \int& \dPhi(k',k) \bra{\psi} a_\eta(k) a_{\eta'}(k') iT \ket{\psi} \\
& \hspace{30pt} \times \kb^\vmu_{[\mu} \polvhconj{\eta}_{\nu]} \kb'_{[\rho} \polvprimehconj{\eta'}_{\sigma]} \, e^{-i (\kb \cdot x+ \kb' \cdot y)} \,,
\]
up to a purely quantum single-photon effect~\cite{Cristofoli:2021vyo}.
Expanding the wavefunctions, we encounter the matrix element $\bra{p_1', p_2'} a_\eta(k) a_{\eta'}(k') T \ket{p_1, p_2}$: a six-point tree amplitude. The classical limit is determined precisely as in the previous section with the result 
\[
\hspace{-3pt}
\langle \psi | S^\dagger \Fop_{\mu\nu}(x)  \Fop_{\rho\sigma}(y) S| \psi \rangle &= - 2 \hbar^5 \Re \sum_{\hel, \hel'} \Lexp \int \dPhi(\kb',\kb) \, \dd^4 \qb_1 \dd^4 \qb_2 \, \del(2 p_1 \cdot \qb_1) \del(2 p_2 \cdot \qb_2)
 \\
& \hspace{-20pt}\times \, i\mathcal{A}_{6,0}
 \,  \del^4(\kb + \kb' + \qb_1 + \qb_2)  \, \kb^\vmu_{[\mu} \polvhconj{\eta}_{\nu]}\kb'_{[\rho} \polvprimehconj{\eta'}_{\sigma]} \, e^{-i (\kb \cdot x + \kb' \cdot y+\qb_1 \cdot b)} \Rexp \,.
 \label{eq:twoFs}
\]
The amplitude is shown in figure~\ref{fig:sixpoints}. 

Similarly, in gravity, we find
\[
\langle \psi | S^\dagger &\Rop_{\mu\nu\rho\sigma}(x)  \Rop_{\alpha\beta\gamma\delta}(y) S | \psi \rangle \\
&= - 2 \hbar^5 \Re \left(-i \frac{\kappa}{2} \right)^2 \sum_{\hel, \hel'} \Lexp \int \dPhi(\kb',\kb) \, \dd^4 \qb_1 \dd^4 \qb_2 \, \del(2 p_1 \cdot \qb_1) \del(2 p_2 \cdot \qb_2) \, \mathcal{M}_{6,0}
 \\
& \hspace{0pt}\times
 \,  \del^4(\kb + \kb' + \qb_1 + \qb_2)  \, \kb^\vmu_{[\mu} \polvhconj{\eta}_{\nu]}\kb_{[\rho} \polvhconj{\eta}_{\sigma]} \, \kb'_{[\alpha} \polvprimehconj{\eta'}_{\beta]}\kb'_{[\gamma} \polvprimehconj{\eta'}_{\delta]}\, e^{-i (\kb \cdot x + \kb' \cdot y+\qb_1 \cdot b)} \Rexp \,.
\label{eq:twoRs}
\]
In both cases, the expectation of two field strengths is given to leading order in $g$ by a tree-level six-point amplitude.

\subsection{Negligible variance?}

We have now seen explicitly that the expectation of a single field strength is determined by a five-point amplitude, while the expectation of two field strengths is a six-point tree amplitude at lowest order in the coupling $g$. But for the uncertainty in the field strength to be negligible, we need the variance to be negligible. How can this happen? 

Let us count powers of the coupling in the electromagnetic variance. The product of two field-strength expectations is
\[
\bra {\psi} S^\dagger \Fop_{\mu\nu}(x) S \ket{\psi} \bra{\psi} S^\dagger \Fop_{\rho\sigma}(y) S \ket{\psi} \sim (\mathcal{A}_{5,0})^2 \sim (g^3)^2 \,.
\]
while the expectation of two field strengths is
\[
\bra{\psi} S^\dagger \Fop_{\mu\nu}(x) \Fop_{\rho \sigma}(y) S \ket{\psi} \sim \mathcal{A}_{6,0} \sim g^4 \,.
\]
Thus the situation seems to very bad: the variance 
\[
\bra {\psi} S^\dagger \Fop_{\mu\nu}(x) S \ket{\psi} \bra{\psi} S^\dagger \Fop_{\rho\sigma}(y) S \ket{\psi} - \bra{\psi} S^\dagger \Fop_{\mu\nu}(x) \Fop_{\rho \sigma}(y) S \ket{\psi}
\] 
seems to be dominated by the expectation of two field-strength operators! For the classical limit to emerge as expected, we need the six-point tree amplitude to be suppressed somehow. 

\begin{figure}
\centering
\begin{tikzpicture}[scale=1] 
\begin{feynman}
  \vertex (o1) {};
  \vertex [right=5 of o1] (o2) {};
  \vertex [below=4 of o1] (i1) {};
  \vertex [below=4 of o2] (i2) {};
  \vertex [below right=2 and 1.5 of o1] (v1);
  \vertex [below left=2 and 1.5 of o2] (v2);
  \draw [thick] (i1) -- (v1) node [midway, shape=coordinate] (v3);
  \draw [thick] (v1) -- (o1) node [midway, shape=coordinate] (v4);
  \draw [thick] (i2) -- (v2) -- (o2);
  \vertex [above=2 of v1] (g1) {};
  \vertex [below left=1.8 and 0.4 of o1] (g2) {};
  \diagram* {
    (v1) -- [photon] (v2);
    (v3) -- [photon] (g2);
    (v4) -- [photon] (g1);
  };
  \filldraw [fill=black] (v1) circle [radius=1pt];
  \filldraw [fill=black] (v2) circle [radius=1pt];
  \filldraw [fill=black] (v3) circle [radius=1pt];
  \filldraw [fill=black] (v4) circle [radius=1pt];
  \node [below right=-0.1 and -0.1 of v3] {$\hbar^{-\frac12}$};
  \node [left=0.1 of v4] {$\hbar^{-\frac12}$};
  \node [left=0.1 of v1] {$\hbar^{-\frac12}$};
  \node [right=0.1 of v2] {$\hbar^{-\frac12}$};
  \node [below right=0.1 and 0.75 of v1] {$\hbar^{-2}$};
  \node [below right=0.05 and 0.3 of v4] (t) {$\hbar^{-1}$};
  \node [below=1.2 of t] {$\hbar^{-1}$};
\end{feynman}
\end{tikzpicture} 
\caption{A sample Feynman diagram contributing to the six-point tree amplitude, indicating powers of $\hbar$ assigned by naive power counting to the propagators and vertices.}
\label{fig:sixpointdiagram}
\end{figure}
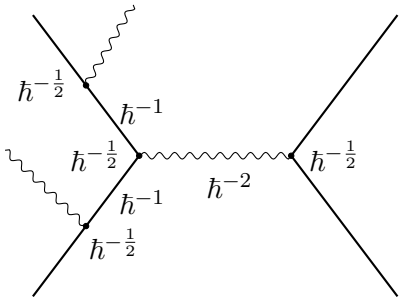

One possibility is that it is suppressed by powers of $\hbar$, but naively that is not the case. Consider, for example, the 6-point Feynman diagram shown in figure~\ref{fig:sixpointdiagram}. We can count the powers of $\hbar$ associated to the diagram as follows. Each vertex contributes a factor $\hbar^{-\frac12}$; this is because the dimensionless coupling is $e/\sqrt{\hbar}$ or $\kappa / \sqrt{\hbar}$. Each messenger propagator contributes a factor $\hbar^{-2}$ because messenger momenta are of order $\hbar$. Meanwhile each massive propagator contributes a factor $\hbar^{-1}$; this arises since the momenta flowing through these propagators are a sum of a massive on-shell momentum $p$ and a messenger momentum $k$, so that the propagator denominator is $(p+k)^2 - m^2 \simeq 2 \hbar \, p \cdot \kb$. We conclude that the six-point tree amplitude contains terms of order $\hbar^{-6}$. Referring back to eq.~\eqref{eq:twoFs} or eq.~\eqref{eq:twoRs} for the expectation of two field strengths, we see that the observables contribute a total of $\hbar^{+5}$. Based on this counting, the observable seems to scale as $\hbar^{-1}$, which would be a serious obstruction to the emergence of a classical limit. Evidently there is more to understand here: the powers of $\hbar$ don't make sense.

It is a familiar story that power counting Feynman diagrams can be misleading: upon combining diagrams to evaluate an amplitude, there can be cancellations. In fact this already happens in the case of the five-point tree; there, naive power counting suggests that the amplitude scales as $\hbar^{-9/2}$ but in fact the leading term in the amplitude is of order $\hbar^{-7/2}$~\cite{Luna:2017dtq}\footnote{In reference~\cite{Luna:2017dtq}, the analysis of the five-point tree amplitude was performed in gravity using the large mass expansion, which is equivalent to expanding in small $\hbar$.}. The question, then, of the fate of the six-point tree amplitude in the expectation value of two field strengths becomes a question of the overall $\hbar$ scaling of six-point tree amplitudes in QED and gravity. We will shortly demonstrate explicitly that the QED amplitude in fact scales as $\hbar^{-4}$; the gravitational case will be discussed in reference~\cite{Britto:2021pud}. Two powers of $\hbar$ cancel; consequently the contribution of the six-point tree to the variance is entirely at the quantum level. 

It is amusing that at next-to-leading order in the perturbative coupling $g$, namely order $g^6$, the expectation value of two field strengths is sensitive to one-loop six-point amplitudes and to products of two tree five-point amplitudes\footnote{We learned from G. Veneziano that the use of a cut 6-point function to compute observables which are quadratic in the fields (such as number or energy densities in inclusive cross-sections) goes back to Mueller's generalized optical theorem from the seventies (see e.g. \cite{PhysRevD.4.906} and references therein).}. These products of tree-level five-point amplitudes can be viewed as the cut of a six-point one-loop amplitude. It is easy to check that these scale as $\hbar^{-5}$, so in this sense they are \emph{enhanced} relative to the six-point tree amplitude. This is as desired for negligible uncertainty: 
\[
\bra {\psi} &S^\dagger \Fop_{\mu\nu}(x) S \ket{\psi} \bra{\psi} S^\dagger  \Fop_{\rho \sigma}(y) S \ket{\psi} \sim (\mathcal{A}_{5,0})^2 \sim (g^3)^2 \,; \\
\bra{\psi} & S^\dagger \Fop_{\mu\nu}(x) \Fop_{\rho \sigma}(y) S \ket{\psi} \sim \mathcal{A}_{6, 1} \sim (\mathcal{A}_{5,0})^2 \sim (g^3)^2 \,.
\]
Thus $\bra{\psi} S^\dagger \Fop_{\mu\nu}(x) \Fop_{\rho \sigma}(y)  S\ket{\psi} \sim \bra {\psi} S^\dagger \Fop_{\mu\nu}(x) S \ket{\psi} \bra{\psi} S^\dagger \Fop_{\rho \sigma}(y) S \ket{\psi}$, as one would expect classically.

\subsection{Explicit six-point tree amplitudes}
\label{sec:6_point_tree}
We now compute the leading contribution in $\hbar$ of the six-point tree amplitude. We will discuss the computation explicitly in electromagnetism, and explain only the mechanism for cancellation of apparent excess powers of $\hbar$ in gravity (see also \cite{Britto:2021pud}).

Suppose particle 1 has charge $Q_1$ while particle 2 has charge $Q_2$. Then there are three gauge-invariant six-point tree partial amplitudes: 
\[
\mathcal{A}_{6,0}(p_1 + q_1, p_2 + q_2 \rightarrow p_1, p_2, k_1, k_2) = Q_1^3 Q_2 \, {A}_{(3,1)} + Q_1^2 Q_2^2 \, {A}_{(2,2)} + Q_1 Q_2^3 \, {A}_{(1,3)} \,. 
\label{eq:amp_charge_sectors}
\]
The ``charge-ordered'' partial amplitudes $A_{(3,1)}$, $A_{(2,2)}$, and ${A}_{(1,3)}$ are analogues of color-ordered amplitudes in gauge theory, which motivates our choice of notation.

Evidently there can be no cancellation of powers of $\hbar$ between these partial amplitudes because of the different powers of the charges. Thus the problem reduces to computing the leading-in-$\hbar$ terms in these amplitudes. There are two partial amplitudes to consider, since ${A}_{(1,3)}$ can be obtained from ${A}_{(3,1)}$ by trivially swapping the labels 1 and 2.

We have performed two separate computations of these partial amplitudes. Firstly, we made use of standard automated tools to directly compute the amplitude in full detail. Specifically, we used FeynArts~\cite{Hahn:2000kx} to create a model for scalar QED, extracting the Feynman rules directly from the Lagrangian. FeynCalc~\cite{Shtabovenko:2016sxi,Shtabovenko:2020gxv} can automatically generate all the topologies relevant for the calculation of $\amp_{6,0}$. There are 42 diagrams to be computed, and FeynCalc provided direct automatic expressions for each of these. We processed our expressions further in Mathematica with the tensor package xAct~\cite{xAct}, which helps to extract the classical limit. The final result for $\amp_{6,0}$ at leading order in the $\hbar$ expansion is provided as an attachment to this paper and is indeed of order $\hbar^{-4}$.

To gain further insight we performed a separate computation of the partial amplitudes $A_{(3,1)}$ and $A_{(2,2)}$ in a convenient gauge which greatly reduced the labour necessary to see that two powers of $\hbar$ cancel.
The gauge we chose (referring to the momentum routing in equation~\eqref{eq:amp_charge_sectors}) is
\[
p_1\cdot \varepsilon(k_i) = 0 \,\, \text{for}\,\, i = 1,2,
\]
where $p_1$ is the momentum of particle 1 while $k_i$ is the outgoing momentum of photon $i$.

The effect of this choice is twofold; it removes many diagrams from the calculation and those that remain get an $\hbar$ enhancement from each emission vertex. For example, consider $A_{(3,1)}$: in this case, the emitted photon is radiated from particle 1. With our momentum labelling convention the emission vertices will produce factors of the form $(2p_1+\hbar  \bar Q))\cdot \varepsilon(k_i)$, where $\bar Q$ is some combination of wavenumbers. The first part vanishes leaving only  the $\hbar$ enhanced $\varepsilon(k_i)\cdot \bar Q$ term. Any diagram with a photon emitted from the outgoing line of particle 1 vanishes, as it is proportional to $(2p_1+\hbar \bar k _i)\cdot \varepsilon(k_i) = 0$. In what follows we will write $\varepsilon(k_i) = \varepsilon_i$. We will also suppress the $i\epsilon$ factors in the propagators\footnote{The $i \epsilon$ factors are often important --- and will play an important role in section~\ref{sec:one-loop} --- but in this computation they are spectators.}; they all implicitly come with $+i\epsilon$. Finally we will refer to each diagram contributing to the amplitude by, for example, $D{(3,1)}$ so that  
\[iA_{(3,1)} = iD_{(3,1)cubic}+iD_{(3,1)quartic}.\]
Some of the sub-amplitudes are given by a single diagram, whereas others are made up of multiple diagrams.

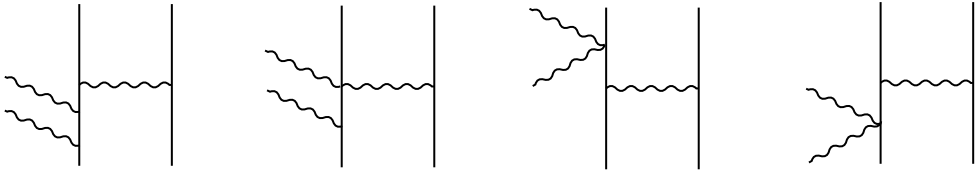
\begin{figure}[t]
\begin{center}
\tikzset{every picture/.style={line width=0.75pt}}

\begin{tikzpicture}[x=0.75pt,y=0.75pt,yscale=-1,xscale=1]

\draw    (350.37,3599.73) -- (350.37,3681.06) ;
%Straight Lines [id:da5869845021412388] 
\draw    (396.58,3599.73) -- (396.58,3681.06) ;
%Straight Lines [id:da5081599851872434] 
\draw    (350.37,3640.4) .. controls (352.04,3638.73) and (353.7,3638.73) .. (355.37,3640.4) .. controls (357.04,3642.07) and (358.7,3642.07) .. (360.37,3640.4) .. controls (362.04,3638.73) and (363.7,3638.73) .. (365.37,3640.4) .. controls (367.04,3642.07) and (368.7,3642.07) .. (370.37,3640.4) .. controls (372.04,3638.73) and (373.7,3638.73) .. (375.37,3640.4) .. controls (377.04,3642.07) and (378.7,3642.07) .. (380.37,3640.4) .. controls (382.04,3638.73) and (383.7,3638.73) .. (385.37,3640.4) .. controls (387.04,3642.07) and (388.7,3642.07) .. (390.37,3640.4) .. controls (392.04,3638.73) and (393.7,3638.73) .. (395.37,3640.4) -- (396.58,3640.4) -- (396.58,3640.4) ;
%Straight Lines [id:da8583439364706429] 
\draw    (349.65,3618.36) .. controls (347.42,3619.14) and (345.92,3618.42) .. (345.14,3616.2) .. controls (344.36,3613.97) and (342.86,3613.25) .. (340.63,3614.03) .. controls (338.41,3614.81) and (336.91,3614.09) .. (336.13,3611.87) .. controls (335.35,3609.64) and (333.85,3608.92) .. (331.62,3609.7) .. controls (329.4,3610.48) and (327.9,3609.76) .. (327.11,3607.54) .. controls (326.34,3605.31) and (324.84,3604.59) .. (322.61,3605.37) .. controls (320.39,3606.15) and (318.89,3605.43) .. (318.1,3603.21) .. controls (317.32,3600.98) and (315.82,3600.26) .. (313.59,3601.04) -- (312.11,3600.33) -- (312.11,3600.33) ;
%Straight Lines [id:da5155974451130667] 
\draw    (349.65,3618.36) .. controls (349.04,3620.64) and (347.6,3621.48) .. (345.32,3620.87) .. controls (343.05,3620.26) and (341.61,3621.1) .. (341,3623.37) .. controls (340.39,3625.65) and (338.95,3626.49) .. (336.67,3625.88) .. controls (334.39,3625.27) and (332.95,3626.11) .. (332.35,3628.39) .. controls (331.74,3630.67) and (330.3,3631.51) .. (328.02,3630.9) .. controls (325.74,3630.29) and (324.3,3631.12) .. (323.69,3633.4) .. controls (323.09,3635.68) and (321.65,3636.52) .. (319.37,3635.91) .. controls (317.09,3635.3) and (315.65,3636.14) .. (315.04,3638.42) -- (313.72,3639.18) -- (313.72,3639.18) ;
%Straight Lines [id:da2995332576906975] 
\draw    (487.37,3596.96) -- (487.37,3678.29) ;
%Straight Lines [id:da987046981229442] 
\draw    (533.58,3596.96) -- (533.58,3678.29) ;
%Straight Lines [id:da34880017176969336] 
\draw    (487.37,3637.62) .. controls (489.04,3635.95) and (490.7,3635.95) .. (492.37,3637.62) .. controls (494.04,3639.29) and (495.7,3639.29) .. (497.37,3637.62) .. controls (499.04,3635.95) and (500.7,3635.95) .. (502.37,3637.62) .. controls (504.04,3639.29) and (505.7,3639.29) .. (507.37,3637.62) .. controls (509.04,3635.95) and (510.7,3635.95) .. (512.37,3637.62) .. controls (514.04,3639.29) and (515.7,3639.29) .. (517.37,3637.62) .. controls (519.04,3635.95) and (520.7,3635.95) .. (522.37,3637.62) .. controls (524.04,3639.29) and (525.7,3639.29) .. (527.37,3637.62) .. controls (529.04,3635.95) and (530.7,3635.95) .. (532.37,3637.62) -- (533.58,3637.62) -- (533.58,3637.62) ;
%Straight Lines [id:da7406050080861872] 
\draw    (487.66,3656.91) .. controls (487.05,3659.18) and (485.6,3660.01) .. (483.33,3659.4) .. controls (481.05,3658.79) and (479.6,3659.62) .. (478.99,3661.9) .. controls (478.38,3664.17) and (476.93,3665) .. (474.66,3664.39) .. controls (472.39,3663.78) and (470.94,3664.61) .. (470.32,3666.88) .. controls (469.71,3669.15) and (468.26,3669.98) .. (465.99,3669.37) .. controls (463.71,3668.76) and (462.26,3669.59) .. (461.65,3671.87) .. controls (461.04,3674.14) and (459.59,3674.97) .. (457.32,3674.36) .. controls (455.05,3673.75) and (453.6,3674.58) .. (452.99,3676.85) -- (451.61,3677.64) -- (451.61,3677.64) ;
%Straight Lines [id:da47390134516779336] 
\draw    (487.66,3657.95) .. controls (485.45,3658.76) and (483.94,3658.05) .. (483.13,3655.84) .. controls (482.32,3653.63) and (480.81,3652.92) .. (478.6,3653.73) .. controls (476.39,3654.54) and (474.87,3653.83) .. (474.06,3651.62) .. controls (473.25,3649.41) and (471.74,3648.7) .. (469.53,3649.51) .. controls (467.32,3650.32) and (465.81,3649.61) .. (465,3647.4) .. controls (464.19,3645.19) and (462.68,3644.48) .. (460.47,3645.29) .. controls (458.26,3646.1) and (456.74,3645.39) .. (455.93,3643.18) .. controls (455.12,3640.97) and (453.61,3640.26) .. (451.4,3641.07) -- (450.21,3640.52) -- (450.21,3640.52) ;
%Straight Lines [id:da7467897550723124] 
\draw    (218.37,3598.73) -- (218.37,3680.06) ;
%Straight Lines [id:da8279059716437647] 
\draw    (264.58,3598.73) -- (264.58,3680.06) ;
%Straight Lines [id:da05972711908959227] 
\draw    (218.37,3639.4) .. controls (220.04,3637.73) and (221.7,3637.73) .. (223.37,3639.4) .. controls (225.04,3641.07) and (226.7,3641.07) .. (228.37,3639.4) .. controls (230.04,3637.73) and (231.7,3637.73) .. (233.37,3639.4) .. controls (235.04,3641.07) and (236.7,3641.07) .. (238.37,3639.4) .. controls (240.04,3637.73) and (241.7,3637.73) .. (243.37,3639.4) .. controls (245.04,3641.07) and (246.7,3641.07) .. (248.37,3639.4) .. controls (250.04,3637.73) and (251.7,3637.73) .. (253.37,3639.4) .. controls (255.04,3641.07) and (256.7,3641.07) .. (258.37,3639.4) .. controls (260.04,3637.73) and (261.7,3637.73) .. (263.37,3639.4) -- (264.58,3639.4) -- (264.58,3639.4) ;
%Straight Lines [id:da44782634737056926] 
\draw    (217.65,3639.36) .. controls (215.42,3640.14) and (213.92,3639.42) .. (213.14,3637.2) .. controls (212.36,3634.97) and (210.86,3634.25) .. (208.63,3635.03) .. controls (206.41,3635.81) and (204.91,3635.09) .. (204.13,3632.87) .. controls (203.35,3630.64) and (201.85,3629.92) .. (199.62,3630.7) .. controls (197.4,3631.48) and (195.9,3630.76) .. (195.11,3628.54) .. controls (194.34,3626.31) and (192.84,3625.59) .. (190.61,3626.37) .. controls (188.39,3627.15) and (186.89,3626.43) .. (186.1,3624.21) .. controls (185.32,3621.98) and (183.82,3621.26) .. (181.59,3622.04) -- (180.11,3621.33) -- (180.11,3621.33) ;
%Straight Lines [id:da055404480629124064] 
\draw    (87.37,3597.96) -- (87.37,3679.29) ;
%Straight Lines [id:da24152746394509816] 
\draw    (133.58,3597.96) -- (133.58,3679.29) ;
%Straight Lines [id:da308065949544837] 
\draw    (87.37,3638.62) .. controls (89.04,3636.95) and (90.7,3636.95) .. (92.37,3638.62) .. controls (94.04,3640.29) and (95.7,3640.29) .. (97.37,3638.62) .. controls (99.04,3636.95) and (100.7,3636.95) .. (102.37,3638.62) .. controls (104.04,3640.29) and (105.7,3640.29) .. (107.37,3638.62) .. controls (109.04,3636.95) and (110.7,3636.95) .. (112.37,3638.62) .. controls (114.04,3640.29) and (115.7,3640.29) .. (117.37,3638.62) .. controls (119.04,3636.95) and (120.7,3636.95) .. (122.37,3638.62) .. controls (124.04,3640.29) and (125.7,3640.29) .. (127.37,3638.62) .. controls (129.04,3636.95) and (130.7,3636.95) .. (132.37,3638.62) -- (133.58,3638.62) -- (133.58,3638.62) ;
%Straight Lines [id:da8937170501440526] 
\draw    (87.66,3651.95) .. controls (85.45,3652.76) and (83.94,3652.05) .. (83.13,3649.84) .. controls (82.32,3647.63) and (80.81,3646.92) .. (78.6,3647.73) .. controls (76.39,3648.54) and (74.87,3647.83) .. (74.06,3645.62) .. controls (73.25,3643.41) and (71.74,3642.7) .. (69.53,3643.51) .. controls (67.32,3644.32) and (65.81,3643.61) .. (65,3641.4) .. controls (64.19,3639.19) and (62.68,3638.48) .. (60.47,3639.29) .. controls (58.26,3640.1) and (56.74,3639.39) .. (55.93,3637.18) .. controls (55.12,3634.97) and (53.61,3634.26) .. (51.4,3635.07) -- (50.21,3634.52) -- (50.21,3634.52) ;
%Straight Lines [id:da16006761852049878] 
\draw    (218.65,3659.36) .. controls (216.42,3660.14) and (214.92,3659.42) .. (214.14,3657.2) .. controls (213.36,3654.97) and (211.86,3654.25) .. (209.63,3655.03) .. controls (207.41,3655.81) and (205.91,3655.09) .. (205.13,3652.87) .. controls (204.35,3650.64) and (202.85,3649.92) .. (200.62,3650.7) .. controls (198.4,3651.48) and (196.9,3650.76) .. (196.11,3648.54) .. controls (195.34,3646.31) and (193.84,3645.59) .. (191.61,3646.37) .. controls (189.39,3647.15) and (187.89,3646.43) .. (187.1,3644.21) .. controls (186.32,3641.98) and (184.82,3641.26) .. (182.59,3642.04) -- (181.11,3641.33) -- (181.11,3641.33) ;
%Straight Lines [id:da9490817167483869] 
\draw    (87.66,3668.95) .. controls (85.45,3669.76) and (83.94,3669.05) .. (83.13,3666.84) .. controls (82.32,3664.63) and (80.81,3663.92) .. (78.6,3664.73) .. controls (76.39,3665.54) and (74.87,3664.83) .. (74.06,3662.62) .. controls (73.25,3660.41) and (71.74,3659.7) .. (69.53,3660.51) .. controls (67.32,3661.32) and (65.81,3660.61) .. (65,3658.4) .. controls (64.19,3656.19) and (62.68,3655.48) .. (60.47,3656.29) .. controls (58.26,3657.1) and (56.74,3656.39) .. (55.93,3654.18) .. controls (55.12,3651.97) and (53.61,3651.26) .. (51.4,3652.07) -- (50.21,3651.52) -- (50.21,3651.52) ;
\end{tikzpicture}
\end{center}
\caption{Diagrams in $Q_1^3Q_2$ sector.}
\label{Fig:Q1^3Q_2Diagrams}
\end{figure}

Our gauge choice is most powerful in the case of $A_{(3,1)}$, so we discuss this case in most detail. The Feynman diagrams that constitute this amplitude can be split into 3 classes: the first involves single photon emissions coming from cubic vertices, the second has precisely one photon emitted into the final state from a quartic vertex, while the third class has two photons emitted from the same quartic vertex. These classes are shown in figure \ref{Fig:Q1^3Q_2Diagrams}, after removing diagrams which vanish by gauge choice. The first diagram is an example of the first class, the second an example of the second class and the last two diagrams are in the third class.

We choose a particular ordering of $k_1$ and $k_2$ for the calculation, and include the permuted case by swapping $k_1\leftrightarrow k_2$.
For the first class there is a single diagram to compute after gauge fixing, and it is trivial to write down the leading term in the amplitude and see that it has the desired scaling. This diagram is

\[
D_{(3,1)}|_{cubic}  =& \frac{ 1 }{\hbar^4 \bar q_2^2}\left[\frac{2(\varepsilon_2\cdot \barq_1  )(\varepsilon_1\cdot (2\barq_1-\bar k_2))(2p_2 +\hbar \barq_2)\cdot (2p_1 - \hbar \barq_2)}{(-2p_1\cdot \barq_2 + \hbar \barq_2^2 )(2p_1\cdot( \barq_1-\bar k_1) + \hbar ( \barq_1-\bar k_1)^2 )}\right]
\\
=& \frac{ 1 }{\hbar^4 \bar q_2^2}\left[\frac{4p_1\cdot p_2(\varepsilon_2\cdot \barq_1  )(\varepsilon_1\cdot (2\barq_1-\bar k_2))}{2(-p_1\cdot \barq_2  )(p_1\cdot( \barq_1-\bar k_1)} +\mathcal{O}(\hbar)\right]
\\
=& \frac{2 p_1\cdot p_2 }{\hbar^4 \bar q_2^2}\left[\frac{(\varepsilon_2\cdot \barq_1  )(\varepsilon_1\cdot (\barq_1-\bar q_2))}{(p_1\cdot (\bar k_1 +\bar k_2))(p_1\cdot \bar k_1)}\right] +\mathcal{O}(\hbar^{-3}).
\label{eq:q13q2cubic}
\]
We  used momentum conservation $q_1+q_2 = k_1+k_2$ to write our expressions in terms of just the $k_i$ or the $q_i$. The choice here is most natural for obtaining a similarly simple expression for the amplitude with $k_1$ and $k_2$ swapped.

The second class is actually tractable without fixing a gauge, as there is only a single cancellation to show. However with our gauge fixing this class becomes trivial, and there is only a single diagram to compute. This is
\[\label{eq:q13q2cubicquartic}
D_{(3,1)}|_{cubic/quartic} = \frac{4(p_2\cdot \varepsilon_2) (\bar q_1 \cdot \varepsilon_1)}{\hbar^4 \bar q_2^2 p_1\cdot \bar k_1}  .
\]

The final class is unaffected by our choice of gauge as it is proportional to $\varepsilon_1 \cdot \varepsilon_2$. After gauge fixing this is naively $\hbar^{-5}$, so we must find a single cancellation. The mechanism of the cancellation is
identical to the case of the five-point tree amplitude~\cite{Luna:2017dtq,KMOC}. The key step is to make use of the on-shell conditions
\[\label{eq:shell}
(p_i+q_i)^2=p_i^2=m_i^2, \,\,\,\, i=1,2, 
\]
which allows  us, after the $\hbar$ rescaling, to replace $2p_i\cdot \bar q_i \rightarrow - \hbar \bar q_i^2$ in the propagators. Lastly, we Taylor expand. There are two diagrams to compute which are,
\[
D_{(3,1)}|_{
quartic,1} =& - \frac{2\varepsilon_1\cdot\varepsilon_2}{\hbar^5 \bar q_2^2} \left[\frac{ (2p_1+\hbar (2\barq_1 +\barq_2))\cdot (2p_2 + \hbar \barq_2) }{2p_1\cdot(\bark_1+\bark_2) + \hbar (\bark_1 + \bark_2)^2}\right]
\\
=&- \frac{2\varepsilon_1\cdot\varepsilon_2}{\hbar^5 \bar q_2^2} \left[\frac{4p_1\cdot p_2  }{2p_1\cdot(\bark_1+\bark_2)}
\right. \\ &\qquad \qquad \quad \left. 
+ \hbar\left( \frac{4p_2 \cdot \qb_1 + 2p_1\cdot \barq_2}{2p_1\cdot(\bark_1+\bark_2)}-\frac{4p_1 \cdot p_2 (\bark_1+\bark_2)^2}{(2p_1\cdot(\bark_1+\bark_2))^2} \right)\right]
\]
and \[
D_{(3,1)}|_{
quartic,2} 
=& - \frac{2\varepsilon_1\cdot\varepsilon_2}{\hbar^5 \bar q_2^2} \left[\frac{4p_1\cdot p_2  }{-2p_1\cdot(\bark_1+\bark_2)}\right. 
\\
&\qquad \qquad \quad +\left. \hbar\left( \frac{4p_1 \cdot \qb_2 }{-2p_1\cdot(\bark_1+\bark_2)}+\frac{4p_1 \cdot p_2 ( (\bark_1+\bark_2)^2 -\qb_1^2)}{(2p_1\cdot(\bark_1+\bark_2))^2} \right)\right].
\]
Notice the most singular terms are equal up to a sign, and so cancel. Combining the remaining terms we obtain
\[\label{eq:e1e2amp}
D_{(3,1)}|_{
quartic} =- \frac{\varepsilon_1\cdot\varepsilon_2}{\hbar^4 \bar q_2^2} \left[ \frac{4(p_1\cdot p_2)( \barq_2 \cdot (\bar k_1 + \bar k_2))}{(p_1\cdot (\bar k_1 + \bar k_2))^2} + \frac{4p_2 \cdot \barq_1 +2p_1 \cdot \barq_2}{p_1 \cdot (\bar k_1 + \bar k_2)}\right] \,.
\]
These can be combined as $A_{(3,1)} = D_{(3,1)cubic}+D_{(3,1)quartic}$ yielding,
\begin{equation}
\begin{aligned}
A_{(3,1)}  =& \frac{ 1  }{\hbar^4 \bar q_2^2}\Bigg[\frac{4p_1\cdot p_2(\varepsilon_2\cdot \barq_1  )(\varepsilon_1\cdot (\barq_1-\bar q_2))}{2(p_1\cdot (\bar k_1 +\bar k_2))(p_1\cdot \bar k_1)}
+\frac{4p_1\cdot p_2(p_2\cdot \varepsilon_2) (\bar q_1 \cdot \varepsilon_1)}{p_1\cdot \bar k_1} 
\\
&-  (\varepsilon_1\cdot\varepsilon_2)\left( \frac{4(p_1\cdot p_2)( \barq_2 \cdot (\bar k_1 + \bar k_2))}{(p_1\cdot (\bar k_1 + \bar k_2))^2} + \frac{4p_2 \cdot \barq_1 +2p_1 \cdot \barq_2}{p_1 \cdot (\bar k_1 + \bar k_2)}\right)\Bigg]
\\ & +(k_1\leftrightarrow k_2).
\end{aligned}
\label{eq:ampq13q2full}
\end{equation}

The story is very similar for $A_{(2,2)}$. We can split into the same 3 classes, use gauge fixing to get rid of one factor of $\hbar$ and then massage using the on-shell constraints to show the final cancellation. Here we just quote the result and give details in the appendix.  The result is 
\begin{equation}
\label{eq:ampq12q22final}
\begin{aligned}
A_{(2,2)} =& \frac{4}{\hbar^4 (\barq_2 - \bar k_2)^2 }
\Bigg[4 \varepsilon_1\cdot \varepsilon_2 
+\frac{(\varepsilon_1\cdot p_2 )(\varepsilon_2\cdot \barq _2)}{p_2\cdot \bar k_2}
 - \frac{(\varepsilon_2\cdot p_2)(\varepsilon_1\cdot(\barq_2 + \bar q_1))}{2p_2\cdot \bar k_2} 
 \\&
   \frac{ p_1\cdot p_2(\varepsilon_1\cdot\barq_1)( \varepsilon_2 \cdot \barq_2  ) }{(p_2\cdot \bar k_2) (p_1\cdot \bar k_1)} 
  -\frac{ (p_1\cdot \bar k_2)(\varepsilon_1\cdot \barq_1)(\varepsilon_2\cdot p_2)}{(p_2\cdot \bar k_2) (p_1\cdot \bar k_1)}
\\
& - \frac{(\varepsilon_1 \cdot p_2)(\varepsilon_2\cdot p_2)(\barq_2 \cdot \bar k_2)}{(p_2\cdot \bar k_2)^2}     - \frac{ p_1\cdot p_2(\varepsilon_1\cdot \barq_1)(\varepsilon_2\cdot p_2)\bar q_2\cdot \bar k_2}{(p_2\cdot \bar k_2)^2 (p_1\cdot \bar k_1)} 
\Bigg] 
\\&+(k_1\leftrightarrow k_2).
\end{aligned}
\end{equation}
Finally $A_{(1,3)}$ can be obtained by swapping the labels $1\leftrightarrow 2$ in the expression for $A_{(3,1)}$ (written in the gauge where $p_2\cdot \varepsilon_i = 0$. Of course the partial amplitudes themselves are gauge-invariant.) In all cases, the $\hbar$ scaling is as required from negligible variance.

\section{Zero-variance relations}
\label{sec:infinity_relations}

Our discussion so far reveals that scattering amplitudes, viewed as Laurent series in $\hbar$, obey certain properties which permit the emergence of a classical limit through negligible uncertainty. This Laurent expansion can also be viewed as an expansion in small momentum transfers divided by the centre-of-mass energy $\sqrt{s}$ (we will refer to this expansion as a transfer expansion below).
In fact, the emergence of the classical limit imposes an \emph{infinite} set of these relationships, which we will call ``zero-variance relations'' on scattering amplitudes. 
In this section we will describe the origin of these relations, and explicitly demonstrate a non-trivial example at one loop and five points.

\subsection{Mixed variances}
\label{sec:fivepointrelation}

To see where these relationships are coming from, recall that the double field-strength expectation~\eqref{eq:twoFs} depends on a six-point amplitude. We have seen that the dominant term is actually the six-point one-loop amplitude, occurring at next-to-leading order in the expansion in $g$. At this order an additional term contributes to the double field-strength expectation; this term is the product of two five-point amplitudes. Now, negligible uncertainty demands that the complete double field-strength expectation must be the product of two single field-strength expectations. At leading order in the coupling, and leading non-trivial order in $\hbar$, we conclude that there must exist a relationship between the leading-in-$\hbar$ six-point one-loop amplitude and the product of two five-point trees.

Further examples of relationships between amplitudes can be obtained by considering expectations of three (or more) field strengths, leading to relationships between seven- (or higher-) point loop amplitudes and products of three (or more) five-point amplitudes. 

Yet more relationships occur by considering expectations of products of operators including field strengths and momenta. For example,
consider the variance
\[
V_{\mu\nu\rho} \equiv \bra{\psi}  S^\dagger \Fop_{\nu\rho}(x)  S \Pop_\mu \ket \psi - \bra{\psi}  S^\dagger \Fop_{\nu\rho}(x)  S \ket\psi \bra\psi \Pop_\mu \ket \psi \,.
\]
This is the variance in a measurement of the initial momentum and the future field strength; it must be negligible in the classical regime. In a quantum-first approach, however, this variance will not vanish. Indeed it need not be real:
\[
V_{\mu\nu\rho}^* = \bra{\psi}  \Pop_\mu S^\dagger \Fop_{\nu\rho}(x)  S  \ket \psi - \bra{\psi}  S^\dagger \Fop_{\nu\rho}(x)  S \ket\psi \bra\psi \Pop_\mu \ket \psi \neq V_{\mu\nu\rho}.
\]
We can derive an interesting constraint on the five-point one-loop amplitude by demanding that imaginary part of this variance vanishes in the classical approximation. We therefore define
\[
\mathcal{O}_{\mu\nu\rho} &= i (V_{\mu\nu\rho}^* - V_{\mu\nu\rho}) \\
&= i  \bra{\psi}  \Pop_\mu S^\dagger \Fop_{\nu\rho}(x)  S - S^\dagger \Fop_{\nu\rho}(x)  S \Pop_\mu \ket \psi \,.
\label{eq:O3def}
\]
Expanding the states as usual, we easily find 
\[
\mathcal{O}_{\mu\nu\rho} = \int \dPhi(p_1', p_2', p_1, p_2)& \, \phi_b^*(p_1', p_2') \phi_b(p_1, p_2) i (p'_{1\mu} - p_{1\mu}) \\
& \times \bra{p'_1p'_2} i (\Fop_{\nu\rho}(x) T - T^\dagger \Fop_{\nu\rho}(x)) + T^\dagger \Fop_{\nu\rho}(x) T \ket{p_1p_2} \,.
\]
The factor $i (p'_{1\mu} - p_{1\mu})$ is important here: working at leading perturbative order, this factor is of order $\hbar$. It is also worth noting that we may write the expectation of the field strength itself as
 \[
\bra{\psi} S^\dagger \Fop_{\nu\rho} S \ket{\psi} = \int \dPhi(p_1', p_2',& p_1, p_2) \, \phi_b^*(p_1', p_2') \phi_b(p_1, p_2)\\
& \hspace{-20pt}\times \bra{p'_1p'_2} i (\Fop_{\nu\rho}(x) T - T^\dagger \Fop_{\nu\rho}(x)) + T^\dagger \Fop_{\nu\rho}(x) T \ket{p_1p_2} \,.
\]
Thus the $i (p'_{1\mu} - p_{1\mu}) \sim \hbar$ factor in the variance is the key distinction between the variance, which vanishes classically, and the field strength which of course should not vanish classically. As we have already seen that the field strength is related to five-point amplitudes, it is now clear that the condition of vanishing $\mathcal{O}_{\mu\nu\rho}$ will become a condition on five-point amplitudes.

It is useful to break the variance $\mathcal{O}_{\mu\nu\rho}$ up into two structures:
\[
\mathcal{O}_{\mu\nu\rho}^{(1)} = \int \dPhi(p_1', p_2', p_1, p_2)& \, \phi_b^*(p_1', p_2') \phi_b(p_1, p_2) i (p'_{1\mu} - p_{1\mu}) \\
& \times \bra{p'_1p'_2} i (\Fop_{\nu\rho}(x) T - T^\dagger \Fop_{\nu\rho}(x)) \ket{p_1p_2} \,,
\]
and
\[
\mathcal{O}_{\mu\nu\rho}^{(2)} = \int \dPhi(p_1', p_2', p_1, p_2)& \, \phi_b^*(p_1', p_2') \phi_b(p_1, p_2) i (p'_{1\mu} - p_{1\mu}) 
\\& \times 
\bra{p'_1p'_2} T^\dagger \Fop_{\nu\rho}(x) T \ket{p_1p_2} \,.
\]
Both of these objects are real, which is convenient in terms of keeping the expressions simple.

We may simplify these structures using the explicit expression for the field strength given in equation~\eqref{eq:explicitF}. For $\mathcal{O}^{(1)}$ we find
\[
\mathcal{O}_{\mu\nu\rho}^{(1)} &= 2 \Re \frac1{\sqrt{\hbar}} \sum_{\hel} \int \dPhi(p_1', p_2', p_1, p_2, k) \phi_b^*(p_1', p_2') \phi_b(p_1, p_2) \times \\
& \hspace{100pt}\times i (p_{1\mu}' - p_{1\mu}) \bra{k^\eta, p_1', p_2' } T \ket{p_1, p_2} \, \kb^\vmu_{[\nu} \polvhconj{\eta}_{\rho]} \, e^{-i \kb \cdot x} \,,
\]
which should be compared to equation~\eqref{eq:F_expectation}. Again we see that the crucial new ingredient is a factor $i (p_{1\mu}' - p_{1\mu})$. In the classical regime, we may write this term as
\[
\mathcal{O}_{\mu\nu\rho}^{(1)} &= 2 \Re \hbar^{9/2} \sum_{\hel} \Lexp \int \dPhi(\kb) \, \dd^4 \qb_1 \dd^4 \qb_2 \, \del(2 p_1 \cdot \qb_1) \del(2 p_2 \cdot \qb_2) \,   \\
&\hspace{10pt} \times i \qb_\mu  \mathcal{A}_{5}(p_1 p_2 \rightarrow p_1+q_1, p_2+q_2, k^\eta) \,  \del^4(\kb + \qb_1 + \qb_2)  \, \kb^\vmu_{[\nu} \polvhconj{\eta}_{\rho]} \, e^{-i (\kb \cdot x +\qb_1 \cdot b)} \Rexp \,.
\label{eq:O1expanded}
\]
Referring back once more to equation~\eqref{eq:F_expectation2}, the additional $\hbar$ suppression is now manifest.

In order to control the $\hbar$ expansion of scattering amplitudes, it is useful to introduce some further notation. Let us write the amplitudes as explicit Laurent series in $\hbar$ building on equation~\eqref{eq:seriesTaster}. For example, at five-points we may write
\[
\label{eq:hbarExpansion}
\mathcal{A}_{5,0}(i \rightarrow f) = \hbar^{-7/2} \left( \mathcal{A}_{5,0}^{(0)}(i \rightarrow f) + \hbar \mathcal{A}_{5,0}^{(1)}(i \rightarrow f) + \cdots \right) \,,\\
\mathcal{A}_{5,1}(i \rightarrow f) = \hbar^{-9/2} \left( \mathcal{A}_{5,1}^{(0)}(i \rightarrow f) + \hbar \mathcal{A}_{5,1}^{(1)}(i \rightarrow f) + \cdots \right) \,.\\
\]
We have scaled out the dominant (inverse) power of $\hbar$; the quantities $\mathcal{A}_{n,L}^{(p)}$ are $\hbar$-independent gauge-invariant sub-amplitudes; it is precisely these quantities that are related by our reasoning. 
This expansion defines an infinite set of objects $\mathcal{A}_{n,L}^{(p)}$ which could in principle be reassembled into the full amplitude.
They are a kind of partial amplitude, but distinct from the usual use of this term. 
We will therefore refer to them as ``fragmentary amplitudes'', or simply as ``fragments.''

Since $\hbar$ is dimensionful, it is useful to view these fragmentary amplitudes in a slightly different
way. 
Amplitudes are functions of Mandelstam invariants; in the semi-classical region, we are expanding in powers of momentum transfers, such as $q^2 = \hbar^2 \qb^2$ at four points, divided by Mandelstam $s = (p_1 + p_2)^2$. The semiclassical expansion is an expansion in powers of $\hbar \sqrt{-\qb^2 / s}$. More general amplitudes involve a richer set of momentum transfers $\qb_{ij}^2$; our expansion is in powers of $\hbar \sqrt{-\qb_{ij}^2 / s}$. 
We only consider amplitudes with two incoming massive particles. 

We may also view the expansion as being in (inverse) powers of the large mass of the scattering particles~\cite{Luna:2017dtq,Damgaard:2019lfh,Aoude:2020onz,Haddad:2020tvs}. 
This makes contact with effective field theory, especially heavy quark effective theory or, more generally, heavy particle effective theories as has been
emphasised in references~\cite{Damgaard:2019lfh,Aoude:2020onz,Haddad:2020tvs}.
Our fragmentary amplitudes correspond in this context to the standard HQET expansion in inverse powers of heavy masses.
It seems likely that a study of the properties of amplitudes in these theories would illuminate the structure of the fragmentary amplitudes.

Notice that this expansion is analogous, but different, to a soft expansion. In the soft expansion we take the momentum of an individual particle
soft.
In this transfer expansion we take the momenta in all messenger lines to be of the same order, and small compared to the incoming centre of
mass energy.
It is possible to perform the transfer expansion and then, in a second stage, to single out some line, say an outgoing photon, and take its momentum to be softer than all other messenger lines. This yields the soft limit of the transfer expansion. 
It corresponds to the low-frequency limit in the classical approximation. 
Interesting classical physics, including memory effects, appear in this region~\cite{Laddha:2018rle,Laddha:2018myi,Laddha:2018vbn,Sahoo:2018lxl,Bautista:2019tdr,Laddha:2019yaj,Saha:2019tub,Sahoo:2021ctw,Bautista:2021llr}.

Now at classical order ($\hbar^0$) the tree level amplitude $\mathcal{A}_{5,0}$ does not appear in $\mathcal{O}_{\mu\nu\rho}^{(1)}$ on account of the explicit factor $ \hbar^{9/2}$ in equation~\eqref{eq:O1expanded}. The leading in $g$, non-trivial, classical contribution arises from the fragment $\mathcal{A}_{5,1}^{(0)}$. We conclude then that
\[
\mathcal{O}_{\mu\nu\rho}^{(1)} &= 2 \Re \sum_{\hel} \Lexp \int \dPhi(\kb) \, \dd^4 \qb_1 \dd^4 \qb_2 \, \del(2 p_1 \cdot \qb_1) \del(2 p_2 \cdot \qb_2) \,  \times \\
& \times i \qb_\mu  \mathcal{A}_{5,1}^{(0)}(p_1 p_2 \rightarrow p_1+q_1, p_2+q_2, k^\eta) \,  \del^4(\kb + \qb_1 + \qb_2)  \, \kb^\vmu_{[\nu} \polvhconj{\eta}_{\rho]} \, e^{-i (\kb \cdot x +\qb_1 \cdot b)} \Rexp \,.
\label{eq:O1ampl}
\]
The relevant fragmentary amplitude is the leading-in-$\hbar$ five-point one-loop amplitude, sometimes known as the ``superclassical'' part of the one-loop amplitude. Of course in this context this fragment of the amplitude is contributing precisely at classical order.

Now the full $\mathcal{O}_{\mu\nu\rho}$ should vanish at classical order. Since $\mathcal{O}_{\mu\nu\rho}^{(1)}  \neq 0$, it must be that the second structure $\mathcal{O}_{\mu\nu\rho}^{(2)}$ cancels the contribution of equation~\eqref{eq:O1ampl}. We find that 
\[
\mathcal{O}_{\mu\nu\rho}^{(2)} &= 2 \Re \sum_{\hel} \Lexp \int \dPhi(\kb) \, \dd^4 \qb_1 \dd^4 \qb_2 \dd^4 \bar{w}_1 \dd^4 \bar{w}_2 \, \del(2 p_1 \cdot \qb_1) \del(2 p_2 \cdot \qb_2)\del(2 p_1 \cdot \bar{w}_1) \del(2 p_2 \cdot \bar{w}_2)  \\
&\times  \qb_\mu  \,  \del^4(\kb + \qb_1 + \qb_2) \del^4(\qb_1 + \qb_2 - \bar{w}_1 - \bar{w}_2)  \, \kb^\vmu_{[\nu} \polvhconj{\eta}_{\rho]} \, e^{-i (\kb \cdot x +\qb_1 \cdot b)} \\
&\times \mathcal{A}_{5,0}^{(0)}(p_1 p_2 \rightarrow p_1 + w_1, p_2 + w_2, k^\eta)\mathcal{A}_{4,0}^{(0)}(p_1 + w_1, p_2 + w_2 \rightarrow p_1 + q_1, p_2 + q_2) \Rexp \,.
\label{eq:O2ampl}
\]
Comparing equations~\eqref{eq:O1ampl} and~\eqref{eq:O2ampl}, the condition for vanishing $\mathcal{O}$ is
\[
i & \mathcal{A}_{5,1}^{(0)}(p_1 p_2 \rightarrow p_1 + q_1, p_2 +q_2, k^\eta) \\
&= -\int \dd^4 \bar{w}_1 \dd^4 \bar{w}_2 \, \del(2 p_1 \cdot \bar{w}_1) \del(2 p_2 \cdot \bar{w}_2) \del^4(\qb_1 + \qb_2 - \bar{w}_1 - \bar{w}_2) \\
&\hspace{10pt}\times\mathcal{A}_{5,0}^{(0)}(p_1 p_2 \rightarrow p_1 + w_1, p_2 + w_2, k^\eta)\mathcal{A}_{4,0}^{(0)}(p_1 + w_1, p_2 + w_2 \rightarrow p_1 + q_1, p_2 + q_2) \,.
\label{eq:fivePointRelation}
\]
Thus the dominant part of the five-point one-loop amplitude is given by the tree five-point and tree four-point amplitudes; we will check this relation explicitly in the next subsection.

Clearly this explicit example is one among an infinite set of relationships. Variances involving one field strength operator and two momenta will lead to relationships among two-loop five-point amplitudes and the product of one five-point tree and two four-point trees. We can continue, in principle, as far as we wish generating similar relations.
These negligible uncertainty relations generalise the well-known relations between multiloop four-point amplitudes required for eikonal exponentiation. Indeed consideration of expectations such as
\[
\bra{\psi}  S^\dagger \Pop_{\mu_1}\Pop_{\mu_2} \cdots \Pop_{\mu_n} S \ket{\psi} \simeq \bra{\psi}  S^\dagger \Pop_{\mu_1} \ket{\psi} \bra{\psi}  S^\dagger \Pop_{\mu_2} \ket{\psi} \cdots \bra{\psi}  S^\dagger \Pop_{\mu_n} \ket{\psi} \,,
\]
shows that there must be a relationship between the $n-1$ loop four-point amplitude and the product of $n$ tree amplitudes.

Thus we find a remarkable abundance of relationships between multiloop, multileg amplitudes, considered as Laurent series in $\hbar$, forced on us by the \emph{absence} of uncertainty in the classical regime. In the next section of this article we will interpret these relationships in terms of a radiative generalisation of the eikonal exponentiation. 

As well as finding explicit relations between different fragmentary amplitudes, we can use similar ideas to determine the $\hbar$ scaling associated with fragments in the transfer expansion. 
The kinds of multiple cancellations of $\hbar$ powers we saw at six points must continue to occur at higher points. The reason again follows from considering expectations of products of more than two field-strength operators.

The arguments are based simply on counting powers of coupling and $\hbar$. We know that for the single expectation, at leading order, we have 
\[
\langle \Fop_{\mu\nu} \rangle \sim g^3 .
\] 
This means that we must also have $\langle \Fop ^n \rangle \sim (g^3)^n$. Now we perform the KMOC analysis of $\langle \Fop ^n \rangle $. Following the steps of the calculation earlier in section \ref{sec:expectation_of_two_F} we find, schematically, that 
\[
\langle \Fop ^n \rangle \sim \hbar^{3n/2 +2} \int \amp_{4+n}.
\]
These relations allow us to deduce two things. Firstly the relevant fragment of the complete amplitude $\amp_{4+n}$ must scale as $\hbar ^{-3n/2 -2}$. Secondly this fragment must have $3n$ powers of the coupling $g$ --- this corresponds to having $n-1$ loops. From this we can also infer the scaling of all other loop and tree amplitudes, in the classical limit, since each loop contributes an extra factor of $\hbar^{-1}$. In particular the tree scaling will be 
\[
\amp_{4+n,0} \sim \hbar^{-n/2-3}.
\]
This is consistent with the scaling we computed above for six points $(n=2)$, and we have also checked explicitly at seven points.
It is interesting to see how these two very simple power counting arguments have completely constrained the $\hbar$ scaling of all $2\rightarrow 2+n$ amplitudes.

\subsection{One loop factorisation }
\label{sec:one-loop}
We now turn to verifying equation~\eqref{eq:fivePointRelation}. For simplicity we focus on the case of scalar QED, though the general nature of our arguments indicates that the result should also hold in gravity and in Yang-Mills theory. In order to keep the computational labour to the minimum necessary, we take advantage of lessons we learned in the context of the six-point tree amplitude in section~\ref{sec:6_point_tree}. First, we note that the scalar QED five-point amplitudes can be reduced to gauge-invariant partial amplitudes analogous to colour-ordered amplitudes in Yang-Mills theory. In particular we write
\[
\label{eq:expicitAs}
\mathcal{A}_{4,0}^{(0)}(p_1p_2 \rightarrow p_1 + w_1, p_2 + w_2) &= Q_1 Q_2 A_{(1,1)} \,,\\
\mathcal{A}_{5,0}^{(0)}(p_1p_2 \rightarrow p_1 + q_1, p_2 + q_2, k) &= Q_1Q_2^2 A_{(1,2)} + Q_1^2 Q_2 A_{(2,1)} \,,\\
\mathcal{A}_{5,1}^{(0)}(p_1p_2 \rightarrow p_1 + q_1, p_2 + q_2, k) &= Q_1^2Q_2^3 A_{(2,3)} + Q_1^3 Q_2^2 A_{(3,2)} \,.
\]
In view of the symmetry between $A_{2,3}$ and $A_{3,2}$ we may compute just one choice: we choose to focus on the charge sector $Q_1^2 Q_2^3$.

Second, we find it useful to choose an explicit gauge, namely
\[\label{gaugee}
\varepsilon_\hel (\bar{k}) \cdot  p_2=0 \,.
\]
This choice drastically reduces the relevant number of terms in the $\hbar$ expansion.
It is trivial to determine the tree partial amplitudes in this gauge, which are
\begin{equation}
\label{A4}
A_{(1,1)} =e^2 \frac{4 p_1\cdot {p}_2}{\bar{w}_1 ^2} \,, 
\end{equation}
and
\[
\label{A5} 
A_{(1,2)} =\frac{4e^3}{\bar{q}_1^2}&\left(
 p _1\cdot \varepsilon_\hel  +\frac{ p _1\cdot  p_2\,  \varepsilon_\hel\cdot \bar{q}_1}{ p_2\cdot \bar{k}}
\right)
\,.
\]

The anatomy of \eqref{eq:fivePointRelation} is that the leading-in-$\hbar$ fragment of the one-loop five-point amplitude $A_{(2,3)}$ will organise itself  into a product of $A_{(1,1)}$ times $A_{(1,2)}$. 
Our strategy will be to isolate these terms one by one.
Let us start gathering the relevant diagrams of $A_{(2,3)}$.
At one loop, and in the classical limit, we need to consider the transfer  expansion of the following five-point diagrams with a photon emitted in the final state:
\begin{center}
\begin{figure}[H]

\tikzset{every picture/.style={line width=0.75pt}} %set default line width to 0.75pt        

\begin{tikzpicture}[x=0.75pt,y=0.75pt,yscale=-.9,xscale=.99]
%uncomment if require: \path (0,8388); %set diagram left start at 0, and has height of 8388

%Shape: Ellipse [id:dp9809922788625585] 
\draw  [fill={rgb, 255:red, 74; green, 144; blue, 226 }  ,fill opacity=1 ] (82.64,7848.38) .. controls (74.15,7839.24) and (74.56,7824.84) .. (83.56,7816.22) .. controls (92.56,7807.6) and (106.73,7808.02) .. (115.21,7817.16) .. controls (123.7,7826.3) and (123.29,7840.7) .. (114.29,7849.32) .. controls (105.29,7857.94) and (91.12,7857.52) .. (82.64,7848.38) -- cycle ;
%Straight Lines [id:da2900269360832519] 
\draw    (83.56,7816.22) -- (55.35,7778.51) ;
\draw [shift={(69.45,7797.36)}, rotate = 413.2] [fill={rgb, 255:red, 0; green, 0; blue, 0 }  ][line width=0.08]  [draw opacity=0] (5.36,-2.57) -- (0,0) -- (5.36,2.57) -- cycle    ;
%Straight Lines [id:da03716835035992161] 
\draw    (115.21,7817.16) -- (141.22,7776.89) ;
\draw [shift={(128.22,7797.02)}, rotate = 482.86] [fill={rgb, 255:red, 0; green, 0; blue, 0 }  ][line width=0.08]  [draw opacity=0] (5.36,-2.57) -- (0,0) -- (5.36,2.57) -- cycle    ;
%Straight Lines [id:da5008412318055209] 
\draw    (60.71,7884.64) -- (82.64,7848.38) ;
\draw [shift={(71.67,7866.51)}, rotate = 481.16] [fill={rgb, 255:red, 0; green, 0; blue, 0 }  ][line width=0.08]  [draw opacity=0] (5.36,-2.57) -- (0,0) -- (5.36,2.57) -- cycle    ;
%Straight Lines [id:da8385205711782695] 
\draw    (135.66,7882.85) -- (114.29,7849.32) ;
\draw [shift={(124.97,7866.08)}, rotate = 417.5] [fill={rgb, 255:red, 0; green, 0; blue, 0 }  ][line width=0.08]  [draw opacity=0] (5.36,-2.57) -- (0,0) -- (5.36,2.57) -- cycle    ;
%Straight Lines [id:da010847250160856747] 
\draw    (98.64,7809.88) .. controls (96.99,7808.2) and (97.01,7806.53) .. (98.69,7804.88) .. controls (100.37,7803.23) and (100.38,7801.56) .. (98.73,7799.88) .. controls (97.08,7798.2) and (97.1,7796.53) .. (98.78,7794.88) .. controls (100.46,7793.23) and (100.47,7791.56) .. (98.82,7789.88) .. controls (97.17,7788.2) and (97.19,7786.53) .. (98.87,7784.88) .. controls (100.55,7783.23) and (100.56,7781.56) .. (98.91,7779.88) .. controls (97.26,7778.2) and (97.27,7776.53) .. (98.95,7774.88) .. controls (100.63,7773.23) and (100.65,7771.56) .. (99,7769.88) -- (99,7769.67) -- (99,7769.67) ;
%Straight Lines [id:da9136130074011992] 
\draw    (255,7802.45) .. controls (256.67,7800.78) and (258.33,7800.78) .. (260,7802.45) .. controls (261.67,7804.12) and (263.33,7804.12) .. (265,7802.45) .. controls (266.67,7800.78) and (268.33,7800.78) .. (270,7802.45) .. controls (271.67,7804.12) and (273.33,7804.12) .. (275,7802.45) .. controls (276.67,7800.78) and (278.33,7800.78) .. (280,7802.45) .. controls (281.67,7804.12) and (283.33,7804.12) .. (285,7802.45) .. controls (286.67,7800.78) and (288.33,7800.78) .. (290,7802.45) -- (294.11,7802.45) -- (294.11,7802.45) ;
%Straight Lines [id:da6203634462561736] 
\draw    (255,7772.1) .. controls (256.67,7770.43) and (258.33,7770.43) .. (260,7772.1) .. controls (261.67,7773.77) and (263.33,7773.77) .. (265,7772.1) .. controls (266.67,7770.43) and (268.33,7770.43) .. (270,7772.1) .. controls (271.67,7773.77) and (273.33,7773.77) .. (275,7772.1) .. controls (276.67,7770.43) and (278.33,7770.43) .. (280,7772.1) .. controls (281.67,7773.77) and (283.33,7773.77) .. (285,7772.1) .. controls (286.67,7770.43) and (288.33,7770.43) .. (290,7772.1) -- (294.11,7772.1) -- (294.11,7772.1) ;
%Straight Lines [id:da20266369209882718] 
\draw    (255,7747) -- (255,7827.95) ;
%Straight Lines [id:da22576660894701717] 
\draw    (294.11,7747) -- (294.11,7827.95) ;
%Straight Lines [id:da12578882747337805] 
\draw    (294.5,7755.1) .. controls (296.19,7753.46) and (297.86,7753.49) .. (299.5,7755.18) .. controls (301.14,7756.87) and (302.81,7756.9) .. (304.5,7755.26) .. controls (306.19,7753.62) and (307.86,7753.65) .. (309.5,7755.34) .. controls (311.14,7757.03) and (312.81,7757.06) .. (314.5,7755.42) -- (314.84,7755.42) -- (314.84,7755.42) ;
%Straight Lines [id:da9321746785910134] 
\draw    (335.57,7802.45) .. controls (337.24,7800.78) and (338.9,7800.78) .. (340.57,7802.45) .. controls (342.24,7804.12) and (343.9,7804.12) .. (345.57,7802.45) .. controls (347.24,7800.78) and (348.9,7800.78) .. (350.57,7802.45) .. controls (352.24,7804.12) and (353.9,7804.12) .. (355.57,7802.45) .. controls (357.24,7800.78) and (358.9,7800.78) .. (360.57,7802.45) .. controls (362.24,7804.12) and (363.9,7804.12) .. (365.57,7802.45) .. controls (367.24,7800.78) and (368.9,7800.78) .. (370.57,7802.45) -- (374.69,7802.45) -- (374.69,7802.45) ;
%Straight Lines [id:da15615021130428697] 
\draw    (335.57,7772.1) .. controls (337.24,7770.43) and (338.9,7770.43) .. (340.57,7772.1) .. controls (342.24,7773.77) and (343.9,7773.77) .. (345.57,7772.1) .. controls (347.24,7770.43) and (348.9,7770.43) .. (350.57,7772.1) .. controls (352.24,7773.77) and (353.9,7773.77) .. (355.57,7772.1) .. controls (357.24,7770.43) and (358.9,7770.43) .. (360.57,7772.1) .. controls (362.24,7773.77) and (363.9,7773.77) .. (365.57,7772.1) .. controls (367.24,7770.43) and (368.9,7770.43) .. (370.57,7772.1) -- (374.69,7772.1) -- (374.69,7772.1) ;
%Straight Lines [id:da4209012550341089] 
\draw    (335.57,7747) -- (335.57,7827.95) ;
%Straight Lines [id:da6505319972934505] 
\draw    (374.69,7747) -- (374.69,7827.95) ;
%Straight Lines [id:da11123821940249257] 
\draw    (374.69,7787.48) .. controls (376.38,7785.84) and (378.04,7785.87) .. (379.68,7787.56) .. controls (381.32,7789.25) and (382.99,7789.28) .. (384.68,7787.64) .. controls (386.37,7786) and (388.04,7786.03) .. (389.68,7787.72) .. controls (391.32,7789.41) and (392.99,7789.44) .. (394.68,7787.8) -- (395.02,7787.8) -- (395.02,7787.8) ;
%Straight Lines [id:da7919375227346557] 
\draw    (416.15,7802.45) .. controls (417.82,7800.78) and (419.48,7800.78) .. (421.15,7802.45) .. controls (422.82,7804.12) and (424.48,7804.12) .. (426.15,7802.45) .. controls (427.82,7800.78) and (429.48,7800.78) .. (431.15,7802.45) .. controls (432.82,7804.12) and (434.48,7804.12) .. (436.15,7802.45) .. controls (437.82,7800.78) and (439.48,7800.78) .. (441.15,7802.45) .. controls (442.82,7804.12) and (444.48,7804.12) .. (446.15,7802.45) .. controls (447.82,7800.78) and (449.48,7800.78) .. (451.15,7802.45) -- (455.26,7802.45) -- (455.26,7802.45) ;
%Straight Lines [id:da1922367699161771] 
\draw    (416.15,7772.1) .. controls (417.82,7770.43) and (419.48,7770.43) .. (421.15,7772.1) .. controls (422.82,7773.77) and (424.48,7773.77) .. (426.15,7772.1) .. controls (427.82,7770.43) and (429.48,7770.43) .. (431.15,7772.1) .. controls (432.82,7773.77) and (434.48,7773.77) .. (436.15,7772.1) .. controls (437.82,7770.43) and (439.48,7770.43) .. (441.15,7772.1) .. controls (442.82,7773.77) and (444.48,7773.77) .. (446.15,7772.1) .. controls (447.82,7770.43) and (449.48,7770.43) .. (451.15,7772.1) -- (455.26,7772.1) -- (455.26,7772.1) ;
%Straight Lines [id:da5604147263751147] 
\draw    (416.15,7747) -- (416.15,7827.95) ;
%Straight Lines [id:da5211961544927726] 
\draw    (455.26,7747) -- (455.26,7827.95) ;
%Straight Lines [id:da25480011623741294] 
\draw    (454.87,7815.81) .. controls (456.56,7814.17) and (458.23,7814.2) .. (459.87,7815.89) .. controls (461.51,7817.58) and (463.18,7817.61) .. (464.87,7815.97) .. controls (466.56,7814.33) and (468.23,7814.36) .. (469.86,7816.05) .. controls (471.5,7817.74) and (473.17,7817.77) .. (474.86,7816.13) -- (475.21,7816.14) -- (475.21,7816.14) ;
%Straight Lines [id:da1871708537174117] 
\draw    (255,7897.17) .. controls (255.41,7894.84) and (256.77,7893.88) .. (259.09,7894.29) .. controls (261.41,7894.69) and (262.77,7893.73) .. (263.18,7891.41) .. controls (263.58,7889.09) and (264.94,7888.13) .. (267.26,7888.53) .. controls (269.58,7888.93) and (270.94,7887.97) .. (271.35,7885.65) .. controls (271.76,7883.33) and (273.12,7882.37) .. (275.44,7882.77) .. controls (277.76,7883.17) and (279.12,7882.21) .. (279.53,7879.89) .. controls (279.94,7877.57) and (281.3,7876.61) .. (283.62,7877.01) .. controls (285.94,7877.41) and (287.3,7876.45) .. (287.7,7874.13) .. controls (288.11,7871.81) and (289.47,7870.85) .. (291.79,7871.26) -- (293.33,7870.17) -- (293.33,7870.17) ;
%Straight Lines [id:da8855223117304349] 
\draw    (255,7866.81) .. controls (257.34,7866.52) and (258.66,7867.54) .. (258.95,7869.88) .. controls (259.24,7872.22) and (260.56,7873.24) .. (262.9,7872.94) .. controls (265.24,7872.65) and (266.56,7873.67) .. (266.85,7876.01) .. controls (267.14,7878.35) and (268.46,7879.37) .. (270.8,7879.07) .. controls (273.14,7878.78) and (274.46,7879.8) .. (274.75,7882.14) .. controls (275.04,7884.48) and (276.36,7885.5) .. (278.7,7885.2) .. controls (281.04,7884.91) and (282.36,7885.93) .. (282.65,7888.27) .. controls (282.94,7890.61) and (284.26,7891.63) .. (286.6,7891.33) .. controls (288.94,7891.04) and (290.26,7892.06) .. (290.55,7894.4) -- (294.11,7897.17) -- (294.11,7897.17) ;
%Straight Lines [id:da8187698259743739] 
\draw    (255,7841.71) -- (255,7922.67) ;
%Straight Lines [id:da6804766194416041] 
\draw    (294.11,7841.71) -- (294.11,7922.67) ;
%Straight Lines [id:da06115739792844077] 
\draw    (294.5,7849.81) .. controls (296.19,7848.17) and (297.86,7848.2) .. (299.5,7849.89) .. controls (301.14,7851.58) and (302.81,7851.61) .. (304.5,7849.97) .. controls (306.19,7848.33) and (307.86,7848.36) .. (309.5,7850.05) .. controls (311.14,7851.74) and (312.81,7851.77) .. (314.5,7850.13) -- (314.84,7850.14) -- (314.84,7850.14) ;
%Straight Lines [id:da32774471607438427] 
\draw    (335.57,7897.98) .. controls (335.98,7895.65) and (337.34,7894.69) .. (339.66,7895.1) .. controls (341.98,7895.5) and (343.34,7894.54) .. (343.75,7892.22) .. controls (344.16,7889.9) and (345.52,7888.94) .. (347.84,7889.34) .. controls (350.16,7889.74) and (351.52,7888.78) .. (351.92,7886.46) .. controls (352.33,7884.14) and (353.69,7883.18) .. (356.01,7883.58) .. controls (358.33,7883.98) and (359.69,7883.02) .. (360.1,7880.7) .. controls (360.51,7878.38) and (361.87,7877.42) .. (364.19,7877.82) .. controls (366.51,7878.22) and (367.87,7877.26) .. (368.28,7874.94) .. controls (368.68,7872.62) and (370.04,7871.66) .. (372.36,7872.07) -- (373.9,7870.98) -- (373.9,7870.98) ;
%Straight Lines [id:da4855848699816465] 
\draw    (335.57,7867.62) .. controls (337.91,7867.32) and (339.23,7868.34) .. (339.52,7870.68) .. controls (339.81,7873.02) and (341.13,7874.04) .. (343.47,7873.75) .. controls (345.81,7873.46) and (347.13,7874.48) .. (347.42,7876.82) .. controls (347.71,7879.16) and (349.03,7880.18) .. (351.37,7879.88) .. controls (353.71,7879.59) and (355.03,7880.61) .. (355.32,7882.95) .. controls (355.61,7885.29) and (356.93,7886.31) .. (359.27,7886.01) .. controls (361.61,7885.72) and (362.93,7886.74) .. (363.22,7889.08) .. controls (363.51,7891.42) and (364.83,7892.44) .. (367.17,7892.14) .. controls (369.51,7891.85) and (370.83,7892.87) .. (371.12,7895.21) -- (374.69,7897.98) -- (374.69,7897.98) ;
%Straight Lines [id:da5247382717160549] 
\draw    (335.57,7842.52) -- (335.57,7923.48) ;
%Straight Lines [id:da8444578806891945] 
\draw    (374.69,7842.52) -- (374.69,7923.48) ;
%Straight Lines [id:da20220773560975225] 
\draw    (374.69,7883) .. controls (376.38,7881.36) and (378.04,7881.39) .. (379.68,7883.08) .. controls (381.32,7884.77) and (382.99,7884.8) .. (384.68,7883.16) .. controls (386.37,7881.52) and (388.04,7881.55) .. (389.68,7883.24) .. controls (391.32,7884.93) and (392.99,7884.96) .. (394.68,7883.32) -- (395.02,7883.33) -- (395.02,7883.33) ;
%Straight Lines [id:da5348804458046716] 
\draw    (416.93,7895.55) .. controls (417.34,7893.22) and (418.7,7892.26) .. (421.02,7892.67) .. controls (423.34,7893.07) and (424.7,7892.11) .. (425.1,7889.79) .. controls (425.51,7887.47) and (426.87,7886.51) .. (429.19,7886.91) .. controls (431.51,7887.31) and (432.87,7886.35) .. (433.28,7884.03) .. controls (433.69,7881.71) and (435.05,7880.75) .. (437.37,7881.15) .. controls (439.69,7881.55) and (441.05,7880.59) .. (441.46,7878.27) .. controls (441.86,7875.95) and (443.22,7874.99) .. (445.54,7875.39) .. controls (447.86,7875.8) and (449.22,7874.84) .. (449.63,7872.52) .. controls (450.04,7870.2) and (451.4,7869.24) .. (453.72,7869.64) -- (455.26,7868.55) -- (455.26,7868.55) ;
%Straight Lines [id:da11390765292048055] 
\draw    (416.93,7865.19) .. controls (419.27,7864.9) and (420.59,7865.92) .. (420.88,7868.26) .. controls (421.17,7870.6) and (422.49,7871.62) .. (424.83,7871.32) .. controls (427.17,7871.03) and (428.49,7872.05) .. (428.78,7874.39) .. controls (429.07,7876.73) and (430.39,7877.75) .. (432.73,7877.45) .. controls (435.07,7877.16) and (436.39,7878.18) .. (436.68,7880.52) .. controls (436.97,7882.86) and (438.29,7883.88) .. (440.63,7883.58) .. controls (442.97,7883.29) and (444.29,7884.31) .. (444.58,7886.65) .. controls (444.87,7888.99) and (446.19,7890.01) .. (448.53,7889.72) .. controls (450.87,7889.42) and (452.19,7890.44) .. (452.48,7892.78) -- (456.04,7895.55) -- (456.04,7895.55) ;
%Straight Lines [id:da7152834100861238] 
\draw    (416.93,7840.1) -- (416.93,7921.05) ;
%Straight Lines [id:da5077281202246724] 
\draw    (456.04,7840.1) -- (456.04,7921.05) ;
%Straight Lines [id:da5776352459676277] 
\draw    (455.65,7904.86) .. controls (457.34,7903.22) and (459.01,7903.25) .. (460.65,7904.94) .. controls (462.29,7906.63) and (463.96,7906.66) .. (465.65,7905.02) .. controls (467.34,7903.38) and (469.01,7903.41) .. (470.65,7905.1) .. controls (472.29,7906.79) and (473.96,7906.82) .. (475.65,7905.18) -- (475.99,7905.18) -- (475.99,7905.18) ;
%Straight Lines [id:da2205012041801131] 
\draw    (255,7993.5) .. controls (256.67,7991.83) and (258.33,7991.83) .. (260,7993.5) .. controls (261.67,7995.17) and (263.33,7995.17) .. (265,7993.5) .. controls (266.67,7991.83) and (268.33,7991.83) .. (270,7993.5) .. controls (271.67,7995.17) and (273.33,7995.17) .. (275,7993.5) .. controls (276.67,7991.83) and (278.33,7991.83) .. (280,7993.5) .. controls (281.67,7995.17) and (283.33,7995.17) .. (285,7993.5) .. controls (286.67,7991.83) and (288.33,7991.83) .. (290,7993.5) -- (294.11,7993.5) -- (294.11,7993.5) ;
%Straight Lines [id:da4712425261062618] 
\draw    (255,7963.14) .. controls (256.67,7961.47) and (258.33,7961.47) .. (260,7963.14) .. controls (261.67,7964.81) and (263.33,7964.81) .. (265,7963.14) .. controls (266.67,7961.47) and (268.33,7961.47) .. (270,7963.14) .. controls (271.67,7964.81) and (273.33,7964.81) .. (275,7963.14) .. controls (276.67,7961.47) and (278.33,7961.47) .. (280,7963.14) .. controls (281.67,7964.81) and (283.33,7964.81) .. (285,7963.14) .. controls (286.67,7961.47) and (288.33,7961.47) .. (290,7963.14) -- (294.11,7963.14) -- (294.11,7963.14) ;
%Straight Lines [id:da12310092119553828] 
\draw    (255,7938.05) -- (255,8019) ;
%Straight Lines [id:da26196159617219017] 
\draw    (294.11,7938.05) -- (294.11,8019) ;
%Straight Lines [id:da3323959156908294] 
\draw    (294.11,7963.14) .. controls (295.8,7961.5) and (297.47,7961.53) .. (299.11,7963.22) .. controls (300.75,7964.91) and (302.42,7964.94) .. (304.11,7963.3) .. controls (305.8,7961.66) and (307.47,7961.69) .. (309.11,7963.38) .. controls (310.75,7965.07) and (312.42,7965.1) .. (314.11,7963.46) -- (314.45,7963.47) -- (314.45,7963.47) ;
%Straight Lines [id:da6638303696687318] 
\draw    (335.57,7993.5) .. controls (337.24,7991.83) and (338.9,7991.83) .. (340.57,7993.5) .. controls (342.24,7995.17) and (343.9,7995.17) .. (345.57,7993.5) .. controls (347.24,7991.83) and (348.9,7991.83) .. (350.57,7993.5) .. controls (352.24,7995.17) and (353.9,7995.17) .. (355.57,7993.5) .. controls (357.24,7991.83) and (358.9,7991.83) .. (360.57,7993.5) .. controls (362.24,7995.17) and (363.9,7995.17) .. (365.57,7993.5) .. controls (367.24,7991.83) and (368.9,7991.83) .. (370.57,7993.5) -- (374.69,7993.5) -- (374.69,7993.5) ;
%Straight Lines [id:da0650083926797358] 
\draw    (335.57,7963.14) .. controls (337.24,7961.47) and (338.9,7961.47) .. (340.57,7963.14) .. controls (342.24,7964.81) and (343.9,7964.81) .. (345.57,7963.14) .. controls (347.24,7961.47) and (348.9,7961.47) .. (350.57,7963.14) .. controls (352.24,7964.81) and (353.9,7964.81) .. (355.57,7963.14) .. controls (357.24,7961.47) and (358.9,7961.47) .. (360.57,7963.14) .. controls (362.24,7964.81) and (363.9,7964.81) .. (365.57,7963.14) .. controls (367.24,7961.47) and (368.9,7961.47) .. (370.57,7963.14) -- (374.69,7963.14) -- (374.69,7963.14) ;
%Straight Lines [id:da7842944197244812] 
\draw    (335.57,7938.05) -- (335.57,8019) ;
%Straight Lines [id:da012446744040506497] 
\draw    (374.69,7938.05) -- (374.69,8019) ;
%Straight Lines [id:da8505405272977846] 
\draw    (374.69,7993.5) .. controls (376.38,7991.86) and (378.04,7991.89) .. (379.68,7993.58) .. controls (381.32,7995.27) and (382.99,7995.3) .. (384.68,7993.66) .. controls (386.37,7992.02) and (388.04,7992.05) .. (389.68,7993.74) .. controls (391.32,7995.43) and (392.99,7995.46) .. (394.68,7993.82) -- (395.02,7993.83) -- (395.02,7993.83) ;
%Straight Lines [id:da4277406591331179] 
\draw    (416.93,7992.69) .. controls (417.22,7990.35) and (418.54,7989.33) .. (420.88,7989.62) .. controls (423.22,7989.92) and (424.54,7988.9) .. (424.83,7986.56) .. controls (425.12,7984.22) and (426.44,7983.2) .. (428.78,7983.49) .. controls (431.12,7983.79) and (432.44,7982.77) .. (432.73,7980.43) .. controls (433.02,7978.09) and (434.34,7977.07) .. (436.68,7977.36) .. controls (439.02,7977.66) and (440.34,7976.64) .. (440.63,7974.3) .. controls (440.92,7971.96) and (442.24,7970.94) .. (444.58,7971.23) .. controls (446.92,7971.53) and (448.24,7970.51) .. (448.53,7968.17) .. controls (448.82,7965.83) and (450.14,7964.81) .. (452.48,7965.1) -- (456.04,7962.33) -- (456.04,7962.33) ;
%Straight Lines [id:da9090890123174797] 
\draw    (416.93,7962.33) .. controls (419.27,7962.04) and (420.59,7963.06) .. (420.88,7965.4) .. controls (421.17,7967.74) and (422.49,7968.76) .. (424.83,7968.46) .. controls (427.17,7968.17) and (428.49,7969.19) .. (428.78,7971.53) .. controls (429.07,7973.87) and (430.39,7974.89) .. (432.73,7974.6) .. controls (435.07,7974.3) and (436.39,7975.32) .. (436.68,7977.66) .. controls (436.97,7980) and (438.29,7981.02) .. (440.63,7980.73) .. controls (442.97,7980.43) and (444.29,7981.45) .. (444.58,7983.79) .. controls (444.87,7986.13) and (446.19,7987.15) .. (448.53,7986.86) .. controls (450.87,7986.56) and (452.19,7987.58) .. (452.48,7989.92) -- (456.04,7992.69) -- (456.04,7992.69) ;
%Straight Lines [id:da933402350072936] 
\draw    (416.93,7937.24) -- (416.93,8018.19) ;
%Straight Lines [id:da9085599104257132] 
\draw    (456.04,7937.24) -- (456.04,8018.19) ;
%Straight Lines [id:da5174048307841594] 
\draw    (456.04,7962.33) .. controls (457.73,7960.69) and (459.4,7960.72) .. (461.04,7962.41) .. controls (462.68,7964.1) and (464.35,7964.13) .. (466.04,7962.49) .. controls (467.73,7960.85) and (469.4,7960.88) .. (471.04,7962.57) .. controls (472.68,7964.26) and (474.35,7964.29) .. (476.04,7962.65) -- (476.38,7962.66) -- (476.38,7962.66) ;
%Straight Lines [id:da07722024856430121] 
\draw    (495.55,7992.07) .. controls (495.84,7989.73) and (497.16,7988.71) .. (499.5,7989.01) .. controls (501.84,7989.3) and (503.16,7988.28) .. (503.45,7985.94) .. controls (503.74,7983.6) and (505.06,7982.58) .. (507.4,7982.87) .. controls (509.74,7983.17) and (511.06,7982.15) .. (511.35,7979.81) .. controls (511.64,7977.47) and (512.96,7976.45) .. (515.3,7976.74) .. controls (517.64,7977.04) and (518.96,7976.02) .. (519.25,7973.68) .. controls (519.54,7971.34) and (520.86,7970.32) .. (523.2,7970.61) .. controls (525.54,7970.91) and (526.86,7969.89) .. (527.15,7967.55) .. controls (527.44,7965.21) and (528.76,7964.19) .. (531.1,7964.48) -- (534.66,7961.71) -- (534.66,7961.71) ;
%Straight Lines [id:da3684293971768404] 
\draw    (495.55,7961.71) .. controls (497.89,7961.42) and (499.21,7962.44) .. (499.5,7964.78) .. controls (499.79,7967.12) and (501.11,7968.14) .. (503.45,7967.85) .. controls (505.79,7967.55) and (507.11,7968.57) .. (507.4,7970.91) .. controls (507.69,7973.25) and (509.01,7974.27) .. (511.35,7973.98) .. controls (513.69,7973.68) and (515.01,7974.7) .. (515.3,7977.04) .. controls (515.59,7979.38) and (516.91,7980.4) .. (519.25,7980.11) .. controls (521.59,7979.81) and (522.91,7980.83) .. (523.2,7983.17) .. controls (523.49,7985.51) and (524.81,7986.53) .. (527.15,7986.24) .. controls (529.49,7985.95) and (530.81,7986.97) .. (531.1,7989.31) -- (534.66,7992.07) -- (534.66,7992.07) ;
%Straight Lines [id:da8461244800390422] 
\draw    (495.55,7936.62) -- (495.55,8017.57) ;
%Straight Lines [id:da678594609465774] 
\draw    (534.66,7936.62) -- (534.66,8017.57) ;
%Straight Lines [id:da8302351989142132] 
\draw    (534.66,7992.07) .. controls (536.35,7990.43) and (538.02,7990.46) .. (539.66,7992.15) .. controls (541.3,7993.84) and (542.97,7993.87) .. (544.66,7992.23) .. controls (546.35,7990.59) and (548.02,7990.62) .. (549.66,7992.31) .. controls (551.3,7994) and (552.97,7994.03) .. (554.66,7992.39) -- (555,7992.4) -- (555,7992.4) ;
%Straight Lines [id:da16836772723990046] 
\draw    (106,7795.67) -- (106.15,7776.3) ;
\draw [shift={(106.17,7773.3)}, rotate = 450.43] [fill={rgb, 255:red, 0; green, 0; blue, 0 }  ][line width=0.08]  [draw opacity=0] (5.36,-2.57) -- (0,0) -- (5.36,2.57) -- cycle    ;

% Text Node
\draw (30,7760.67) node [anchor=north west][inner sep=0.75pt]    {$ p_{1} + q_1 $};
% Text Node
\draw (118,7760.67) node [anchor=north west][inner sep=0.75pt]    {$ p _{2} + q_2  $};
% Text Node
\draw (93,7750.67) node [anchor=north west][inner sep=0.75pt]    {$k$};
% Text Node
\draw (50,7891.67) node [anchor=north west][inner sep=0.75pt]    {$ p_{1}$};
% Text Node
\draw (132,7890.67) node [anchor=north west][inner sep=0.75pt]    {$  p_{2}  $};
% Text Node
\draw (184,7821) node [anchor=north west][inner sep=0.75pt]    {$=$};
% Text Node
\draw (306,7779.33) node [anchor=north west][inner sep=0.75pt]    {$+$};
% Text Node
\draw (394,7778.33) node [anchor=north west][inner sep=0.75pt]    {$+$};
% Text Node
\draw (475,7778.33) node [anchor=north west][inner sep=0.75pt]    {$+$};
% Text Node
\draw (306,7872.33) node [anchor=north west][inner sep=0.75pt]    {$+$};
% Text Node
\draw (475,7872.33) node [anchor=north west][inner sep=0.75pt]    {$+$};
% Text Node

\draw (306,7972.33) node [anchor=north west][inner sep=0.75pt]    {$+$};

% Text Node
\draw (393,7872.33) node [anchor=north west][inner sep=0.75pt]    {$+$};
% Text Node
\draw (393,7972.33) node [anchor=north west][inner sep=0.75pt]    {$+$};
% Text Node
\draw (475.21,7972.5) node [anchor=north west][inner sep=0.75pt]    {$+$};
% Text Node
\draw (555.21,7965.5) node [anchor=north west][inner sep=0.75pt]    {$+\, \cdots$};
\end{tikzpicture}
\end{figure}
\end{center}
The ellipsis indicate purely quantum diagrams which are not relevant for us.
 
As in section~\ref{sec:6_point_tree} we tidy our expressions up by making use of the on-shell conditions. Using the momentum labelling in the figure above these read 
\begin{equation}\label{sciell}
 p _1\cdot \bar{q}_1=p_2\cdot \bar{q}_2=\mathcal{O}(\hbar).
\end{equation}

Keeping this in mind we start to compute the diagrams. It is helpful to compute the first six diagrams in the figure above, which involve only three-point vertices, separately from the rest of the diagrams (which involved contact four-point vertices). We therefore write
\[
A_{(2,3)} = A_{(2,3)}^\textrm{3pt} + A_{(2,3)}^\textrm{4pt}\,. 
\]
We focus first on the six diagrams which constitute $A_{(2,3)}^\textrm{3pt}$. For clarity, let us begin by describing the contribution from the pentagon diagram, in our gauge~\eqref{gaugee} in detail. We choose the momentum routing shown in figure~\ref{fig:pentagon}. On the support of the momentum-conserving delta function 
$\hat{\delta}^{4}(q_1+ q_2+k)$, its contribution to $A_{(2,3)}$ is
\begin{equation}
i e^5\int \frac{\hat{{\d}}^4 \bar{l}}{\bar{l}^2(\bar{l}-\bar{q}_1)^2}\frac{(4 p_1\cdot p_2)^2(- 2\, \varepsilon_\hel\cdot \bar{l})}{(2p_1\cdot\bar{l})(-2p_2\cdot\bar{l})(-2p_2\cdot (\bar{k}+\bar{l}))}.
\end{equation}
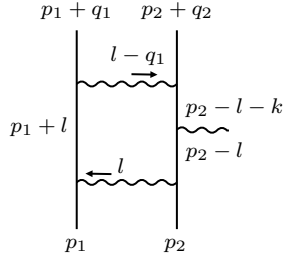
\begin{figure}[t]
\begin{center}
\tikzset{every picture/.style={line width=0.75pt}} %set default line width to 0.75pt        

\begin{tikzpicture}[x=0.75pt,y=0.75pt,yscale=-1,xscale=1]
%uncomment if require: \path (0,8388); %set diagram left start at 0, and has height of 8388

%Straight Lines [id:da8653684822705865] 
\draw    (84,5367.68) .. controls (85.67,5366.01) and (87.33,5366.01) .. (89,5367.68) .. controls (90.67,5369.35) and (92.33,5369.35) .. (94,5367.68) .. controls (95.67,5366.01) and (97.33,5366.01) .. (99,5367.68) .. controls (100.67,5369.35) and (102.33,5369.35) .. (104,5367.68) .. controls (105.67,5366.01) and (107.33,5366.01) .. (109,5367.68) .. controls (110.67,5369.35) and (112.33,5369.35) .. (114,5367.68) .. controls (115.67,5366.01) and (117.33,5366.01) .. (119,5367.68) .. controls (120.67,5369.35) and (122.33,5369.35) .. (124,5367.68) .. controls (125.67,5366.01) and (127.33,5366.01) .. (129,5367.68) .. controls (130.67,5369.35) and (132.33,5369.35) .. (134,5367.68) -- (134,5367.68) ;
%Straight Lines [id:da8510637105674375] 
\draw    (84,5318.18) .. controls (85.67,5316.51) and (87.33,5316.51) .. (89,5318.18) .. controls (90.67,5319.85) and (92.33,5319.85) .. (94,5318.18) .. controls (95.67,5316.51) and (97.33,5316.51) .. (99,5318.18) .. controls (100.67,5319.85) and (102.33,5319.85) .. (104,5318.18) .. controls (105.67,5316.51) and (107.33,5316.51) .. (109,5318.18) .. controls (110.67,5319.85) and (112.33,5319.85) .. (114,5318.18) .. controls (115.67,5316.51) and (117.33,5316.51) .. (119,5318.18) .. controls (120.67,5319.85) and (122.33,5319.85) .. (124,5318.18) .. controls (125.67,5316.51) and (127.33,5316.51) .. (129,5318.18) .. controls (130.67,5319.85) and (132.33,5319.85) .. (134,5318.18) -- (134,5318.18) ;
%Straight Lines [id:da3865356923261385] 
\draw    (84,5291.18) -- (84,5391.18) ;
%Straight Lines [id:da5032520573553434] 
\draw    (134,5291.18) -- (134,5391.18) ;
%Straight Lines [id:da4297581371610062] 
\draw    (91.33,5363.31) -- (101.33,5363.5) ;
\draw [shift={(88.33,5363.25)}, rotate = 1.1] [fill={rgb, 255:red, 0; green, 0; blue, 0 }  ][line width=0.08]  [draw opacity=0] (3.57,-1.72) -- (0,0) -- (3.57,1.72) -- cycle    ;
%Straight Lines [id:da38939518364816905] 
\draw    (111.33,5313.25) -- (121.33,5313.44) ;
\draw [shift={(124.33,5313.5)}, rotate = 181.1] [fill={rgb, 255:red, 0; green, 0; blue, 0 }  ][line width=0.08]  [draw opacity=0] (3.57,-1.72) -- (0,0) -- (3.57,1.72) -- cycle    ;
%Straight Lines [id:da15810726817885734] 
\draw    (134,5341.18) .. controls (135.69,5339.54) and (137.36,5339.57) .. (139,5341.26) .. controls (140.65,5342.95) and (142.31,5342.97) .. (144,5341.33) .. controls (145.69,5339.69) and (147.36,5339.72) .. (149,5341.41) .. controls (150.64,5343.1) and (152.31,5343.13) .. (154,5341.49) .. controls (155.69,5339.85) and (157.36,5339.88) .. (159,5341.57) -- (160,5341.58) -- (160,5341.58) ;

% Text Node
\draw (64.83,5275.5) node [anchor=north west][inner sep=0.75pt]  [font=\scriptsize]  {$p_{1} + q_1$};
% Text Node
\draw (76.83,5394.83) node [anchor=north west][inner sep=0.75pt]  [font=\scriptsize]  {$p_{1}$};
% Text Node
\draw (126.33,5394.83) node [anchor=north west][inner sep=0.75pt]  [font=\scriptsize]  {$p_{2}$};
% Text Node
\draw (114.83,5275.5) node [anchor=north west][inner sep=0.75pt]  [font=\scriptsize]  {$p_{2} + q_2$};
% Text Node
\draw (102.83,5354.01) node [anchor=north west][inner sep=0.75pt]  [font=\scriptsize]  {$l$};
% Text Node
\draw (99.33,5298.01) node [anchor=north west][inner sep=0.75pt]  [font=\scriptsize]  {$l-q_1$};
% Text Node
\draw (137,5346.18) node [anchor=north west][inner sep=0.75pt]  [font=\scriptsize]  {$p_{2} -l$};
% Text Node
\draw (49.5,5333.68) node [anchor=north west][inner sep=0.75pt]  [font=\scriptsize]  {$p_{1} +l$};
% Text Node
\draw (137,5324.18) node [anchor=north west][inner sep=0.75pt]  [font=\scriptsize]  {$p_{2} -l-k$};
% Text Node

\end{tikzpicture}
\end{center}
\caption{Momentum routing for the pentagon contribution to the five-point one-loop amplitude.}
\label{fig:pentagon}
\end{figure}
The sum of the cubic diagrams yields 
\begin{equation} \label{eq:boxes}
\begin{split}
A_{(2,3)}^\textrm{3pt}&= ie^5 \int \frac{\hat{{\d}}^4 \bar{l}}{\bar{l}^2(\bar{l}-\bar{q}_1)^2} (4p_1\cdot p_2)^2 \times  \\& \left(
\frac{ -2\varepsilon_\hel\cdot\bar{l}
}{(2p_1\cdot\bar{l})(-2p_2\cdot \bar{l})(-2p_2\cdot(\bar{k}+\bar{l}))} \right.
-  
\frac{2  \varepsilon_\hel \cdot \bar{q}_1}{(2p_1\cdot\bar{l})(-2p_2\cdot \bar{l})(2p_2\cdot \bar{k})}
\\&+  
\frac{ 2\varepsilon_\hel\cdot(\bar{l}-\bar{q}_1)
}{(2p_1\cdot\bar{l})(2p_2\cdot (\bar{k}+\bar{l}))(2p_2\cdot\bar{l})}
-\left.   
\frac{2\varepsilon_\hel\cdot\bar{q}_1
}{(2p_1\cdot\bar{l})(2p_2\cdot(\bar{k}+\bar{l}))(2p_2\cdot\bar{k})}\right).
\end{split}
\end{equation}
Note that the signs in the linearised propagators are important! We use them to indicate the hidden $i\epsilon$'s,
\[
\frac{1}{\pm p \cdot\bar{l}}\equiv\frac{1}{\pm p\cdot\bar{l}+ i\epsilon} \neq \pm\frac{1}{ p \cdot\bar{l}}=\frac{1}{\pm p \cdot\bar{l}\pm i\epsilon}.
\] 
which allows us to make use of the following identity 
\[-i\hat{\delta}(p\cdot\bar{l})=\frac{1}{p\cdot\bar{l}}+\frac{1}{-p\cdot\bar{l}}.
\label{eq:delta}\]
In order to make use of this identity we apply a change of variables $l \to \bar q_1 -l$ in the last two terms in \eqref{eq:boxes}. This, along with the on-shell conditions, allows pairs of terms to take an almost identical form --- denominators differ only by a sign in the $p_1\cdot l$ term which is precisely what is needed to apply \eqref{eq:delta}. The amplitude then reduces to
\begin{equation}\label{eq:finbox}
\begin{split}
A_{(2,3)}^\textrm{3pt}&=4 e^5 (p_1\cdot p_2)^2 \int \frac{\hat{{\d}}^4 \bar{l}}{\bar{l}^2(\bar{l}-\bar{q}_1)^2} 
\frac{\hat{\delta}(p_1\cdot\bar{l})}{-p_2\cdot\bar{l}}\left(-\frac{\varepsilon_\hel\cdot\bar{q}_1 
}{p_2\cdot\bar{k}} -\frac{ \varepsilon_\hel\cdot\bar{l} 
}{-p_2\cdot(\bar{k}+\bar{l})}\right) \,.
\end{split}
\end{equation}  
It is possible to expose a second delta function in this expression by writing $-\varepsilon_\hel\cdot\bar{q}_1 = \varepsilon_\hel\cdot (\bar{l}-\bar{q}_1) - \varepsilon_\hel\cdot \bar{l}$. The two terms involving $\varepsilon_\hel\cdot \bar{l}$ under the integral sign can be simplified by a partial fraction, yielding
\[
\label{eq:finbox2}
A_{(2,3)}^\textrm{3pt}&=4 e^5 (p_1\cdot p_2)^2 \int \frac{\hat{{\d}}^4 \bar{l}}{\bar{l}^2(\bar{l}-\bar{q}_1)^2} 
\frac{\hat{\delta}(p_1\cdot\bar{l})}{p_2\cdot\kb}\left(\frac{ \varepsilon_\hel\cdot(\bar{l}-\bar{q}_1)
}{-p_2\cdot\bar{l}} - \frac{ \varepsilon_\hel\cdot\bar{l} 
}{-p_2\cdot(\bar{l}-\bar{q}_1)}\right) \\
&=4i e^5 (p_1\cdot p_2)^2 \int \frac{\hat{{\d}}^4 \bar{l}}{\bar{l}^2(\bar{l}-\bar{q}_1)^2}  \hat{\delta}(p_1\cdot\bar{l}) \hat{\delta}(p_2\cdot(\bar{l} -\qb_1))
\frac{\varepsilon_\hel\cdot\bar{l}}{p_2\cdot\kb} \,.
\]
To obtain the second of these equalities, we redefined the variable of integration to $\bar{l}' = -(\bar{l}-\qb_1)$ in the first term, and dropped the prime.

Next, we address the remaining diagrams contributing to $A_{(2,3)}$ which now involve four-point vertices.
After a straightforward computation, we find
\[  \label{eq:fincntct}
A_{(2,3)}^\textrm{4pt}&= 
{4ie^5}p_1\cdot p_2 \, \varepsilon_\hel \cdot p_1\int \frac{\hat{\d}^4 \bar{l}}{\bar{l}^2(\bar{l}-\bar{q}_1)^2}\hat{\delta}(p_1\cdot \bar{l}) \hat{\delta}(p_2\cdot (\bar{l}-\qb_1)) \, .
\]
At this stage we can see the structure of the required factorisation --- we have exposed the delta functions present in equation~\eqref{eq:fivePointRelation}.

Combining the contact terms of equation~\eqref{eq:fincntct} with the rest of the diagrams, equation~\eqref{eq:finbox2}, we find that the total expression for the amplitude fragment is
\[
A_{(2,3)}&= {4ie^5} \, p_1 \cdot p_2 \int \frac{\hat{\d}^4 \bar{l}}{\bar{l}^2(\bar{l}-\bar{q}_1)^2} \hat{\delta}(p_1\cdot \bar{l}) \hat{\delta}(p_2\cdot (\bar{l}-\qb_1)) \left( \varepsilon_\hel \cdot p_1 
 +p_1\cdot p_2\frac{\varepsilon_\hel\cdot\bar{l} }{p_2\cdot\kb} \right) \,.
 \label{eq:A23final}
\]

The final step is to compare this result with the prediction for $A_{(2,3)}$ from equation~\eqref{eq:fivePointRelation}. Using the amplitudes of equation~\eqref{eq:expicitAs}, it is easy to see that the prediction is
\[
A_{(2,3)}&= \int \dd^4 \bar{w}_1 \dd^4 \bar{w_2} \, \del(2 p_1 \cdot \bar{w}_1) \del(2 p_2 \cdot \bar{w}_2) \del^4(\qb_1 + \qb_2 - \bar{w}_1 - \bar{w}_2) \\
&\hspace{40pt}\times \frac{4e^3}{\bar{w}_1^2} \left( p_1 \cdot \varepsilon_\eta + \frac{p_1 \cdot p_2 \, \varepsilon_\eta \cdot \bar{w}_1}{p_2 \cdot \bar{k}} \right) 
\frac{4 e^2}{(\qb_1- \bar{w}_1)^2}p_1 \cdot p_2 \,.
\]
Upon performing the integral over $\bar{w}_2$ using the four-fold delta function, relabelling $\bar{w}_1 = \bar{l}$ and recognising that $\kb = -\qb_1 -\qb_2$, we immediately recover equation~\eqref{eq:A23final}.

\section{Generalising the Eikonal}
\label{sec:eikonal-review}

We have now seen that the uncertainty, or the variance, in the measurement of a scattering observable can be computed in terms of amplitudes and, moreover, that the classical absence of uncertainty leads to an infinite set of relationships among fragments of amplitudes expanded in powers of 
momentum transfer, which is a Laurent series in $\hbar$. 
In a purely conservative limit, these relationships can be understood in terms of eikonal exponentiation. Our goal now is to review the eikonal formula, emphasising its connection to final state dynamics. We will then build on this eikonal state to incorporate radiative dynamics as a kind of coherent state so that the variance is naturally small.

\subsection{Eikonal final state}
\label{sec:KMOC_eikonal}

Eikonal methods have long been used to extract classical physics from quantum mechanics. Recent years have seen a renewed surge of interest in this approach, especially in the context of gravitational scattering~\cite{Saotome:2012vy,Akhoury:2013yua,Ciafaloni:2015xsr,Bjerrum-Bohr:2016hpa,Luna:2016idw,Bjerrum-Bohr:2018xdl,Ciafaloni:2018uwe,DiVecchia:2019kta,DiVecchia:2019myk,KoemansCollado:2019ggb,Bern:2020gjj,Parra-Martinez:2020dzs,DiVecchia:2020ymx,Mogull:2020sak,DiVecchia:2021bdo,DiVecchia:2021ndb,Jakobsen:2021smu,Shi:2021qsb,Heissenberg:2021tzo,Haddad:2021znf,Herrmann:2021tct,Bjerrum-Bohr:2021din,Damgaard:2021ipf,Emond:2021lfy}, though this has roots in earlier 
work~\cite{Amati:1990xe,Kabat:1992tb,Amati:2007ak}. Originally born out of the study of high energy/Regge scattering~\cite{PhysRev.143.1194,Abarbanel:1969ek,Cheng:1969eh,PhysRev.186.1611,WALLACE1973190,PhysRev.186.1656,Wallace:1977ae, Cheng1981ConsequencesOT} where the Feynman diagrammatics dramatically simplify, eikonal physics now have much wider application. The simplification in this regime allows diagrams, expressed in impact parameter space rather than momentum space, to be summed exactly to an exponential form. This exponential depends on the $2 \rightarrow 2$ scattering amplitude, and contains information about classical quantities such as the deflection angle. There are rich connections to soft/IR physics and Wilson lines \cite{Korchemskaya:1994qp,Laenen:2008gt,Laenen:2010uz,White:2011yy,Melville:2013qca,White:2015wha} which lead to a formal proof of the exponentiation quite generally \cite{Laenen:2008gt}. Nowadays the exponentiation is taken as a starting point and applied to various scattering regimes.

In this section our goal is to explain the link between eikonal methods and the KMOC approach. Firstly it is worth noting that the small $\hbar$ expansion in the KMOC formalism is essentially the same as the soft expansion in the eikonal literature; the $\hbar$ scaling counts the order of softness. The key connection is to compute the final state using the methods of KMOC instead of computing observables directly. We will see that this final, outgoing, state is controlled by the usual eikonal function. In this section we restrict to a purely conservative scattering scenario: then eikonal exponentiation is exact. We take two incoming particles and (since the scattering is conservative) assume that the outgoing state is also an element of the two-particle Hilbert space.

We begin with the standard definition of the eikonal as the transverse Fourier transform of the four-point amplitude
\[
e^{i\chi(x; \, s)/ \hbar}(1+i \Delta(x; \, s)) - 1 = i\int \dd ^4 q \, \del(2p_1\cdot q)\del(2p_2\cdot q) e^{-i q \cdot x / \hbar} \amp_4(s,q^2)  \;,
\label{eq:eikonal_def}
\]
where $\chi(x; \, s)$ is the eikonal function and $\Delta(x; \, s)$ is the so-called quantum remainder which takes into account contributions that do not exponentiate (see for example~\cite{DiVecchia:2021bdo}). This remainder is important for computing the eikonal function --- but it will play no role in the remainder of our article, so we will omit it. Meanwhile the two Dirac delta functions appearing in equation~\eqref{eq:eikonal_def} ensure that we integrate only over the components of $q$ transverse to the momenta. This is often just written instead as $\dd ^{D-2}q_{\perp}$ (times a Jacobian factor). Indeed, the parameter $x$ should be thought of as an element of the $D-2$ dimensional spatial slice perpendicular to $p_1$ and $p_2$: this is most evident on the right-hand-side of the equation, where the Dirac delta functions project away any components of $q$ in the (timelike) $p_1$ and $p_2$ directions. Consequently no components of $x$ in the space spanned by $p_1$ and $p_2$ enter the dot product $q\cdot x$.

The eikonal can be written as an expansion in powers of the generic coupling $g$
\[\label{eq:eikonalExpansion}
\chi(x; \, s)=\sum_{n=0}^{\infty}\chi_{n}(x; \, s) \quad , \quad \chi_{n}(x; \, s) \sim g^{2n} \,,
\] 
and we will, as others have (see for example~\cite{KoemansCollado:2018hss,KoemansCollado:2019ggb,DiVecchia:2019myk,Bern:2020gjj,Parra-Martinez:2020dzs,DiVecchia:2021bdo}), assume that this holds to all orders. The structure has been formally proven at leading order, (see for example \cite{Laenen:2008gt,Akhoury:2013yua}), however an all orders proof has not been given (to the best of our knowledge).

Starting with the in state \eqref{eq:psi} from section \ref{sec:expectation_of_two_F}, we obtain the final state by acting with the $S$ matrix. Writing $S = 1+i T$ and inserting a complete set of states, we have
\[
S \ket{\psi} =& \ket{\psi} + i \int \dPhi(p_1',p_2',p_1,p_2) \, \phi_{b}(p_1,p_2) \ket{p_1', p_2'} \bra{p_1',p_2'} T \ket{p_1, p_2} 
\\
=& \ket {\psi} + i \int \dPhi(p_1',p_2',p_1,p_2) \, \phi_{b} (p_1,p_2)  \amp_4(s, q^2) 
\delta^{(4)}(p_1 \!+ \! p_2\! -\! p_1'\! - \!p_2') \ket{p_1', p_2'} \,.
\]
Notice that we made explicit use of our assumption of conservative scattering by restricting the complete set of states to the two-particle Hilbert space.

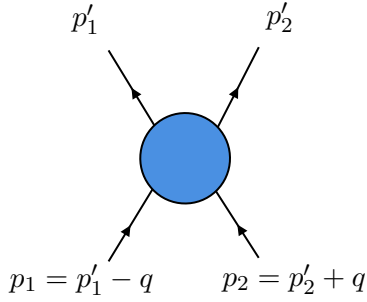
\begin{figure}[t]
\begin{center}

\tikzset{every picture/.style={line width=0.75pt}} %set default line width to 0.75pt

\begin{tikzpicture}[x=0.75pt,y=0.75pt,yscale=-1,xscale=1]
%uncomment if require: \path (0,516); %set diagram left start at 0, and has height of 516

%Shape: Ellipse [id:dp46332193163472657]
\draw [fill={rgb, 255:red, 74; green, 144; blue, 226 } ,fill opacity=1 ] (87.64,433.04) .. controls (79.15,423.91) and (79.56,409.51) .. (88.56,400.89) .. controls (97.56,392.27) and (111.73,392.69) .. (120.21,401.83) .. controls (128.7,410.96) and (128.29,425.36) .. (119.29,433.98) .. controls (110.29,442.6) and (96.12,442.18) .. (87.64,433.04) -- cycle ;
%Straight Lines [id:da25724527320075596]
\draw (88.56,400.89) -- (65.7,363.17) ;
\draw [shift={(77.13,382.03)}, rotate = 58.78] [fill={rgb, 255:red, 0; green, 0; blue, 0 } ][line width=0.08] [draw opacity=0] (5.36,-2.57) -- (0,0) -- (5.36,2.57) -- cycle ;
%Straight Lines [id:da5677841483207073]
\draw (120.21,401.82) -- (140.7,361.55) ;
\draw [shift={(130.46,381.69)}, rotate = 116.96] [fill={rgb, 255:red, 0; green, 0; blue, 0 } ][line width=0.08] [draw opacity=0] (5.36,-2.57) -- (0,0) -- (5.36,2.57) -- cycle ;
%Straight Lines [id:da7710269791476811]
\draw (65.71,469.3) -- (87.64,433.04) ;
\draw [shift={(76.67,451.17)}, rotate = 121.16] [fill={rgb, 255:red, 0; green, 0; blue, 0 } ][line width=0.08] [draw opacity=0] (5.36,-2.57) -- (0,0) -- (5.36,2.57) -- cycle ;
%Straight Lines [id:da19080369401651964]
\draw (140.66,467.51) -- (119.29,433.98) ;
\draw [shift={(129.97,450.75)}, rotate = 57.5] [fill={rgb, 255:red, 0; green, 0; blue, 0 } ][line width=0.08] [draw opacity=0] (5.36,-2.57) -- (0,0) -- (5.36,2.57) -- cycle ;

% Text Node
\draw (46,338.33) node [anchor=north west][inner sep=0.75pt] {$p_{1} '$};
% Text Node
\draw (143,337.33) node [anchor=north west][inner sep=0.75pt] {$p'_{2}$};
% Text Node
\draw (15,469.33) node [anchor=north west][inner sep=0.75pt] {$p_{1} =p'_{1} -q$};
% Text Node
\draw (121,469) node [anchor=north west][inner sep=0.75pt] {$p_{2} =p'_{2} +q$};

\end{tikzpicture}

\end{center}

\caption{Momentum labelling at four points}
\label{fig:fourpoints}
\end{figure}

With the momentum labelling in figure \ref{fig:fourpoints} we can convert the $p_1$ and $p_2$ phase space integrals to integrals over $q$.
Doing so, we may write 
\[
&S\ket \psi  = \ket {\psi} \\
& \quad + i \int \dPhi(p_1',p_2') \, \dd^4 q \, \del (2p_1'\!\cdot \! q \!- \!q^2)\del (2p_2'\!\cdot \!q \!+ \!q^2) \, \phi_{b} (p_1'\!-\!q , p_2'\!+\!q) %\\& \quad\times
\amp_4(s, q^2)  \ket{p_1', p_2'}.
\label{eq:stepToFinalState}
\]
At this point, the $q$ integral is tantalising similar to the $q$ integral in the eikonal formula~\eqref{eq:eikonal_def}. However there is a key difference in the nature of the delta functions: those in equation~\eqref{eq:stepToFinalState} involve $q^2$ terms which are absent in equation~\eqref{eq:eikonal_def}. This issue appeared in refs.~\cite{Brandhuber:2021kpo,Brandhuber:2021eyq}: there the authors proceeded using a ``HEFT'' phase, which is analogous to yet distinct from the eikonal phase. We will instead continue with the eikonal phase.

It may be worth remarking that the $q^2$ terms in these delta functions are suppressed in specific examples. One such example is the leading order impulse~\cite{KMOC}. Nevertheless the impulse at NLO does indeed involve the full delta functions (for the loop integrals)~\cite{KMOC}.

To incorporate the full delta functions, we follow the route described in~\cite{Parra-Martinez:2020dzs}. We introduce new momentum variables $\tilde{p}$ as 
\[\label{exactonshell}
\tilde{p}_1 = p_1' -\frac{q}{2} \qquad \tilde{p}_2 = p'_2 +\frac{q}{2}.
\]
Now, rather than using the eikonal equation~\eqref{eq:eikonal_def} directly we can take advantage of its inverse Fourier transform in the following form\footnote{As noted above, we dropped the quantum remainder $\Delta$ which plays no role in our analysis.}
\[
i\del(2 \tilde{p}_1\cdot q)\del(2 \tilde{p}_2\cdot q) \amp_4(s, q^2) = \frac{1}{\hbar^4} \int \d^4 x \, e^{iq\cdot x / \hbar}\left\{e^{i\chi (x_{\perp}; \, s)/ \hbar}-1\right\},
\label{eq:finalStateStep}
\]
where we have written $x_\perp$ as one of the arguments of the eikonal function $\chi(x_{\perp}; \, s)$ to emphasise that $\chi(x_{\perp}; \, s)$ only depends on components of $x$ which are orthogonal to the space spanned by $\tilde{p}_1$ and $\tilde{p}_2$.  Indeed, integrating over the two components of $x$ which \emph{are} in the space spanned by $\tilde{p}_1$ and $\tilde{p}_2$, one recovers the two Dirac delta functions on the left-hand-side of equation~\eqref{eq:finalStateStep}. In this way, we find that the final state is
\[
\label{eq:eikonalState}
S\ket \psi  = \int \dPhi(p_1',p_2')\, \ket{p_1', p_2'} \frac{1}{\hbar^4} \int \dd^4  q \, \d^4 x  \,e^{iq\cdot x /\hbar }\, e^{i\chi(x_{\perp}; \, s) / \hbar} \, {\phi}_{b} ( p_1'-q,p_2'+q) \,.
\]
It is worth emphasising once again that $x_{\perp}$ is perpendicular to $\tilde p _i$, rather than to $p_i$, so that 
\[
\tilde p_1 \cdot x_\perp = 0 = \tilde p_2 \cdot x_\perp.
\]
In particular, $x_\perp$ depends on $q$.

\subsection{The impulse from the eikonal}
\label{sec:conservativeImpulse}

In this subsection, we will recover one beautiful result from the literature on the eikonal function: the scattering angle can be extracted from the eikonal function using a stationary phase argument. As our interest is not so much in this conservative case but rather in its radiative generalisation (which we discuss below), we wish to emphasise that it is, in fact, possible to extract the \emph{full} final momentum from the eikonal using stationary phase. Later, in section~\ref{sec:rad-react}, we will use the same ideas to extract the final momentum in the case of radiation --- with radiation, of course, knowledge of the direction of the final momentum is insufficient to recover the full momentum.

The impulse is the observable
\[
\Delta p_1^\mu &\equiv \bra{\psi}  S^\dagger \Pop_1^\mu S \ket{\psi} - \bra{\psi}  \Pop_1^\mu  \ket{\psi} \\
&= \bra{\psi}  S^\dagger [\Pop_1^\mu, S] \ket{\psi} \,.
\]
It is convenient to focus on
\[
[\Pop_1^\mu, S] \ket{\psi} 
\,.
\]
As we shall see, in essence the operator $[\Pop_1^\mu, S]$ pulls out a factor of the momentum transfer multiplying $S \ket{\psi}$. We will
evaluate $[\Pop_1^\mu, S]\ket{\psi} $ by stationary phase; it is then trivial to determine $\bra \psi S^\dagger$ in the same way.

We begin with our expression for the final-state wavepacket, equation~\eqref{eq:eikonalState}, quickly finding
\[
\label{eq:beginningImpulse}
[\Pop_1^\mu, S] \ket{\psi} =\int \dPhi(p_1',p_2') \,  e^{i b \cdot p_1' / \hbar} \,& \ket{p_1', p_2'}
\frac{1}{\hbar^4} \int \dd^4  q \,\d^4 x \,e^{iq \cdot x/ \hbar} e^{-i b \cdot q / \hbar}  \\
&\times e^{i\chi(x_{\perp}; \, s)/ \hbar} \,\, {\phi} ( p_1'-q,p_2'+q) \, q^\mu
\,.
\]
To obtain formulae for the impulse, we apply the stationary phase approximation to the $x$ and $q$ integrals. (Our approach is 
very similar to that of Ciafaloni and Colferai~\cite{Ciafaloni:2014esa} who previously discussed wavepacket dynamics, the eikonal, and stationary phase.)

The stationary phase condition for $x$ is\footnote{One can find this expression also by promoting the eikonal phase to an operator $\hat{\chi}(x_{\perp},s)$: the impulse on particle $i$ in the transverse plane $\hat{Q}^{\mu}_{\perp i}$ is then related to the standard commutator $[\hat{Q}^{\mu}_{\perp i}, e^{i\hat{\chi}(\hat{x}_{\perp}; \, s)/ \hbar}] = e^{i\hat{\chi}(\hat{x}_{\perp}; \, s)/ \hbar} \partial^{\mu}_{\perp i} \hat{\chi}(\hat{x}_{\perp},s)$ and the result follows from unitarity.} 
\[
\label{eq:stationaryImpulse}
q_{\mu} = -\frac{\partial}{\partial x^{\mu}} \chi (x_{\perp},s) \,.
\] 
One thing to note immediately is that this $q^\mu$ is \emph{not} of order $\hbar$: it is a classical momentum, of order $g^2$. Further, it is useful to note that $\chi$ is actually a function of $x_\perp^2$ (since the four-point function is a function of $s$ and $q^2$.) Therefore we may write the momentum transfer as
\[
q^{\mu} = -2 \chi'(x_\perp^2, s) x_{\perp}^\mu \,,
\]
where $\chi'(x_\perp^2, s)$ is the derivative of $\chi$ with respect to its first argument. Since $\tilde p_1 \cdot x_\perp = 0 = \tilde p_2 \cdot x_\perp$, it now follows that $\tilde p_1 \cdot q = 0 = \tilde p_2 \cdot q$: thus the on-shell delta functions in equation~\eqref{eq:finalStateStep} have reappeared, now as ``equations of motion'' following from the stationary phase conditions\footnote{Since some of the $x$ integrations may be performed exactly to yield
delta functions, we are slightly abusing terminology by referring to all of the conditions on $x$ and $q$ as ``stationary phase conditions''. 
Some of the conditions arise approximately by stationary phase, and some are exact conditions due to the delta functions. 
Nevertheless it is most convenient to treat all the conditions as one set.}.

The second stationary phase condition, associated with $q$, is
\[
x^\mu - b^\mu + \frac{\partial}{\partial q_\mu} \chi(x_\perp^2, s) = 0 \,.
\label{eq:b_x_relation}
\]
The $q$ derivative is non-vanishing because $x_\perp$ depends on $q$. 
It is often useful to introduce a particular notation for the variables $q$ and $x$ when they satisfy the stationary phase conditions: we will
denote these by $q_*$ and $x_*$.
Armed with this notation, we may use equation~\eqref{eq:stationaryImpulse} to write equation~\eqref{eq:b_x_relation} as
\[
x^\mu = b^\mu + q_{*\nu} \frac{\partial}{\partial q_\mu} x_\perp^\nu \,.
\]
Performing the derivative is straightforward, but requires some notation which we relegate to appendix~\ref{app:projector}. The result may be expressed in the form
\[
x_{*\perp}^\mu = b^\mu - \tilde{N}_q (p_1^\mu - p_2^\mu) - \tilde{N}_{0q} (p_1^\mu + p_2^\mu) \, ,
\]
where $\tilde{N}_q$ and $\tilde{N}_{0q}$ can be interpreted as Lagrange multipliers ensuring that $x_\perp \cdot \tilde p_1 = 0 = x_\perp \cdot \tilde p_2$. 

Our result for $x_{*\perp}$ takes a familiar form in the CM frame. Then $x_{*\perp}^0 = 0$ and, using, $\v{p_2} = -\v{p_1}$, we have
\[\label{eq:x-eikonal}
\v{x}_{*\perp} &= \v{b} - 2 \tilde{N}_q  \v{p} \\
\Rightarrow \v b \cdot \v{x}_{*\perp} &= \v b^2 \,.
\]
Denoting the scattering angle by $\Psi$, we may write this result\footnote{In this context, the quantity $x_\perp$ is sometimes referred to as the ``eikonal'' impact parameter.} as
\[
|\v b| = |\v{x}_{*\perp}| \cos(\Psi/2) \,,
\]
where $\Psi$ is the scattering angle.

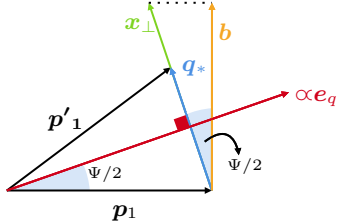
\begin{figure}[H]
\begin{center}

\tikzset{every picture/.style={line width=0.75pt}} %set default line width to 0.75pt        

\begin{tikzpicture}[x=0.75pt,y=0.75pt,yscale=-1,xscale=1]
%uncomment if require: \path (0,8388); %set diagram left start at 0, and has height of 8388

%Shape: Arc [id:dp9635008973095371] 
\draw  [draw opacity=0][fill={rgb, 255:red, 74; green, 144; blue, 226 }  ,fill opacity=0.22 ] (178.53,8286.48) .. controls (180.38,8290.75) and (181.55,8295.33) .. (181.9,8300.07) -- (140,8300) -- cycle ; \draw  [color={rgb, 255:red, 255; green, 255; blue, 255 }  ,draw opacity=1 ] (178.53,8286.48) .. controls (180.38,8290.75) and (181.55,8295.33) .. (181.9,8300.07) ;
%Straight Lines [id:da505222356584839] 
\draw [color={rgb, 255:red, 0; green, 0; blue, 0 }  ,draw opacity=1 ]   (140,8300) -- (219.44,8239.89) ;
\draw [shift={(221.83,8238.08)}, rotate = 142.89] [fill={rgb, 255:red, 0; green, 0; blue, 0 }  ,fill opacity=1 ][line width=0.08]  [draw opacity=0] (3.57,-1.72) -- (0,0) -- (3.57,1.72) -- cycle    ;
%Straight Lines [id:da45394720241734676] 
\draw [color={rgb, 255:red, 208; green, 2; blue, 27 }  ,draw opacity=1 ]   (140,8300) -- (278,8251.58) ;
\draw [shift={(280.83,8250.58)}, rotate = 160.66] [fill={rgb, 255:red, 208; green, 2; blue, 27 }  ,fill opacity=1 ][line width=0.08]  [draw opacity=0] (3.57,-1.72) -- (0,0) -- (3.57,1.72) -- cycle    ;
%Straight Lines [id:da49219037940313437] 
\draw [color={rgb, 255:red, 126; green, 211; blue, 33 }  ,draw opacity=1 ]   (242.36,8300) -- (211.78,8208.43) ;
\draw [shift={(210.83,8205.58)}, rotate = 71.54] [fill={rgb, 255:red, 126; green, 211; blue, 33 }  ,fill opacity=1 ][line width=0.08]  [draw opacity=0] (3.57,-1.72) -- (0,0) -- (3.57,1.72) -- cycle    ;
%Straight Lines [id:da6583532833060661] 
\draw [color={rgb, 255:red, 74; green, 144; blue, 226 }  ,draw opacity=1 ]   (242.36,8300) -- (226.11,8251.33) -- (222.75,8240.94) ;
\draw [shift={(221.83,8238.08)}, rotate = 72.11] [fill={rgb, 255:red, 74; green, 144; blue, 226 }  ,fill opacity=1 ][line width=0.08]  [draw opacity=0] (3.57,-1.72) -- (0,0) -- (3.57,1.72) -- cycle    ;
%Straight Lines [id:da8482000792017059] 
\draw [color={rgb, 255:red, 245; green, 166; blue, 35 }  ,draw opacity=1 ]   (242.36,8300) -- (242.4,8208.58) ;
\draw [shift={(242.4,8205.58)}, rotate = 90.02] [fill={rgb, 255:red, 245; green, 166; blue, 35 }  ,fill opacity=1 ][line width=0.08]  [draw opacity=0] (3.57,-1.72) -- (0,0) -- (3.57,1.72) -- cycle    ;
%Straight Lines [id:da40173973053963286] 
\draw  [dash pattern={on 0.84pt off 2.51pt}]  (210.83,8205.58) -- (242.4,8205.58) ;
%Curve Lines [id:da31543988226115305] 
\draw [color={rgb, 255:red, 0; green, 0; blue, 0 }  ,draw opacity=1 ]   (238.78,8275.44) .. controls (249.42,8267.18) and (253.71,8273.86) .. (255.2,8279.37) ;
\draw [shift={(255.78,8282.28)}, rotate = 262.87] [fill={rgb, 255:red, 0; green, 0; blue, 0 }  ,fill opacity=1 ][line width=0.08]  [draw opacity=0] (3.57,-1.72) -- (0,0) -- (3.57,1.72) -- cycle    ;
%Straight Lines [id:da16669278862596948] 
\draw [color={rgb, 255:red, 0; green, 0; blue, 0 }  ,draw opacity=1 ]   (140,8300) -- (239.36,8300) ;
\draw [shift={(242.36,8300)}, rotate = 180] [fill={rgb, 255:red, 0; green, 0; blue, 0 }  ,fill opacity=1 ][line width=0.08]  [draw opacity=0] (3.57,-1.72) -- (0,0) -- (3.57,1.72) -- cycle    ;
%Shape: Arc [id:dp3026068220119731] 
\draw  [draw opacity=0][fill={rgb, 255:red, 74; green, 144; blue, 226 }  ,fill opacity=0.22 ] (228.69,8261.52) .. controls (232.96,8259.65) and (237.54,8258.47) .. (242.28,8258.1) -- (242.36,8300) -- cycle ; \draw  [color={rgb, 255:red, 255; green, 255; blue, 255 }  ,draw opacity=1 ] (228.69,8261.52) .. controls (232.96,8259.65) and (237.54,8258.47) .. (242.28,8258.1) ;
%Straight Lines [id:da0642774018752541] 
\draw [color={rgb, 255:red, 74; green, 144; blue, 226 }  ,draw opacity=1 ]   (242.36,8300) -- (226.11,8251.33) -- (222.75,8240.94) ;
\draw [shift={(221.83,8238.08)}, rotate = 72.11] [fill={rgb, 255:red, 74; green, 144; blue, 226 }  ,fill opacity=1 ][line width=0.08]  [draw opacity=0] (3.57,-1.72) -- (0,0) -- (3.57,1.72) -- cycle    ;
%Straight Lines [id:da1914312134448728] 
\draw [color={rgb, 255:red, 208; green, 2; blue, 27 }  ,draw opacity=1 ]   (140,8300) -- (278,8251.58) ;
\draw [shift={(280.83,8250.58)}, rotate = 160.66] [fill={rgb, 255:red, 208; green, 2; blue, 27 }  ,fill opacity=1 ][line width=0.08]  [draw opacity=0] (3.57,-1.72) -- (0,0) -- (3.57,1.72) -- cycle    ;
%Straight Lines [id:da24350726789392918] 
\draw [color={rgb, 255:red, 245; green, 166; blue, 35 }  ,draw opacity=1 ]   (242.36,8300) -- (242.4,8208.58) ;
\draw [shift={(242.4,8205.58)}, rotate = 90.02] [fill={rgb, 255:red, 245; green, 166; blue, 35 }  ,fill opacity=1 ][line width=0.08]  [draw opacity=0] (3.57,-1.72) -- (0,0) -- (3.57,1.72) -- cycle    ;
%Straight Lines [id:da4839340371290204] 
\draw [color={rgb, 255:red, 208; green, 2; blue, 27 }  ,draw opacity=1 ]   (140,8300) -- (278,8251.58) ;
\draw [shift={(280.83,8250.58)}, rotate = 160.66] [fill={rgb, 255:red, 208; green, 2; blue, 27 }  ,fill opacity=1 ][line width=0.08]  [draw opacity=0] (3.57,-1.72) -- (0,0) -- (3.57,1.72) -- cycle    ;
%Curve Lines [id:da25151732113063363] 
\draw [color={rgb, 255:red, 0; green, 0; blue, 0 }  ,draw opacity=1 ]   (238.78,8275.44) .. controls (249.42,8267.18) and (253.71,8273.86) .. (255.2,8279.37) ;
\draw [shift={(255.78,8282.28)}, rotate = 262.87] [fill={rgb, 255:red, 0; green, 0; blue, 0 }  ,fill opacity=1 ][line width=0.08]  [draw opacity=0] (3.57,-1.72) -- (0,0) -- (3.57,1.72) -- cycle    ;
%Shape: Square [id:dp9943658630579271] 
\draw  [color={rgb, 255:red, 208; green, 2; blue, 27 }  ,draw opacity=1 ][fill={rgb, 255:red, 208; green, 2; blue, 27 }  ,fill opacity=1 ] (223.92,8264.51) -- (228.88,8262.74) -- (230.65,8267.7) -- (225.69,8269.47) -- cycle ;

% Text Node
\draw (191.18,8306) node [anchor=north west][inner sep=0.75pt]  [font=\footnotesize,color={rgb, 255:red, 0; green, 0; blue, 0 }  ,opacity=1 ]  {$\textcolor[rgb]{0,0,0}{\boldsymbol{p}}\textcolor[rgb]{0.96,0.65,0.14}{_{\textcolor[rgb]{0,0,0}{1}}{}}\textcolor[rgb]{0.96,0.65,0.14}{}$};
% Text Node
\draw (244.2,8214.36) node [anchor=north west][inner sep=0.75pt]  [font=\footnotesize]  {$\textcolor[rgb]{0.96,0.65,0.14}{\boldsymbol{b}}$};
% Text Node
\draw (159.2,8256.7) node [anchor=north west][inner sep=0.75pt]  [font=\footnotesize,color={rgb, 255:red, 0; green, 0; blue, 0 }  ,opacity=1 ]  {$\boldsymbol{\textcolor[rgb]{0,0,0}{p'}\textcolor[rgb]{0.96,0.65,0.14}{_{\textcolor[rgb]{0,0,0}{1}}{}}\textcolor[rgb]{0.96,0.65,0.14}{}}$};
% Text Node
\draw (178.53,8286.48) node [anchor=north west][inner sep=0.75pt]  [font=\tiny]  {$\Psi /2$};
% Text Node
\draw (248.5,8282.17) node [anchor=north west][inner sep=0.75pt]  [font=\tiny]  {$\Psi /2$};
% Text Node
\draw (225.83,8233.08) node [anchor=north west][inner sep=0.75pt]  [font=\footnotesize,color={rgb, 255:red, 0; green, 0; blue, 0 }  ,opacity=1 ]  {$\boldsymbol{\textcolor[rgb]{0.29,0.56,0.89}{q}}\textcolor[rgb]{0.29,0.56,0.89}{_*}$};
% Text Node
\draw (282.2,8245.2) node [anchor=north west][inner sep=0.75pt]  [font=\footnotesize,color={rgb, 255:red, 0; green, 0; blue, 0 }  ,opacity=1 ]  {$\textcolor[rgb]{0.82,0.01,0.11}{\varpropto }\textcolor[rgb]{0.82,0.01,0.11}{\boldsymbol{e}}\textcolor[rgb]{0.82,0.01,0.11}{_{q}{}}\textcolor[rgb]{0.96,0.65,0.14}{}$};
% Text Node
\draw (197.2,8212.2) node [anchor=north west][inner sep=0.75pt]  [font=\footnotesize,color={rgb, 255:red, 0; green, 0; blue, 0 }  ,opacity=1 ]  {$\textcolor[rgb]{0.49,0.83,0.13}{\boldsymbol{x}}\textcolor[rgb]{0.49,0.83,0.13}{_{\perp }}$};

\end{tikzpicture}

\caption{Geometry of eikonal scattering.}
\end{center}

\end{figure}

In this way, we have performed the integrals over $q$ and $x$. The result has been to evaluate the factor $q^\mu$ as a derivative of the eikonal function. To obtain the full expectation, we simply evaluate $\bra \psi S^\dagger$ using stationary phase in precisely the same way: the only difference (apart from the obvious Hermitian conjugation) is the absence of the $q^\mu$ factor. We then exploit unitarity to conclude that\footnote{We discuss the details
of unitarity more explicitly in appendix~\ref{app:unitarity}, performing all the remaining integrals.} 
\[
\Delta p_1^\mu = q^{\mu} = -\partial^{\mu} \chi (x_{\perp},s) \,,
\label{eq:impulsePerp}
\]
where all quantities are defined using the solution of the stationary phase conditions.

Notice that we have determined the \emph{complete} impulse four-vector, not just the scattering angle. The distinction between these quantities is obviously unimportant at the level of conservative dynamics, but it is important when radiation occurs. The key aspect of our argument which leads to the full impulse rather than the scattering angle is the presence of a perpendicular projector. To see how this works, let us discuss an explicit example: the impulse at next-to-leading order in gravitational fast scattering.

Focusing on the scattering between two massive bodies in general relativity,  the eikonal phase at next-to-leading order in $G$ is (see e.g.~\cite{KoemansCollado:2019ggb})
\[\label{eq:gravEikonal}
\chi=-2G m_{1} m_{2}\left(\frac{\left(2 \gamma^{2}-1\right)}{\sqrt{\gamma^{2}-1}} \log |x_{\perp}|-\frac{3 \pi}{8} \frac{\left(5 \gamma^{2}-1\right)}{\sqrt{\gamma^{2}-1}} \frac{G\left(m_{1}+m_{2}\right)}{|x_{\perp}|}\right) \ ,
\]
where we have defined $\gamma=u_1 \cdot u_2$ as the scalar product between the four-velocities of the particles.  Using equation~\eqref{eq:impulsePerp} and straightforward differentiation, we obtain the following expression in terms of $x_{\perp}^{\mu}$,
 \[
\Delta p_{1}^{\mu}=-\frac{2 G m_{1} m_{2} x^{\mu}_{\perp}}{\left|x^{2}_{\perp}\right|}\left(\frac{\left(2 \gamma^{2}-1\right)}{\sqrt{\gamma^{2}-1}}+\frac{3 \pi}{8} \frac{\left(5 \gamma^{2}-1\right)}{\sqrt{\gamma^{2}-1}} \frac{G\left(m_{1}+m_{2}\right)}{|x_{\perp}|}\right) \ ,
\]
where the four-velocities in $\gamma$ can now be identified with the incoming four-velocities of the particles due to the integrals over the wavepackets that took place to arrive at \eqref{eq:impulsePerp}.
At this point, it is important to remember that $x^{\mu}_{\perp}$ is not quite $b^\mu$.
It is trivial to show that $x^{\mu}_{\perp}$ coincides with $b^{\mu}$ at leading order in the gravitational coupling, but
this is no longer the case at next-to-leading order; then instead 
%change sign in xperp 
\[
x^{\mu}_{\perp}=b^{\mu}+\frac{G(2 \gamma^2-1)}{(\gamma^2-1)^{3/2}}\left(u^{\mu}_{1}(m_2+\gamma m_1)-u^{\mu}_2(m_1+\gamma m_2) \right)\ ,
\]
where
\[
|x_{\perp}^2|=|b^2| \ .
\]
We can now express the impulse in terms of the impact parameter $b^{\mu}$. At the order we are interested in we find,
%minus sign added in G^1 term
\[
\Delta p_{1}^{\mu}= -\frac{2 G m_{1} m_{2} b^{\mu}}{\left|b^{2} \right|}\left(\frac{\left(2 \gamma^{2}-1\right)}{\sqrt{\gamma^{2}-1}}+\frac{3 \pi}{8} \frac{\left(5 \gamma^{2}-1\right)}{\sqrt{\gamma^{2}-1}} \frac{G\left(m_{1}+m_{2}\right)}{|b|}\right) \\ \quad 
 -  \frac{G^{2} m_{1} m_{2}\left(2 \gamma^{2}-1\right)^{2}\left(\left(\gamma m_{1}+m_{2}\right) u_{1}^{\mu}-\left(\gamma m_{2}+m_{1}\right) u_{2}^{\mu}\right)}{\left(\gamma^{2}-1\right)^{2}\left|b^{2}\right|} \ ,
\]
in perfect agreement with the literature~\cite{Herrmann:2021tct}. 

From the perspective of this paper, the key achievement of the eikonal resummation is that negligible variance becomes automatic in the stationary
phase argument. Indeed since the stationary phase condition~\eqref{eq:stationaryImpulse} sets the momentum transfer to a specific, classical, value
it is clear that the expectation value of any polynomial in the momentum operator will evaluate to the classical expectation. 

\subsection{Coherent radiative state}
\label{sec:Rad_State}

In the context of radiation, another exponentiation is well known to lead to small or negligible variance. This occurs when the radiative state is a coherent state with large occupation number. Indeed, in this context the requirement of minimal uncertainty is sometimes referred to as
as complete coherence~\cite{Glauber:1963fi,Glauber:1963tx}. Coherent states have a long history in applications to semi-classical physics~\cite{Yaffe,KlauderJohnR1985Cs:a} as well as more recent interest~\cite{delaCruz:2020bbn,Bern:2020buy, Monteiro:2020plf,Aoude:2021oqj,Cristofoli:2021vyo}. In this subsection, we describe how these states naturally describe the radiation in the final state, working in electrodynamics for simplicity. 

Consider the state 
\[
\label{eq:coherentState}
\ket {\alpha^{\eta}} = \mathcal{N}_{\alpha} \exp\left(\int \dPhi (k) \alpha^{\eta}(k) a^{\dagger}_{\eta}(k)\right) \ket 0 \,.
\]
It is not difficult to verify that $\ket {\alpha^{\eta}}$ is indeed a coherent state satisfying the usual properties: in particular, the minimal uncertainty relation \eqref{eq:variance} is obeyed explicitly if the occupation number is large~\cite{Cristofoli:2021vyo}. Normalising these states to unity requires
\[
\mathcal{N}_\alpha = \exp \left(-\frac{1}{2} \int \dPhi(k)|\alpha^{\eta}(k)|^2\right)\,.
\] 

We would like to understand the final particle distribution for the photons emitted in a scattering process in the classical limit. In particular, we want to show here that for classical radiation the factorization property of expectation values of physical observables is directly connected to coherence. To simplify our discussion, we work in a spacetime box of finite dimension $V$ with periodic boundary conditions so that there will be a finite number $|V^D|$ of allowed momenta $\{k_i\}_{i \in V^D}$ in the dual momentum lattice $V^D$. At the end of the discussion, we will take $V \to +\infty$. In this framework, there is a coherent state for each harmonic oscillator with momentum $\{k_i\}$
\[
\label{eq:coherentRadiation}
\ket {\alpha^{\eta}_{k_i}} = \mathcal{N}_{\alpha} \exp\left( \alpha^{\eta}(k_i) a^{\dagger}_{\eta}(k_i)\right) \ket 0,  \qquad \forall \, k_i \in V^D\,.
\]
It is known that we can write every classical radiation density matrix as a probability distribution in the coherent state space, as proved by Glauber and Sudarshan \cite{Glauber:1963tx,Glauber:1963fi,Sudarshan:1963ts,Zhang:1999is}
\[
\hat{\rho}_{\text{out}} = \sum_{\eta = \pm } \int \prod_{l_i \in V^D} \d^2 \alpha^{\eta}_{l_i}  \, \mathcal{P}^{\eta}(\alpha) \ket{\alpha^{\eta}_{l_i}} \bra{\alpha^{\eta}_{l_i}} \qquad \d^2 \alpha^{\eta}_l := (\d \Re \alpha^{\eta}_l \d \Im \alpha^{\eta}_l)/\pi\,
\label{eq:GS_representation}
\]
where $\mathcal{P}^{\eta}(\alpha)$ is a separable function of $\alpha^{\eta}_{l_i}$ with $i \in V^D$
\[
\mathcal{P}^{\eta}(\alpha) := \prod_{l_i \in V^D} \mathcal{P}^{\eta}(\alpha (l_i)). 
\]
For the classical case $\mathcal{P}^{\eta}(\alpha) \geq 0$, and this is what allows to talk about probability distribution in the standard mathematical sense. Let us stress here that in the notation $\sum_{l_i \in V^D}$ we include not only the summation over the dual lattice vectors but also the appropriate finite-volume on-shell phase-space normalization. 

What is the implication of the exact classical factorization on the final radiation density matrix? Based on the previous discussion of negligible uncertainty we expect that the expectation value $\langle \,\cdot\, \rangle_{\rho_{\text{out}}}$ in the density matrix \eqref{eq:GS_representation} gives
\[
\langle \mathbb{F}_{\mu \nu}(x) \mathbb{F}_{\rho \sigma}(y)\rangle_{\rho_{\text{out}}} \stackrel{\hbar \to 0}{=}  \langle \mathbb{F}_{\mu \nu}(x) \rangle_{\rho_{\text{out}}}  \langle \mathbb{F}_{\rho \sigma}(y)\rangle_{\rho_{\text{out}}} \,.
\]
We use the on-shell mode expansion for the field strength operator
\[
\mathbb{F}_{\mu \nu}(x) = \frac{-i}{\sqrt{\hbar}} \sum_{\eta = \pm} \sum_{k \in V^D} \,\bigl[a_{\eta}(k) \kb^\vmu_{[\mu} \polvhconj{\eta}_{\nu]} e^{-i \kb \cdot x} - \hc\bigr]\,,
\]
and by taking advantage of the completeness relation in the Hilbert space of photons\footnote{For $n=0$, the first term in the sum has to be interpreted as $\ket{0}\bra{0}$. This will be implicitly assumed in the rest of the argument.}
\[
\sum_{n=0}^{+\infty} \frac{1}{n!} \sum_{\eta_1,...,\eta_n = \pm} \sum_{l_1,...,l_n \in V^D} \ket{l_1^{\eta_1} ... l_n^{\eta_n}}\bra{l_1^{\eta_1} ... l_n^{\eta_n}}=1\,,
\]
we have,
\[
\text{Tr}_{\rho_{\text{out}}}(\mathbb{F}_{\mu \nu}(x)) &= \sum_{n=0}^{+\infty} \frac{1}{n!} \sum_{\eta_1,...,\eta_n = \pm} \sum_{l_1,...,l_n \in V^D} \bra{l_1^{\eta_1} ... l_n^{\eta_n}} \mathbb{F}_{\mu \nu}(x) \rho_{\text{out}} \ket{l_1^{\eta_1} ... l_n^{\eta_n}} \\
&= \sum_{n,m=0}^{+\infty} \frac{1}{n!m!} \sum_{\eta_1,\eta'_1,...,\eta_n,\eta'_m = \pm} \sum_{l_1,l'_1,...,l_n,l'_m \in V^D} \\
& \hspace{15pt}\times\bra{l_1^{\eta_1} ... l_n^{\eta_n}} \mathbb{F}_{\mu \nu}(x) \ket{(l'_1)^{\eta_1} ... (l'_m)^{\eta'_m}} \bra{(l'_1)^{\eta_1} ... (l'_m)^{\eta'_m}} \rho_{\text{out}} \ket{l_1^{\eta_1} ... l_n^{\eta_n}} \\
&= -i \hbar^{\frac{3}{2}}  \sum_{n=0}^{+\infty} \frac{1}{n!} \sum_{\eta_1,...,\eta_n = \pm} \sum_{\bar{l}_1,...,\bar{l}_n\in \bar{V}^D} \nonumber \\
& \hspace{15pt}\times \sum_{\bar{k} \in \bar{V}^D} \bigl[\kb^\vmu_{[\mu} \polvhconj{\eta}_{\nu]} \bra{l_1^{\eta_1} ... l_n^{\eta_n} k^{\eta}} \rho_{\text{out}} \ket{l_1^{\eta_1} ... l_n^{\eta_n}}  e^{-i\kb\cdot x} - \hc\bigr]\,.
\]
where in the last line we have restored also the $\hbar$ scaling implicit in the finite volume phase-space normalization. Further manipulations show that
\[
 \sum_{n=0}^{+\infty} \frac{1}{n!} &\sum_{\eta_1,...,\eta_n = \pm} \sum_{\bar{l}_1,...,\bar{l}_n\in \bar{V}^D} \bra{l_1^{\eta_1} ... l_n^{\eta_n} k^{\eta}} \rho_{\text{out}} \ket{l_1^{\eta_1} ... l_n^{\eta_n}} \nonumber \\
 &=   \sum_{n=0}^{+\infty} \frac{1}{n!} \sum_{\eta_1,...,\eta_n = \pm} \sum_{\bar{l}_1,...,\bar{l}_n\in \bar{V}^D} \prod_{i=1}^n \left[\mathcal{N}_{\alpha_{l_i}}^2 \int \d^2 \alpha^{\eta}_{l_i} \,   |\alpha^{\eta_i} (l_i)|^2  \, \mathcal{P}^{\eta_i}(\alpha (l_i))\right] \alpha^{\eta}(k) \nonumber \\
 &= \text{Tr}_{\rho_{\text{out}}}(1) \alpha^{\eta}(k) \nonumber \\
 &= \alpha^{\eta}(k)
\]
where we have used the fact that the density matrix is normalized $\text{Tr}_{\rho_{\text{out}}}(1)=1$. If we demand the expectation value of $\text{Tr}_{\rho_{\text{out}}}(\mathbb{F}_{\mu \nu}(x))$ to be classical, restoring powers of $\hbar$ requires the waveshape to scale as determined in \cite{Cristofoli:2021vyo}
\[
\label{eq:waveform_hbar}
\alpha^{\eta}(k) \to \hbar^{-\frac{3}{2}} \bar{\alpha}^{\eta}(k)  \,.
\]
Using this condition and the classical scaling of the normalization of the final state, we get
\[
\text{Tr}_{\rho_{\text{out}}}(\mathbb{F}_{\mu \nu}(x)) = -i \sum_{\bar{k} \in \bar{V}^D} \sum_{\eta = \pm} \int \d^2 \bar{\alpha}^{\eta}_{k} \, \mathcal{P}^{\eta}(\bar{\alpha}_k)  \,   &\bigl[\kb^\vmu_{[\mu} \polvhconj{\eta}_{\nu]} \bar{\alpha}^{\eta,*}(k)  e^{-i \kb \cdot x} - \hc \bigl] \,.
\]
Using a similar argument, for the expectation value of the product we obtain
\[
\text{Tr}_{\rho_{\text{out}}}(\mathbb{F}_{\mu \nu}(x) \mathbb{F}_{\rho \sigma}(y)) &=  \sum_{n=0}^{+\infty} \frac{1}{n!} \sum_{\eta_1,...,\eta_n = \pm} \sum_{\bar{l}_1,...,\bar{l}_n \in \bar{V}^D} \bra{l_1^{\eta_1} ... l_n^{\eta_n}} \mathbb{F}_{\mu \nu}(x) \mathbb{F}_{\rho \sigma}(y) \rho_{\text{out}}  \ket{l_1^{\eta_1} ... l_n^{\eta_n}} \nonumber \\
&= - \sum_{\bar{k}_1,\bar{k}_2 \in \bar{V}^D} \sum_{\eta_1, \eta_2 = \pm}  \int \d^2 \bar{\alpha}^{\eta_1}_{k_1} \, \d^2 \bar{\alpha}^{\eta_2}_{k_2} \mathcal{P}^{\eta_1,\eta_2}(\bar{\alpha}_{k_1},\bar{\alpha}_{k_2})    \, \nonumber \\
&\hspace{20pt}\times \bigl[ \kb^\vmu_{1,[\mu} \polvhconj{\eta_1}_{\nu]} \kb^\vmu_{2,[\rho} \polvhconj{\eta_2}_{\sigma]} \bar{\alpha}^{\eta_1,*}(k_1) \bar{\alpha}^{\eta_2,*}(k_2)  e^{-i \kb_1\cdot x-i \kb_2\cdot y} \nonumber \\
&\hspace{40pt}\quad+\kb^\vmu_{1,[\mu} \polvh{\eta_1}_{\nu]} \kb^\vmu_{2,[\rho} \polvhconj{\eta_2}_{\sigma]} \bar{\alpha}^{\eta_1}(k_1) \bar{\alpha}^{\eta_2,*}(k_2)  e^{i \kb_1 \cdot x-i \kb_2\cdot y} + \hc \bigl] \,,
\]
up to the commutator term which can be neglected in the $\hbar \to 0$ limit. 

Therefore we are effectively asking whether the product of the averages is equal to the average of the product over a distribution $\mathcal{P}^{\eta}(\alpha)$ for all $k \in V^D$, i.e. we are asking the distribution to have \emph{zero variance} in the coherent state space. But distributions of zero variance are degenerate because it means that the random variable $\alpha^{\eta}(k)$\footnote{Technically this is a random variable $\alpha^{\eta}(k)$ for each value of the momentum in the dual lattice, i.e. there is an harmonic oscillator for each quanta of radiation. But since they are all independent, we can promote this statement to the full expression.} is almost surely constant for each $k \in V^D$ (see page 173 in~\cite{blitzstein2019}). Therefore the distribution has support in a lower-dimensional space, and since we can apply this argument independently both for the real and for the imaginary part of $\alpha^{\eta}(k)$ we have 
\begin{align}
\mathcal{P}^{\eta}(\alpha) = \prod_{k \in V^D} \sum_{j=1}^{+\infty} c^{\eta}_j \delta^2(\alpha^{\eta}(k) - \alpha^{\eta}_j(k)) \,.
\end{align}
What this is essentially saying is that we get a (possibly infinite) sum of discrete distributions with a constant value $\alpha^{\eta}_j(k)$. But then, making use of a crucial result due to Hillery~\cite{HILLERY1985409}, we get 
\begin{align}
\mathcal{P}^{\eta}(\alpha) = \prod_{k \in V^D} \delta^2(\alpha^{\eta}(k) - \alpha^{\eta}_{\star}(k)) \,,
\end{align}
which means that we can describe the final state only with one coherent state for each helicity and for each momentum $k$ in the dual momentum lattice. At this point we can take the large volume limit and what this calculation implies is that we can describe the final state with a single coherent state \eqref{eq:coherentRadiation}, which takes naturally into account the infinite-dimensional superposition of harmonic oscillators of momentum $k$.

This state only describes the gauge particles (photons in this case), but we would like to use this to describe a state with the two massive scattering particles as well as the radiation emitted during the interaction. This is the situation one would generally expect after some scattering event. We will do exactly this later and determine the form of $\alpha$ at leading order. We will find that $\alpha$ depends on the 5-point amplitude, much as the eikonal  depends on the 4-point.

\subsection{Extension with coherent radiation}
\label{sec:eikonalAndRadiation}

The exponentiated eikonal final state of equation~\eqref{eq:eikonalState} beautifully describes semiclassical conservative dynamics, leading to a transparent method for extracting the impulse (or scattering angle) from amplitudes in a manner which automatically enforces minimal uncertainty. We have also seen that coherent states naturally enforce minimal uncertainty for radiation.
Now let us put these two ideas together to form a proposal for an eikonal-type final state in the fully dynamical, radiative, case.

It is very natural to consider a modification of the eikonal formula which includes radiation, and indeed this idea has received attention~\cite{Ciafaloni:2018uwe} in the literature. Given that our motivation is to extend the eikonal while maintaining its
minimal uncertainty property, an obvious way to proceed is to include an additional factor in the eikonal formula which has the structure of a coherent state like~\eqref{eq:coherentRadiation}. If this radiative part of the state has large occupation number, expectations of products of field-strength operators will naturally factorise into products of expectations of the operators.

We will simply propose one possibility for the structure of this final state, depending on a coherent waveshape parameter $\waveshape{\hel}$ (of helicity $\hel$) in addition to an eikonal function $\chi$. We believe there is strong evidence in favour of the basic structure of our proposal, and in particular in the idea that two objects $\chi$ and $\waveshape{\hel}$ suffice to define it; however, it seems possible to implement the idea in somewhat different ways. We will discuss the basic virtues of our proposal in the remainder of this section, leaving it to future work to determine further details. Since we are primarily interested in classical effects, we will continue to neglect the quantum remainder $\Delta$ in this discussion\footnote{Indeed radiative quantum effects will require $\Delta$ to be upgraded to an operator.}.

To describe our proposal, we begin with the eikonal final state~\eqref{eq:eikonalState}. With an eye towards a situation where momentum is lost to radiation, we need a description in which the sum of the momenta of the two final particles differs from the initial momenta. A first step, then, is to Fourier transform the wavepacket to position space:
\[
S\ket \psi|_\textrm{conservative} = \int \dPhi(p_1',p_2') \, \int  \dd^4  \qb \,\d^4 x\, \d^4 x_1 \,& \d^4 x_2 \; \tilde \phi_b ( x_1, x_2) \, e^{i(p_1' \cdot x_1 + p_2' \cdot x_2)/ \hbar}
\\
& \times e^{i [ q\cdot (x - x_1 + x_2) + \chi(x_{\perp}; s)]/ \hbar} \, \ket{p_1', p_2'} \,.
\]
Our proposal is now very straightforward: we simply incorporate a coherent state by assuming that
\[
\hspace{-3pt}
S\ket \psi =  \int &\dPhi(p_1',p_2') \,\int \dd^4  \qb \,\d^4 x \, \d^4 x_1 \, \d^4 x_2 \; \tilde \phi_b ( x_1, x_2) \, e^{i(p_1' \cdot x_1 + p_2' \cdot x_2)/ \hbar}
\\
& \times  \,e^{i [ q\cdot (x - x_1 + x_2) + \chi(x_{\perp}; s)]/ \hbar}  
%\\& \times 
\exp  \left[\sum_\hel \int \dPhi(k)  \waveshape{\hel}(k, x_1, x_2) a^\dagger_\hel(k)\right]
\, \ket{p_1', p_2'} \,.
\label{eq:finalStateProposal1}
\]
This is a minimal proposal: more generally, one could imagine that the coherent waveshape parameter $\waveshape{\hel}$ depends on other variables, for example $x$ or $q$ which appear in the eikonal dynamics. We will nevertheless restrict throughout this article to our minimal proposal. However, it is important that the state is not merely an outer product of a conservative eikonal state with a radiative factor. Some entanglement is necessary so that the radiation can backreact on the motion. In the case of the present proposal, the integrals over the variables $x_1$, $x_2$ and $x$ perform this role. 
We will present a partial derivation of this proposal in section~\ref{sec:Schwinger}, showing that the leading low-frequency classical radiation does indeed exponentiate as anticipated. 
There is a connection between our proposal here and work~\cite{Damgaard:2021ipf} on the exponential structure of the $S$ matrix.

Following the discussion in section~\ref{sec:Rad_State}, we know that the waveshape should be proportional to $\hbar^{-3/2}$ so we may also write the state as
\[
\hspace{-3pt}
S\ket \psi &=  \int \dPhi(p_1',p_2') \,\int \dd^4  \qb \,\d^4 x \, \d^4 x_1 \, \d^4 x_2 \; \tilde \phi_b ( x_1, x_2) \, e^{i(p_1' \cdot x_1 + p_2' \cdot x_2)/ \hbar}
\\
& \times  \,e^{i [ q\cdot (x - x_1 + x_2) + \chi(x_{\perp}; s)]/ \hbar}  
%\\& \times 
\exp  \left[\frac{1}{\hbar^{3/2}}\sum_\hel \int \dPhi(k) \barwaveshape{\hel}(k, x_1, x_2) a^\dagger_\hel(k) \right]
\, \ket{p_1', p_2'} \,.
\label{eq:finalStateProposal1bar}
\]
In this expression, the classical waveshape $\barwaveshape{\hel}$ is \emph{independent} of $\hbar$, just as the eikonal function $\chi$ is independent of $\hbar$.

In order to determine $\waveshape\hel$, we follow the same steps as in sections \ref{sec:expectation_of_two_F} and \ref{sec:KMOC_eikonal};  we act on the incoming state with the $S$ matrix, and then expand in terms of integrals of amplitudes. To isolate the waveshape, we
consider the overlap of our proposed final state with the bra $\bra{p_1'\,p_2' \, k^\hel}$:
\[
\bra{p_1'\, p_2' \, k^\hel} S \ket \psi = \int \dd^4  \qb \,\d^4 x \, \d^4 x_1 \, \d^4 x_2 \; & \tilde \phi_b ( x_1, x_2)  \, e^{i(p_1' \cdot x_1 + p_2' \cdot x_2)/ \hbar}  \\
&\times e^{i [ q\cdot (x - x_1 + x_2) + \chi(x_{\perp}; s)]/ \hbar} \,  \waveshape{\hel}(k, x_1, x_2) \,.
\label{eq:obtainingAlphaStep1}
\]
Next we expand the $S$ matrix as
\[
\bra{p_1'\, p_2' \, k^\hel} S \ket \psi &= \int \dPhi(p_1, p_2) \, \phi_b(p_1, p_2) \, \bra{p_1'\, p_2' \, k^\hel} S \ket{p_1 p_2} \\
&= \int \dPhi(p_1, p_2) \int \d^4 x_1 \, \d^4 x_2 \; \tilde \phi_b(x_1, x_2) \, e^{i(p_1 \cdot x_1 + p_2 \cdot x_2)/ \hbar} \\
& \hspace{50pt}\times i \mathcal{A}_5(p_1 p_2 \rightarrow p_1' p_2' k^\hel) \, \del^4(p_1 + p_2 - p_1' - p_2' - k) \,.
\label{eq:5ptAmpl}
\]
We note that the five-point amplitude appearing here could in principle include disconnected components beginning at order $g$. This order $g$ disconnected term would involve exactly zero-energy photons, and does not contribute to observables
such as the radiated momentum or the asymptotic Newman-Penrose scalar $\psi_4$. We therefore omit this term in the remainder of
this paper.

To continue, it is useful to perform a change of variable in the phase space measures, taking
$q_1 \equiv p_1 - p_1'$ and $q_2 \equiv p_2 - p_2'$ as variables of integration. Neglecting Heaviside theta functions (which will always be unity in the domain of validity of our calculation) we find
\[
\bra{p_1'\, p_2' \, k^\hel} S \ket \psi = & \int \d^4 x_1 \, \d^4 x_2 \; \tilde \phi_b(x_1, x_2) \, e^{i(p'_1 \cdot x_1 + p'_2 \cdot x_2)/ \hbar} \\
&\times \int \dd^4 q_1 \, \dd^4 q_2 \, \del(2p_1' \cdot q_1 + q_1^2) \del(2p_2' \cdot q_2 + q_2^2) \, e^{i (q_1 \cdot x_1 + q_2 \cdot x_2)/ \hbar} \\
& \hspace{50pt} \times i \mathcal{A}_5(p_1'+q_1, p_2'+q_2\to p_1', p_2', k^\hel) \, \del^4(q_1 + q_2 - k) \,.
\label{eq:obtainingAlphaStep2}
\]
Requiring equations~\eqref{eq:obtainingAlphaStep1} and~\eqref{eq:obtainingAlphaStep2} to be equal for any (appropriately classical) initial wavepacket $\tilde \phi_b(x_1, x_2)$ we deduce that
\[
 \waveshape{\hel}(k, x_1, x_2) &=  i \int \dd^4 q_1 \, \dd^4 q_2 \, \del(2p_1' \cdot q_1 + q_1^2) \del(2p_2' \cdot q_2 + q_2^2) \del^4(q_1 + q_2 - k) e^{i (\qb_1 \cdot x_1 + \qb_2 \cdot x_2)}\\
& \times  \mathcal{A}_5(p_1'+q_1, p_2'+q_2\to p_1', p_2', k^\hel) \left[ \int \dd^4 q \, \d^4 x \, e^{i q\cdot x} e^{i q \cdot (x_2 - x_1)/ \hbar} e^{i \chi(x_\perp; s)/ \hbar } \right]^{-1} .
\]
It is easy to use the eikonal equation~\eqref{eq:eikonal_def} to show that equivalently we may write the waveshape as
\[
& \waveshape{\hel}(k, x_1, x_2) = i \int \dd^4 q_1 \, \dd^4 q_2 \, \del(2p_1' \cdot q_1 + q_1^2) \del(2p_2' \cdot q_2 + q_2^2) \del^4(q_1 + q_2 - k) e^{i (\qb_1 \cdot x_1  + \qb_2 \cdot x_2)}  \\
& \times \mathcal{A}_5(p_1'+q_1, p_2'+q_2\to p_1', p_2', k^\hel) 
%\\ & \times
\left[1 +  \int \dd^4 q \, \del ( 2 \tilde p_1 \cdot q) \del (2 \tilde p_2 \cdot q) \, e^{i q \cdot (x_2 - x_1)/ \hbar} i  \mathcal{A}_4(s, q^2) \right]^{-1} \,.
\label{eq:waveshapeAsAmplitudes}
\]
This last expression makes the physical meaning transparent: the waveshape is obtained by removing iterated contributions of four-point amplitudes from the five-point amplitude.

To see this in more detail, it is instructive to expand the $\waveshape\hel$ order-by-order in perturbation theory. We again consider a generic coupling $g$ and expand the waveshape as
\[
\waveshape\hel = \waveshape\hel_0 + \waveshape\hel_1 + \cdots \,.
\]
The leading order term, $\waveshape\hel_0$, follows immediately from equation~\eqref{eq:waveshapeAsAmplitudes}:
\[\label{lowaveshape}
\waveshape{\hel}_0(k, x_1, x_2) =  i \int \dd^4 q_1 \, \dd^4 q_2 \,& \del(2p_1' \cdot q_1 + q_1^2) \del(2p_2' \cdot q_2 + q_2^2) \del^4(q_1 + q_2 - k) \\
& \times e^{i (q_1 \cdot (x_1 + b) + q_2 \cdot x_2)/ \hbar} \mathcal{A}_{5,0}(p_1'+q_1, p_2'+q_2\to p_1', p_2', k^\hel) \,.
\]
It is determined by the tree-level five-point amplitude $\mathcal{A}_{5,0}$, so it is of order $g^3$ in gauge theory and gravity. The fact that the leading-order classical radiation field is intimately related to five-point amplitudes was already discussed in~\cite{Luna:2017dtq,KMOC,Bautista:2019tdr,Cristofoli:2021vyo}. The basic structure of this leading-order waveshape is strikingly reminiscent of a coherent state which describes the \emph{static} Coulomb/Schwarzschild background~\cite{Monteiro:2020plf} on analytic continuation from Minkowski signature to $(2,2)$  signature $(+,+,-,-)$.

More precisely, $\waveshape{\hel}_0$ is really determined by $\mathcal{A}_{5,0}^{(0)}$. This follows by counting powers of $\hbar$. Indeed extracting dominant powers of $\hbar$ using equation~\eqref{eq:hbarExpansion}, we find
\[
\waveshape{\hel}_0(k, x_1, x_2) =  \frac{i}{\hbar^{3/2}} \int \dd^4 \qb_1 \,& \dd^4 \qb_2 \, \del(2p_1' \cdot \qb_1) \del(2p_2' \cdot \qb_2) \del^4(\qb_1 + \qb_2 - \kb) \\
& \times e^{i (\qb_1 \cdot x_1 + \qb_2 \cdot x_2)} \mathcal{A}_{5,0}^{(0)}(p_1'+q_1, p_2'+q_2\to p_1', p_2', k^\hel) \,.
\label{eq:LOwaveshapeExpanded}
\]
Note that the factor $\hbar^{-3/2}$ arises as expected on general grounds. The conclusion is that the leading-in-$\hbar$ part of the five-point tree amplitude determines the radiation. The amplitude itself contains higher order terms in $\hbar$; rather than arising from the radiative factor in our proposal~\eqref{eq:finalStateProposal1bar}, these terms arise from a generalised quantum remainder.

The next-to-leading order correction to the waveshape following from equation~\eqref{eq:waveshapeAsAmplitudes} is
\[
& \waveshape{\hel}_1(k, x_1, x_2) = i \int \dd^4 q_1 \, \dd^4 q_2 \, \del(2p_1' \cdot q_1 + q_1^2) \del(2p_2' \cdot q_2 + q_2^2) \del^4(q_1 + q_2 - k) \\
& \hspace{100pt}\times e^{i (q_1 \cdot x_1 + q_2 \cdot x_2)/ \hbar} \mathcal{A}_{5,1}(p_1'+q_1, p_2'+q_2\to p_1', p_2', k^\hel) \\
& - \waveshape{\hel}_0(k, x_1, x_2) \int \dd^4 q \, \del ( 2 \tilde p_1 \cdot q) \del (2 \tilde p_2 \cdot q) \, e^{i q \cdot (x_2 - x_1)/ \hbar}  i\mathcal{A}_{4,0}(s, q^2) \,.
\label{eq:nloWaveshape}
\]
This correction involves the five-point one-loop amplitude, after subtracting an iteration term. To understand the role of the subtraction, it is instructive to extract the leading-in-$\hbar$ part of $\waveshape{\hel}_1$:
\[
& \waveshape{\hel}_1(k, x_1, x_2) = \frac{i}{\hbar^{5/2}} \int \dd^4 \qb_1 \, \dd^4 \qb_2 \, \del(2p_1' \cdot \qb_1) \del(2p_2' \cdot \qb_2) \del^4(\qb_1 + \qb_2 - \kb) \\
& \hspace{100pt}\times e^{i (\qb_1 \cdot x_1 + \qb_2 \cdot x_2)} \mathcal{A}_{5,1}^{(0)}(p_1'+q_1, p_2'+q_2\to p_1', p_2', k^\hel) \\
& -\frac{1}{\hbar^{5/2}} \barwaveshape{\hel}_0(k, x_1, x_2) \int \dd^4 \qb \, \del ( 2 p'_1 \cdot \qb) \del (2 p'_2 \cdot \qb) \, e^{i \qb \cdot (x_2 - x_1)}  i\mathcal{A}_{4,0}^{(0)}(s, q^2) \\
&+ \mathcal{O}(\hbar^{-3/2}) \,.
\]
At this stage it seems that there is an unwanted order $\hbar^{-5/2}$ term in the NLO waveshape! Consistency with our proposal therefore demands 
\[
\int &\dd^4 \qb_1 \, \dd^4 \qb_2 \, \del(2p_1' \cdot \qb_1) \del(2p_2' \cdot \qb_2) \del^4(\qb_1 + \qb_2 - \kb) 
e^{i (\qb_1 \cdot x_1 + \qb_2 \cdot x_2)} \mathcal{A}_{5,1}^{(0)}(p_1'+q_1, p_2'+q_2\to p_1', p_2', k^\hel) \\
& =- \barwaveshape{\hel}_0(k, x_1, x_2) \int \dd^4 \qb \, \del ( 2 p'_1 \cdot \qb) \del (2 p'_2 \cdot \qb) \, e^{i \qb \cdot (x_2 - x_1)}  \mathcal{A}_{4,0}^{(0)}(s, q^2)  \,.
\]
Since $\barwaveshape{\hel}_0$ is determined by $\mathcal{A}^{(0)}_{5,0}$, this requirement relates $\mathcal{A}^{(0)}_{5,1}$ to 
$\mathcal{A}^{(0)}_{5,0}$ and $\mathcal{A}_{4,0}^{(0)}$. The requirement is nothing but a Fourier transform of the zero-variance relation~\eqref{eq:fivePointRelation} which we encountered in section~\ref{sec:fivepointrelation}.
So we see that the zero-variance relations retain their importance in the context of this eikonal/coherent resummation: their validity
admits the possibility of exponentiation.

In the same vein, it is interesting to project our proposal onto a two-photon final state:
\[
\bra{p_1', p_2', k_1^{\eta_1}, k_2^{\eta_2}}S\ket{\psi} &= \int \dd^4  \qb \,\d^4 x \, \d^4 x_1 \, \d^4 x_2 \;  \tilde \phi_b ( x_1, x_2)  \, e^{i(p_1' \cdot x_1 + p_2' \cdot x_2)/ \hbar}  \\
&\times e^{i [ q\cdot (x - x_1 + x_2) + \chi(x_{\perp}; s)]/ \hbar} \,  \waveshape{\hel_1}(k_1, x_1, x_2)\waveshape{\hel_2}(k_2, x_1, x_2) \,.
\]
Since the waveshape is at least of order $g^3$, it follows that this overlap begins at order $g^6$. However by expanding the $S$ matrix out
directly, we encounter a six-point amplitude. The conclusion is that our proposal does not populate the (order $g^4$) tree-level six-point 
amplitude. Of course this is as it should be: we saw that the six-point tree is suppressed in the classical region in
section~\ref{sec:6_point_tree}. Similarly the seven-point tree and one-loop amplitudes are suppressed, etc.

As a final remark, note that we did not introduce any normalisation factor in our proposal. This may be surprising in view of the necessity to
normalise the coherent state as in equation~\eqref{eq:coherentState}. The origin of the difference is simply that unitarity must already 
guarantee the normalisation of the final state. As is by now well understood, at two loops the eikonal function $\chi$ ceases to be real in the
radiative case. Instead the imaginary part of $\chi$ is related to $\sum_\hel |\waveshape{\hel}|^2$; this supplies the necessary normalisation.

\subsection{Radiation reaction}
\label{sec:rad-react}

Once there is radiation, there must also be radiation reaction: the particle's motion must change in the radiative case relative to the conservative case to account for the loss of momentum to radiation. In this section we will see that the waveshape indeed contributes to the impulse of a particle in the manner required.

We begin by acting on our conjectural final state, equation~\eqref{eq:finalStateProposal1}, with the momentum operator of the field corresponding to particle 1:
\[
\Pop^\mu_1 S \ket{\psi} &=\int \dPhi(p_1',p_2') \,\int \dd^4  \qb \,\d^4 x \, \d^4 x_1 \, \d^4 x_2 \; \tilde \phi_b ( x_1, x_2) \, e^{i(p_1' \cdot x_1 + p_2' \cdot x_2)/ \hbar}
\\
&\hspace{-10pt} \times  \,e^{i [ q\cdot (x - x_1 + x_2) + \chi(x_{\perp}; s)]/ \hbar}  
\exp  \left[\sum_\hel \int \dPhi(k)  \waveshape{\hel}(k, x_1, x_2) a^\dagger_\hel(k)\right]
\, p_1'^\mu \ket{p_1', p_2'} \,.
\label{eq:rRStep}
\]
The operator simply inserts a factor $p_1'^\mu$. We proceed by rewriting this factor in terms of a derivative $-i \hbar \partial/\partial x_{1\mu}$ acting on the exponential factor in the first line of equation~\eqref{eq:rRStep}, and then integrating by parts. Neglecting the boundary term, the result is
\[
\Pop^\mu_1& S \ket{\psi} = \int \dPhi(p_1',p_2') \,\int \dd^4  \qb \,\d^4 x \, \d^4 x_1 \, \d^4 x_2 \; \, e^{i(p_1' \cdot x_1 + p_2' \cdot x_2)/ \hbar} e^{i(q\cdot x + \chi(x_{\perp}; \,s) )  / \hbar} 
\\
& \hspace{-10pt}\times i \hbar \partial_1^\mu \left( \tilde \phi_b ( x_1, x_2) e^{i q \cdot (x_2 - x_1)  / \hbar} \exp \left[ \sum_\hel \int \dPhi(k) \, \waveshape{\hel}(k, x_1, x_2) a^\dagger_\hel(k) \right] \right)
\, \ket{p_1', p_2'} \,.
\]
Expanding out the derivative, we encounter three terms. In the first, the derivative acts on the spatial wavefunction: as usual in quantum mechanics, this term will evaluate (in the expectation value of the final momentum) to the contribution of the initial momentum. The second term arises when the derivative operator acts on $e^{i q \cdot (x_2 - x_1) / \hbar}$, which inserts a factor of $q^\mu$. This term is familiar from equation~\eqref{eq:beginningImpulse} in section~\ref{sec:conservativeImpulse}, and contributes to the impulse as a (suitably projected) derivative of the eikonal function. Only the final term is new: it involves the waveshape, and must then be the origin of radiation reaction in our approach.

Since we have discussed the conservative impulse in detail in section~\ref{sec:conservativeImpulse}, we
focus on the final (new) term here. In this term, the derivative brings down a factor of
\[
\sum_\hel \int \dPhi(k) \, \partial_1^\mu \waveshape{\hel}(k, x_1, x_2) a^\dagger_\hel(k).
\]
Now to extract the momentum observable we multiply by  $\bra \psi S^\dagger$. Since this will introduce yet more integrals, it is helpful to define a modified KMOC style `classical average angle brackets' $\KMOCav {...}$, defined by
\[\label{brackets}
&\KMOCav { ...} = \int \dPhi(p_1',p_2')  \,\int \dd^4  \qb \,\d^4 x \, \dd^4  \bar{Q} \,\d^4 y \,  \d^4 x_1 \, \d^4 x_2  \,\d^4 y_1 \, \d^4 y_2 \,\tilde \phi_b^*(y_1,y_2) \tilde \phi_b( x_1, x_2) \;  \\
& \times e^{i(p_1' \cdot (x_1 - y_1) + p_2' \cdot (x_2 - y_2)/ \hbar} e^{i (q \cdot (x_2 - x_1) - Q \cdot (y_2 - y_1)) / \hbar} e^{i(q\cdot x + \chi(x_{\perp}; \,s) -Q\cdot y - \chi^*(y_{\perp}; \,s) )  / \hbar}  \\
& \times  \exp \left[ \sum_\hel \int \dPhi(k) \, \waveshape{\hel}(k, x_1, x_2)\waveshape{\hel}{}^*(k, y_1, y_2)\right] (...) \,.
\]
The $a ^\dagger (k)$ will act on this left state will produce a delta function and $\alpha^*$. After using the delta function it gives just a factor $\alpha^*$. Putting this together, and using the angle brackets shorthand we get
\[
\langle \Pop_1^\mu \rangle _\textrm{reaction} = \KMOCav{i \sum_\hel \int \dPhi(k) \,\alpha^{(\eta)*}(k,y_1,y_2) \partial_1^\mu \waveshape{\hel}(k, x_1, x_2) } .
\label{deltappp}
\]
Note that we have hidden integrals over the $y_i$ and $x_i$ in the definition of the brackets. While this expression for the momentum is not manifestly real, we anticipate that the result of the integration is real in the classical limit.

Notice that the manipulations leading to equation~\eqref{deltappp} were exact (under the assumption of equation~\eqref{eq:finalStateProposal1}).
However equation~\eqref{deltappp} involves a number of integrals which would need to be performed to arrive at a concrete expression for the impulse. In section~\ref{sec:conservativeImpulse}, we performed these integrals by stationary phase. In the present (radiative) case a similar approach would be possible, but the stationary phase conditions are significantly more complicated. For example, demanding the phase of the $q$ integral to be stationary leads to a condition involving the variables $x_2$ and $x_1$; further demanding that the phases of these $x_i$ integrals should be stationary leads to an equation involving the integral of a quadratic function of the waveshape. In this way the stationary phase conditions involve an intricate interplay of the eikonal and the waveshape. Of course this is as it should be: the complexity of radiation reaction must be captured by the final state.

As a simpler sanity check of our machinery, we evaluate the radiation reaction contribution to the impulse at lowest non-trivial perturbative order in electrodynamics. In \cite{KMOC} the leading order radiation reaction term was written as
\begin{equation}\label{rreact1}
I^\mu_{rad}=e^6 \KMOCav{\int {\d}\Phi(\bar{k})
\prod_{i=1,2} \hat{\d}^4 \bar{q}_i \hat{\d}^4 \bar{q}'_i\, \bar{q}_1^\mu\, \mathcal{Y}(\bar{q}_1, \bar{q}_2, \bar{k})
\mathcal{Y}^*(\bar{q}'_1, \bar{q}'_2, \bar{k})
}\,,
\end{equation}
with 
\begin{equation}
\mathcal{Y}(\bar{q}_1, \bar{q}_2, \bar{k})= \delta(p_1\cdot \bar{q}_1) \delta (p_2\cdot \bar{q}_2)\hat{\delta}^{4}(\bar{q}_1+\bar{q}_2-k)e^{ib\cdot \bar{q}_1} \mathcal{A}_{5, 0}^{(0)}(\bar{q}_1, \bar{k}^\hel).
\end{equation}
It is straightforward to verify that this is equivalent to the right-hand side of equation~\eqref{deltappp}. In fact, recalling the leading order waveshape formula \eqref{eq:LOwaveshapeExpanded},
the match between the two expressions \eqref{rreact1} and \eqref{deltappp} is immediate, once our definition of average over wave packets \eqref{brackets} is taken into account.
Equation~\eqref{rreact1} is still not obviously real, but reality can be shown with some further manipulation (see equations 5.52 and 5.53 in reference~\cite{KMOC}).

We note in passing that the expectation $\bra{\psi} S^\dagger \Fop_{\mu\nu} S \ket{\psi}$ (or $\bra{\psi} S^\dagger \Rop_{\mu\nu\rho\sigma} S \ket{\psi}$) can be determined in a similar way. 
The annihilation operators in the field strength operator immediately bring down a single power of the waveshape. 
At leading non-trivial order in $g$, it is then straightforward to see that the field strength is determined by the five-point
tree amplitude (specifically the leading fragment in $\hbar$) consistent with reference~\cite{Cristofoli:2021vyo}.
Similarly, the momentum radiated into messengers can be computed as the expectation
\[
\bra{\psi}S^\dagger \sum_\hel \int \dPhi(k) \, k^\mu \, a^\dagger_\hel(k) a_\hel(k) S \ket{\psi} \,.
\]
In this case, the creation and annihilation operators bring down the waveshape times its conjugate. 
At leading perturbative order, the momentum radiated is the square of the five-point tree, as observed in~\cite{KMOC}. 
The result is also consistent with classical field theory: in that context, radiation is described by the energy-momentum tensor,
which is quadratic in the field strength. Finally, we note that conservation of momentum holds as discussed in~\cite{KMOC}.

\section{The radiative final state from Schwinger proper time methods}
\label{sec:Schwinger}

In \eqref{eq:finalStateProposal1}, we provided a formula for how an
incoming momentum state evolves into a final state containing both the
two incoming particles, and any additional radiation expressed as a
coherent state. In this section, we note that one may provide a
partial justification of this result by considering the leading soft
emission at low energies. Infrared divergences are a good playground to understand the structure of classical scattering: while \cite{Heissenberg:2021tzo} considered the conservative case, here we extend the discussion to real radiation. This can be done using established methods which have been used in relatively recent literature on the
eikonal approximation~\cite{Laenen:2008gt,White:2011yy}. We start from an incoming state of the form
\[
|\psi_{\text{in}}\rangle= \int  \dPhi(p_1, p_2) \,\int \d^4 x_1 \, \d^4 x_2 \; \tilde \phi ( x_1, x_2) \, e^{i(p_1 \cdot x_1 + p_2 \cdot x_2)/ \hbar}  e^{i (b \cdot p_{1})/ \hbar}\left|p_{1} p_{2}\right\rangle,
\label{eq:KMOC_incoming}
\]
and our aim is to show that this should evolve over time to a state
\begin{align}
  |\psi_{\text{out}}\rangle = &\int \dPhi(p_1',p_2') \,\int \d^4 x_1 \, \d^4 x_2 \; \tilde \phi ( x_1, x_2) \, e^{i(p_1' \cdot x_1 + p_2' \cdot x_2)/ \hbar} \nonumber \\
& \times \int \dd^4  q \,\d^4 x \,e^{iq\cdot x/ \hbar} e^{i b \cdot p_1'/ \hbar} e^{-i b \cdot q / \hbar}  e^{i q \cdot (x_2 - x_1)/ \hbar}e^{i\chi(x_{\perp}; s)/ \hbar} \nonumber \\
& \times \exp  \left[\sum_\hel \int \dPhi(k)  \beta^{(\hel)}(k, x_1, x_2) a^\dagger_\hel(k)\right]
\, \ket{p_1', p_2'} \,,
\label{eq:KMOC_outgoing}
\end{align}
where we use $\beta$ rather than $\alpha$ to refer to the coherent
state parameter, as we will ultimately calculate this only in a
particular limit, namely the forward approximation in which they follow
classical straight-line trajectories. In such limit we are justified
to effectively obtain
 \begin{align}
  |\psi_{\text{out}}\rangle = &\int \dPhi(p_1',p_2') \,\int \d^4 x_1 \, \d^4 x_2 \; \tilde \phi ( x_1, x_2) \, e^{i(p_1' \cdot x_1 + p_2' \cdot x_2)/ \hbar} \nonumber \\
& \times e^{i b \cdot p_1'/ \hbar}  e^{i\chi(x_{\perp} - b_{\perp}; s)/ \hbar} \exp  \left[\sum_\hel \int \dPhi(k)  \beta^{(\hel)}(k, x_1, x_2) a^\dagger_\hel(k)\right]
\, \ket{p_1', p_2'} \,.
\label{eq:KMOC_outgoing2}
\end{align} 
 
The situation being considered here is similar to the analysis of
\cite{Laenen:2010uz}, which considered particles emerging from an
amplitude interacting with soft radiation (see
reference~\cite{White:2011yy} for the gravitational case). This
analysis used path integral methods --- also known as the Schwinger
proper time formalism\footnote{A useful pedagogical review of
Schwinger proper time methods may be found in
ref.~\cite{Schwartz:2014sze}.} --- to write the propagators for the
outgoing particles in terms of explicit sums over their spacetime
trajectories. This provided a very physical picture for describing
soft radiation, allowing the authors to generalise beyond the leading
soft approximation, and to show that certain sets of corrections
exponentiate in perturbation theory. However, only virtual radiation
was included, and thus we must extend such methods to the real
radiative case being considered here.  We restrict our discussion to
the case of QED for simplicity.

Let us first recall a useful result from the Schwinger formalism
(reviewed here in appendix~\ref{app:Schwinger}), namely that the
propagator for a particle in a background gauge field, produced at
position $x_{i}$ and ending up with final momentum $p_{f}$, can be
written as
\[
D_{F}\left(x_{i}, p_{f}\right)=\int_{0}^{\infty} \d T e^{-T \epsilon / \hbar} \, \bra{p_{f}} e^{-i (\hat{H} T) / \hbar}\ket{x_{i}},
\label{eq:DF_propagator}
\]
where
\[
\bra{p_{f}} e^{-i \hat{H} T/ \hbar} \ket{x_{i}}=\int_{x(0)=x_{i}}^{p(T)=p_{f}} \mathcal{D} p \mathcal{D} x \exp \left[\frac{i}{\hbar} p(T) \cdot x(T)+\frac{i}{\hbar} \int_{0}^{T} \d t(-p \cdot \dot{x}-H(p, x))\right]
\label{eq:Schwinger_pathintegral2}
\]
is a double path integral in position and momentum space, subject to
the above boundary conditions, and\footnote{As usual in the context of Schwinger proper time~\cite{Schwartz:2014sze} we choose a convenient
normalisation for ``time'' $t$ and the associated Hamiltonian $H$. In particular the dimensions of $H$ are $\textrm{mass}^2$.}
\[
\hat{H}=-(\hat{p}-e \hat{A})^{2}+m^{2}=-\hat{p}^{2}+e \hat{p} \cdot \hat{A}+e \hat{A} \cdot \hat{p}-e^{2} \hat{A}^{2}+m^{2}\,
\label{HA}
\]
is the Hamiltonian, where $e$ is the coupling constant and
$\hat{A}^{\mu}$ is the gauge field interpreted as an operator in the
quantum mechanical Hilbert space. As discussed in
appendix~\ref{app:Schwinger}, we should Weyl-order the Hamiltonian so
that all momentum operators are to the left. Care must be taken with
the term $\hat{A} \cdot \hat{p}$, where we note that we will be
operating on momentum states to the left:
\[
\langle p| \hat{A} \cdot \hat{p}=\hat{p}_{\mu}(\hat{A}^{\mu})\langle p|+\langle p| \hat{p} \cdot \hat{A}\,.
\]
Here $\hat{p}_{\mu}\left(A^{\mu}\right)$ represents the action of $\hat{p}_{\mu}$ on $\hat{A}^{\mu}$. Using the fact that $\hat{p}=-i \partial_{\mu}$ in position space, we can get rid of this term by using the Lorenz gauge $\partial_{\mu} A^{\mu}=0$, so that our Hamiltonian becomes
\[
\hat{H}=-\hat{p}^{2}+2 e \hat{p} \cdot \hat{A}-e^{2} \hat{A}^{2}+m^{2}.
\]
Plugging this into \eqref{eq:DF_propagator} and using \eqref{eq:Schwinger_pathintegral2}, we obtain a path integral representation for the propagator of a scalar particle in the presence of a gauge field:
\[
\begin{aligned}
D_{F}\left(x_{i}, p_{f} ; A\right)=& \int_{0}^{\infty} \d T e^{-T \epsilon / \hbar} \int_{x(0)=x_{i}}^{p(T)=p_{f}} \mathcal{D} p \mathcal{D} x \\
&\times \exp \left[\frac{i}{\hbar} p(T) \cdot x(T) +\frac{i}{\hbar} \int_{0}^{T} \d t\left(-p \cdot \dot{x}+p^{2}-2 e p \cdot A+e^{2} A^{2}-m^{2}\right)\right].
\label{eq:DF_gauge_pathintegral}
\end{aligned}
\]
It is of course impossible to carry out this path integral in general;
this would amount to exactly solving for a quantum particle moving in
an arbitrary electromagnetic field! But we can evaluate it
approximately in many different cases. Relevant for our purposes is if
a particle has position $x_{i}$ at $t=0$, and follows an approximate
straight-line trajectory, given by
\[
x_{c}^{\mu}=x_{i}^{\mu}+p_{f}^{\mu} t, \quad 0 \leq t \leq T
\]
where $p_{f}$ is the final momentum introduced above. In the sum over
trajectories in \eqref{eq:DF_gauge_pathintegral}, we can then redefine
\[
x(t) \rightarrow x_{c}(t)+x(t), \quad p(t) \rightarrow p_{f}+p(t)\,,
\]
where now $x(t)$ and $p(t)$ are small fluctuations. Substituting this
into \eqref{eq:DF_gauge_pathintegral}, one finds
\[
\hspace{-4pt}
D_{F}&(x_{i}, p_{f} ; A) =\int_{0}^{\infty} \d T e^{-T \epsilon / \hbar} e^{\frac{i}{\hbar} p_{f} \cdot x_{i}+\frac{i}{\hbar} T\left(p_{f}^{2}-m^{2}\right)} \\
& \hspace{-4pt} \times \int_{x(0)=0}^{p(T)=0} \mathcal{D} p \mathcal{D} x \exp \left[\frac{i}{\hbar} \int_{0}^{T} \d t\left(p \cdot\left(p_{f}-\dot{x}\right)+p^{2}-2 e p \cdot A-2 e p_{f} \cdot A+e^{2} A^{2}\right)\right].
\label{eq:dressed_prop_sQED}
\]
This still looks rather formidable, but it will simplify considerably
in what follows.

Let us now apply this to the problem of a pair of propagating
particles, interacting via photon exchange, in the conservative
case. We will consider two different scalar fields $\phi_{1}$ and
$\phi_{2}$, so that the scattering particles can in principle be
different. We may then consider the 4-point Green's function:
\[
G_{4}(\phi_{1}(x_{1}),\phi_{1}(x_{2}),&\phi_{2}(x_{3}),\phi_{2}(x_{4})) \\
&=\int \mathcal{D} A_{\mu} \mathcal{D} \phi_{1} \mathcal{D} \phi_{2} \,\phi_{1}(x_{1}) \phi_{1}(x_{2}) \phi_{2}(x_{3}) \phi_{2}(x_{4}) e^{\frac{i}{\hbar} S(\phi_{1},\phi_{2},A_{\mu})},
\label{eq:4pt_dressed}
\]
which will ultimately be related to S-matrix elements for the
scattering particles, and we may separate the total classical action
$S(\phi_{1},\phi_{2},A_{\mu})$ into three distinct terms:
\[
S(\phi_{1}, \phi_{2}, A_{\mu})=S_{A}(A_{\mu})+S_{1}(\phi_{1}, A_{\mu})+S_{2}(\phi_{2}, A_{\mu}).
\label{eq:Stotal_decomp}
\]
The first term on the right-hand side consists of terms involving only
the gauge field $A_{\mu}$, and $S_{i}\left(\phi_{i}, A_{\mu}\right)$
contains terms involving the individual scalar field $\phi_{i}$ only,
or its coupling to the gauge field. Substituting
\eqref{eq:Stotal_decomp} into \eqref{eq:4pt_dressed}, this may be
rewritten as
\[
G_{4}(\phi_{1}(x_{1}), \phi_{1}(x_{2}), &\phi_{2}(x_{3}), \phi_{2}(x_{4})) \\
&=\int \mathcal{D} A_{\mu} G_{2}\left(\phi_{1}(x_{1}) \phi_{1}(x_{2}) ;A\right) G_{2}\left(\phi_{2}(x_{3}) \phi_{2}(x_{4}); A\right) e^{\frac{i}{\hbar} S_{A}\left(A_{\mu}\right)}\,,
\]
where
\[
G_{2}\left(\phi_{i}(x) \phi_{i}(y) ; A\right)=\int \mathcal{D} \phi_{i} \, \phi_{i}(x) \phi_{i}(y) \, e^{\frac{i}{\hbar} S_{i}\left(\phi_{i}, A\right)}\,,
\label{eq:G2pos}
\]
is the two-point function for the field $\phi_{i}$ in the presence of
a background gauge field. This is simply the propagator, and indeed is
almost exactly the object that we worked out in
\eqref{eq:dressed_prop_sQED}. The only difference is that
\eqref{eq:dressed_prop_sQED} has the scalar particle moving from a
state of given initial position and final momentum, while in
equation~\eqref{eq:G2pos} both initial and final positions are
specified. The form of equation~\eqref{eq:dressed_prop_sQED} is
convenient for our problem given that we wish to consider particles
that are separated by an definite distance $\Delta x$. Let us
therefore consider a pair of particles ``produced'' at positions
$z_{i}$, so that $z_{1}=x_{1}$ and $z_{2}=x_{3}$. Then we can use
translational invariance to set
\[
z_{1}^{\mu}=\Delta x^{\mu}, \quad z_{2}^{\mu}=0\,.
\label{eq:initial_pos}
\]
Each particle $i$ propagates
out to infinity, ending up with a final momentum $p_{i}^{\prime}$, and
so we are interested in the Green's function
\[
G_{4}\left(\phi_{1}(z_{1}),\phi_{1}(p_{1}^{\prime}),\phi_{2}(z_{2}),\phi_{2}(p_{2}^{\prime})\right)=\int \mathcal{D} A_{\mu} \, D_{F}\left(z_{1}, p_{1}^{\prime} ; A\right) D_{F}\left(z_{2}, p_{2}^{\prime} ; A\right) e^{\frac{i}{\hbar} S_{A}\left(A_{\mu}\right)}\,,
\label{eq:4pt-correlator}
\]
where we used the notation for the propagators in
\eqref{eq:dressed_prop_sQED}. To turn this into a transition matrix
element, we need to truncate the free propagators for the final state
particles according to the LSZ prescription. In other words, each of
the full scalar field propagators will be modified
\[
\begin{aligned}
-i\left(p_{i}^{\prime 2}-m_{i}^{2}\right) D_{F}\left(z_{i}, p_{i}^{\prime} ; A\right) &=-i\left(p_{i}^{\prime 2}-m_{i}^{2}\right) \int_{0}^{\infty} \d T e^{-T \epsilon / \hbar} e^{\frac{i}{\hbar} p_{i}^{\prime} \cdot z_{i}+\frac{i}{\hbar} T\left(p_{i}^{\prime 2}-m_{i}^{2}\right)} f_{i}(0,T) \\
&=-e^{\frac{i}{\hbar} p_{i}^{\prime} \cdot z_{i}} \int_{0}^{\infty} \d T e^{-T \epsilon /\hbar} f_{i}(0,T) \frac{d}{d T} e^{\frac{i}{\hbar} T\left(p_{i}^{\prime 2}-m_{i}^{2}\right)}.
\label{eq:LSZ_reduction_cons}
\end{aligned}
\]
Again we used \eqref{eq:dressed_prop_sQED}, and defined
\[
f_{i}(0,T)=\int_{x_i(0)=0}^{p(T)=0} \mathcal{D} p \mathcal{D} x \exp \left[\frac{i}{\hbar} \int_{0}^{T} \d t\left(p \cdot\left(p_{i}^{\prime}-\dot{x}\right)+p^{2}-2 e p \cdot A-2 e p_{i}^{\prime} \cdot A+e^{2} A^{2}\right)\right].
\label{eq:x-pathintegral_consT}
\]
Integrating by parts and enforcing the on shell constraint
${p_{i}^{\prime}}^{2} \rightarrow m_{i}^{2}$ gives 
\[
\begin{aligned}
-i\left(p_{i}^{\prime 2}-m_{i}^{2}\right) D_{F}\left(z_{i}, p_{i}^{\prime} ; A\right) & \rightarrow e^{i (p_{i}^{\prime} \cdot z_{i}) / \hbar} f(0,\infty)\,.
\end{aligned}
\]
Combining this with \eqref{eq:4pt-correlator} and using the initial
positions of \eqref{eq:initial_pos}, we find that the partially
truncated Green's function associated with our chosen process
is
\[
G_4\left(p_{1}^{\prime}, p_{2}^{\prime} ; b\right)\Big|_{3,4}
=\int \mathcal{D} A_{\mu} \, e^{i (\Delta x \cdot p_{1}^{\prime}) / \hbar} 
f_{1}(0,\infty) f_{2}(0,\infty) e^{\frac{i}{\hbar} S_{A}\left(A_{\mu}\right)},
\label{eq:S_evolution_oper}
\]
where the notation on the left-hand side denotes those particles for
which the LSZ reduction has been carried out. After shifting $p
\rightarrow p+e A$ in the integration over $p$,
\eqref{eq:x-pathintegral_consT} implies
\[
&f_{i}(0,\infty)\\
&=\int_{x(0)=0}^{p(T)=0} \!\!\!\mathcal{D} p \mathcal{D} x \exp \left\{\frac{i}{\hbar} \int_{0}^{\infty} \!\!\!\d t\left[\left(p-\frac{\dot{x}}{2}+\frac{p_{i}^{\prime}}{2}\right)^{2}-\frac{\dot{x}^{2}}{4}+\frac{p_{i}^{\prime} \cdot \dot{x}}{2} -\frac{m_{i}^{2}}{4}-e \dot{x} \cdot A-e p_{i}^{\prime} \cdot A\right]\right\}\,.
\]
The $p$ integral is Gaussian and can be absorbed into the overall
normalisation of the path integral of \eqref{eq:x-pathintegral_consT}
(as can the term in $m_{i}^{2}$). One then has
\[
f_{i}(0,\infty)&=\int \mathcal{D} x \exp \left\{\frac{i}{\hbar} \int_{0}^{\infty} \d t\left[-\frac{\dot{x}^{2}}{4}+\frac{1}{2} p_{i}^{\prime} \cdot \dot{x}-e \dot{x} \cdot A-e p_{i}^{\prime} \cdot A\right]\right\}\,.
\label{eq:x-pathintegral_cons}
\]
Equation~(\ref{eq:S_evolution_oper}) has a nice interpretation: to
represent the (half-truncated) Green's function for scalar particles
interacting via a gauge field, one can describe the passage of each
particle by a factor representing the sum over possible trajectories,
weighted by an ``action'' containing the interaction of the particle
with the gauge field. It is possible to carry out the path integral in
\eqref{eq:x-pathintegral_cons} perturbatively, which corresponds to
summing over the various wobbles that the trajectory can have. These
wobbles are caused by interactions with the gauge field, as one
expects: each wobble corresponds to a recoil against an emitted
photon. Then the path integral over the gauge field in
\eqref{eq:S_evolution_oper} sums over all possible photon emissions
between the scalar particle lines.

\begin{figure}
\begin{center}
\scalebox{0.6}{\includegraphics{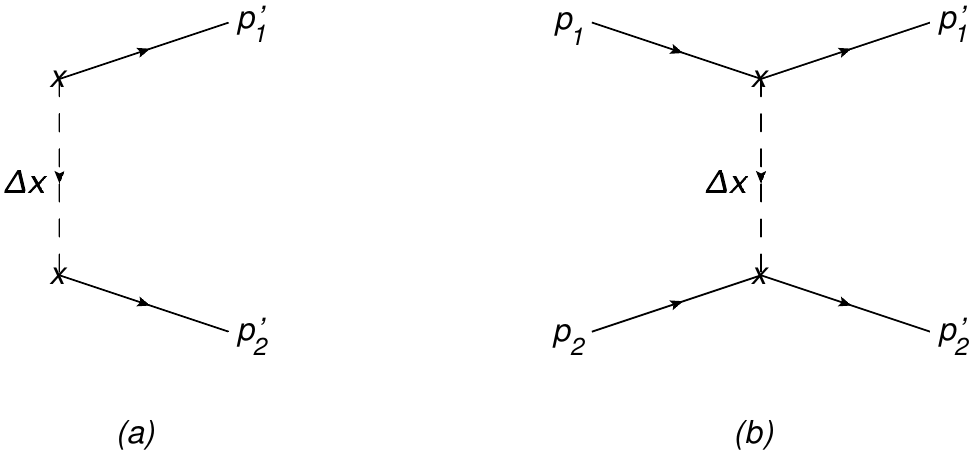}}
\caption{(a) The half-truncated 4-point Green's function
      considered in eq.~(\ref{eq:S_evolution_oper}), in which
      off-shell particles separated by a distance $\Delta x^\mu$
      propagate out to form outgoing states; (b) the complete
      scattering process.}
\label{fig:amphalf}
\end{center}
\end{figure}

Above, we have only carried out the LSZ reduction for the outgoing
particles 3 and 4. We must also carry out the reduction for particles
1 and 2. That this is not as simple as in
eq.~(\ref{eq:LSZ_reduction_cons}) ultimately stems from the fact that
the lower limit of the Schwinger proper time integral in
eq.~(\ref{eq:DF_propagator}) is zero, rather than minus infinity. It
is possible to transform proper time coordinates so that the LSZ
reduction for incoming particles can be carried out in the above
formulae, as argued recently in ref.~\cite{Mogull:2020sak}. An
alternative approach was presented, some time ago, in
ref.~\cite{Fried}. Here we will take a more pedestrian approach, and
simply consider that the above argument has provided only half of the
full four-point Green's function, as shown in
figure~\ref{fig:amphalf}(a). That is, the particles at the origin and
$\Delta x^\mu$ are off-shell and propagate out to form the final
states with momenta $\{p'_i\}$. We can then easily fill in the
remaining half of the scattering process, as shown in
figure~\ref{fig:amphalf}(b), by appending the integrand of
eq.~(\ref{eq:S_evolution_oper}) with two additional $f$-factors for
the incoming particles:
\begin{align}
S\left(\{p_{i}\}, \{p'_{i}\} ; \Delta x\right)
&=\int \mathcal{D} A_{\mu} \, e^{i (\Delta x \cdot p_{1}^{\prime}) / \hbar} 
f_1(-\infty,0)f_{1}(0,\infty) f_2(-\infty,0)
f_{2}(0,\infty) e^{\frac{i}{\hbar} S_{A}\left(A_{\mu}\right)}\notag\\
&=\int \mathcal{D} A_{\mu} \, e^{i (\Delta x \cdot p_{1}^{\prime}) / \hbar} 
f_1(-\infty,\infty) f_2(-\infty,\infty)
e^{\frac{i}{\hbar} S_{A}\left(A_{\mu}\right)},
\label{eq:S-matrix-calc}
\end{align}
where in the second line we have used the definition of the
$f$-factors to combine them into a single factor associated with each
incoming particle. On the left-hand side, we have acknowledged that
the LSZ reduction has now been carried out for the incoming particles,
so that eq.~(\ref{eq:S-matrix-calc}) constitutes an $S$-matrix
element. As such, it includes the case of trivial scattering (which
results if $A_\mu\rightarrow 0$). If we instead want the $T$-matrix
element, we would replace eq.~(\ref{eq:S-matrix-calc}) with
\begin{align}
T\left(\{p_{i}\}, \{p'_{i}\} ; \Delta x\right)
&=e^{i (\Delta x \cdot p_{1}^{\prime}) / \hbar}
\left[\int \mathcal{D} A_{\mu} \,  
f_1(-\infty,\infty) f_2(-\infty,\infty)
e^{\frac{i}{\hbar} S_{A}\left(A_{\mu}\right)}-1\right] \,.
\label{eq:T-matrix-calc}
\end{align}
To make sense of these expressions, let us consider the leading
contribution to the path integral for each incoming particle. This
corresponds to the classical approximation in which the particles
follow classical straight-line trajectories, such that the
fluctuations $x^{\mu}(t)= \dot{x}^{\mu}=0$. Then, one may simplify
\[
f_{i}(-\infty,\infty) \rightarrow \exp \left[-\frac{i}{\hbar}
  e \int_{-\infty}^{\infty} \d t \, p_{i}^{\prime} \cdot A\right]=\Phi_{i}\,.
\label{flim}
\]
The right-hand side is simply a Wilson line operator
\[
\Phi_{i}=\exp \left[-\frac{i}{\hbar} e \int_{-\infty}^{\infty} \d x_{i}^{\prime \mu} A_{\mu}\right]\,,
\label{eq:Wilson_lineop}
\]
evaluated along the path
\[
  {x^{\prime}}_{i}^{\mu}=z_{i}^{\mu}+p_{i}^{\prime \mu} t,\quad
  -\infty< t < \infty.
\]
We have thus found that if our particles start at time $t=0$ separated by an impact parameter $b^{\mu}$, they evolve according to the operator
\[
S\left(p_{1}^{\prime}, p_{2}^{\prime} ; \Delta x\right)=\int \mathcal{D} A_{\mu} e^{i (\Delta x \cdot p_{1}^{\prime}) / \hbar} \Phi_{1} \Phi_{2} e^{\frac{i}{\hbar} S_{A}\left(A_{\mu}\right)}\,.
\label{eq:vev_Wilson}
\]
This amounts to saying that the amplitude is described by a vacuum
expectation value of Wilson line operators, which is by no means a new
observation: the description of the Regge (high energy) limit of $2
\rightarrow 2$ scattering in terms of Wilson lines was given in
QCD~\cite{Korchemskaya:1994qp}, and later generalised to gravity in
\cite{Melville:2013qca,Luna:2016idw} (see also
refs.~\cite{Bonocore:2020xuj,Bonocore:2021qxh}). The QED case,
however, is particularly simple. If we neglect pair production of the
scalar fields, the path-integral in \eqref{eq:x-pathintegral_cons} can
be performed exactly. It may look more familiar if we write the Wilson
line coupling in terms of a current:
\[
-\frac{i}{\hbar} e \int \d x^{\mu} A_{\mu} \equiv-\frac{i}{\hbar} \int \d^4 x J_{i}^{\mu} A_{\mu}, \quad J_{i}^{\mu}=e v_i^{\mu} \delta^{(3)}(\boldsymbol{x}-\boldsymbol{z}(t))\,,
\]
where the delta function localises the current onto the particle's worldline described by $\boldsymbol{z}(t)$. Then one may define an overall current
\[
J=J_{1}+J_{2}\,
\]
so that the path integral of \eqref{eq:vev_Wilson} assumes the familiar form
\[
S\left(p_{1}^{\prime}, p_{2}^{\prime} ; \Delta x\right)=\int \mathcal{D} A_{\mu} e^{i (\Delta x \cdot p_{1}^{\prime}) / \hbar} e^{\frac{i}{\hbar}\left[S_{A}\left(A_{\mu}\right)-A_{\mu} J^{\mu}\right]}\,.
\]
Given $S_{A}\left(A_{\mu}\right)$ is quadratic in QED, one finds
\[
S\left(p_{1}^{\prime}, p_{2}^{\prime} ; \Delta x\right)=e^{i (\Delta x \cdot p_{1}^{\prime}) / \hbar} \exp \left[\frac{i}{\hbar} \int \d^4 x \int \d^4 y J^{\mu}(x) D_{\mu \nu}(x-y) J^{\nu}(y)\right]\,,
\label{eq:current_photon_exchange}
\]
where $D_{\mu \nu}(x-y)$ is the photon propagator. This can be
interpreted as follows. The exponent consists of all possible
one-photon exchanges between the two Wilson lines (including those
diagrams in which the photon may be emitted and absorbed by the same
particle). This one-loop contribution then exponentiates, as we know
it must! We did not have to force this property: it comes out simply
from the formalism we are using. As was shown in
e.g. \cite{Melville:2013qca}, the one-loop VEV of Wilson lines
generates the eikonal phase $\chi$ experienced by two interacting
particles in precisely the situation we are examining. Note that this
eikonal phase is only dependent on the transverse distance which we
can identify with $x_{\perp}-b_{\perp}$. In
general we expect it to be related to the 4-point amplitude via
eq.\eqref{eq:finalStateStep}.  Thus, we can write
\eqref{eq:vev_Wilson} simply as\footnote{We stress that
eq.~(\ref{Sresult}) applies to the $S$-matrix. Substituting this into
eq.~(\ref{eq:T-matrix-calc}) recovers the result that the scattering
amplitude --related to the $T$-matrix -- is given by $e^{i\chi}-1$.}
\[
S\left(p_{1}^{\prime}, p_{2}^{\prime} ; \Delta x\right)=e^{i (\Delta x \cdot p_{1}^{\prime}) / \hbar} e^{i \chi\left(x_{\perp}-b_{\perp},s\right)  / \hbar}.
\label{Sresult}
\]
In turn, this leads to a final state
\[
\int  \dPhi(p_1', p_2') \,\int \d^4 x_1 \, \d^4 x_2 \; \tilde \phi ( x_1, x_2) \, e^{i(p_1' \cdot x_1 + p_2' \cdot x_2)/ \hbar}  e^{i (\Delta x \cdot p'_{1})/ \hbar} e^{i \chi\left(x_{\perp}-b_{\perp},s\right)  / \hbar}\left|p_{1}^{\prime} p_{2}^{\prime}\right\rangle\,.
\label{eq:eikonal-worldline}
\]
At the leading order we're considering, there is no distinction between the impact parameter $b^{\mu}$ and $\Delta x^{\mu}$ and therefore we can equivalently write \eqref{eq:eikonal-worldline} as
\[
&\int  \dPhi(p_1', p_2') \,\int \d^4 x_1 \, \d^4 x_2 \; \tilde \phi ( x_1, x_2) \, e^{i(p_1' \cdot x_1 + p_2' \cdot x_2)/ \hbar}  e^{i (b \cdot p'_{1})/ \hbar} e^{i \chi\left(x_{\perp}-b_{\perp},s\right)  / \hbar}\left|p_{1}^{\prime} p_{2}^{\prime}\right\rangle\, , 
\]
which agrees with \eqref{eq:eikonalState} within our approximation. In general, we need to relate $e^{i \chi\left(x_{\perp}-b_{\perp},s\right)  / \hbar}$ with the four-point amplitude via \eqref{eq:eikonal_def} and this would imply that the relation between $b^{\mu}$ and $\Delta x^{\mu}$ 
is more subtle (see eq.\eqref{eq:b_x_relation}).

This has all been in the conservative regime with no radiation, and now we would
like to extend this to the radiative case.  To this end, we need to go back
and include the contribution of the gauge field insertion in the
Green's function correlator. In principle, we should allow any number
($n$) of photons and therefore we will generalise
eq.~(\ref{eq:4pt_dressed}) to the $(n+4)$-point correlator
\[
G_{n+4}(\phi_1(x_1),\phi_1(x_2),\phi_2(x_3),\phi_2(x_4),\{A_{\sigma_j}(x_j)\}_{j=5,...,n+4}) &  \\
=\int \mathcal{D} A_{\mu} \, \mathcal{D} \phi_{1} \, \mathcal{D} \phi_{2} \, \phi_1(x_1) \phi_1(x_2) \phi_2(x_3) \phi_2(x_4) A_{\sigma_5}(x_5) & ... A_{\sigma_{n+4}}(x_{n+4})  e^{\frac{i}{\hbar} S(\phi_1,\phi_2,A_{\mu})}.
\]
We may then carry out similar steps to above to find the $S$-matrix
element for the radiative corrections to the process of
figure~\ref{fig:amphalf}(b). The result is
(c.f. eq.~(\ref{eq:S-matrix-calc}))
\begin{align}
S\left(\{p_{i}\}, \{p'_{i}\},\{k_i\} ; \Delta x\right)
&=\prod_{j=5}^{n+4} \left[\int \d^4 x_j \, e^{-i (k_j \cdot x_j) / \hbar}  \square_{x_j} \right] e^{i (\Delta x \cdot p_{1}^{\prime}) / \hbar} 
\int \mathcal{D} A_{\mu} \, e^{\frac{i}{\hbar} S_{A}\left(A_{\mu}\right)}
\notag\\
&\times \prod_{j=5}^{n+4} \left[(\varepsilon^{\eta_j}(k_j) \cdot A_{\sigma_j}(x_j)) \right]f_1(-\infty,\infty) f_2(-\infty,\infty),
\label{eq:S-matrix-rad}
\end{align}
where we have included the LSZ reduction for the outgoing photons,
with momenta $\{k_i\}$ and polarisation vectors
$\{\varepsilon(k_i)\}$. Let us clarify this expression by again taking
the leading classical behaviour, such that the two massive particles
are following straight-line trajectories. Then we have, using
\eqref{flim},
\[
S&\left(\{p_{i}\}, \{p'_{i}\},\{k_i\} ; \Delta x\right) \\
&=\prod_{j=5}^{n+4} \left[\int \d^4 x_j \, e^{-i (k_j \cdot x_j) / \hbar}  \square_{x_j} \right] \int \mathcal{D} A_{\mu} \, e^{i (\Delta x \cdot p_{1}^{\prime}) / \hbar} \Phi_{1} \Phi_{2} \prod_{j=5}^{n+4} \left[(\varepsilon^{\eta_j}(k_j) \cdot A_{\sigma_j}(x_j)) \right] e^{\frac{i}{\hbar} S_{A}\left(A_{\mu}\right)}.
\label{eq:vev_Wilsonwithrad}
\]
Since the path integral is Gaussian, it is easy to perform it
analytically and to take the LSZ reduction to get
\[
S&\left(\{p_{i}\}, \{p'_{i}\},\{k_i\} ; \Delta x\right) \nonumber\\
&=\frac{e^{i( \Delta x \cdot p'_1)/\hbar}}{n!} \prod_{j=5}^{n+4} \left[\, \varepsilon^{\eta_j}(k_j) \cdot \tilde{J}(k_j) \right] \exp \left[i \int \d^4 x \int \d^4 y J^{\mu}(x) D_{\mu \nu}(x-y) J^{\nu}(y)\right]\nonumber\\
&=\frac{e^{i( \Delta x \cdot p'_1)/\hbar}}{n!} \prod_{j=5}^{n+4} \left[\, \varepsilon^{\eta_j}(k_j) \cdot \tilde{J}(k_j) \right]e^{i\chi(x_{\perp}-b_{\perp},s)/\hbar},
\label{eq:Smatrix_Wilsonline}
\]
where we have defined the Fourier transform of the current 
\[
\tilde{J}_{\mu}(k_j) = \int \d^4 x_j \, e^{-i (k_j \cdot x_j) / \hbar} J_{\mu}(x_j),
\]
and recognised the eikonal phase from
eqs.~(\ref{eq:current_photon_exchange}, \ref{Sresult}).  We see that
the photon distribution is exactly Poissonian in this approximation,
which is the hallmark of a coherent state. To see this in more detail,
note that we can construct the full final state by summing over
infinitely many photon emissions. Including then the measure for the
incoming wavepackets etc., in our leading soft approximation we obtain
eq.~(\ref{eq:KMOC_outgoing2}) as desired. 

It is worth noticing at this point that a similar coherent state structure 
arise in scattering amplitudes as a consequence of infrared divergences: indeed, this is just a consequence of Weinberg soft theorems \cite{Weinberg1965}. The fact that the scalar QED amplitude factorizes at leading order in the soft expansion for the external photons is a classical phenomenon: indeed, it can be viewed as the soft photon emission from a classical hard current given by the massive particles following straight-line trajectories \cite{Kulish1970AsymptoticCA}. In this setup, there is a classical factorization for observables \cite{Gonzo:2020xza} which now we understand as a simple consequence of the uncertainty principle.

Going beyond the straight-line approximation, we cannot solve the path
integral over particle trajectories exactly, and thus need to rely on
perturbation theory. But still, it is possible to see that the
(perturbatively defined) classical currents associated to massive
particle trajectories are emitting on-shell photons according to a
Poissonian distribution. Therefore, coherent states naturally appear
in this setup. This can be proved by splitting the path integral in
\eqref{eq:vev_Wilsonwithrad} for the gauge field between potential and
radiation modes\footnote{Here, loosely speaking, we call radiation modes the ones corresponding to $k_0 \sim |\v{k}|$ and potential modes the ones with $k_0 \ll |\v{k}|$ \cite{Porto:2016pyg}.}
\[
 \int \mathcal{D} A_{\mu} = \int_{\text{pot}} \mathcal{D} A^{\text{pot}}_{\mu} \int_{\text{rad}} \mathcal{D} A^{\text{rad}}_{\mu}.
\]
For radiation modes we can use the on-shell plane-wave expansion and \eqref{eq:S-matrix-rad} becomes
\[
&S\left(\{p_{i}\}, \{p'_{i}\},\{k_i\} ; \Delta x\right) = \prod_{j=5}^{n+4} \left[\int d^4 x_j \, e^{-i (k_j \cdot x_j) / \hbar}  \square_{x_j} \right]  \int_{\text{pot}} \mathcal{D} A^{\text{pot}}_{\mu} \int_{\text{rad}} \mathcal{D} A^{\text{rad}}_{\mu} \\
& \times  \,\left(\int \mathcal{D} x_1 \, e^{W_1^{\text{pot}} + W_1^{\text{rad}}}\right) \left(\int \mathcal{D} x_2 \,  e^{W_2^{\text{pot}} + W_2^{\text{rad}}}\right) \, \prod_{j=5}^{n+4} \left[(\varepsilon^{\eta_j}(k_j) \cdot A_{\sigma_j}(x_j)) \right] e^{\frac{i}{\hbar} S_A(A_{\mu})}  ,
\label{eq:unitaryevol_rad}
\]
where we have written
\begin{equation}
  f_i(-\infty,\infty)=\int{\cal D}x_i e^{W_i^{\rm pot}+W_i^{\rm rad}},
\end{equation}
and defined
\[
W_i^{\text{pot}} &= \frac{i}{\hbar} \int_{-\infty}^{+\infty} \d t \left(-\frac{1}{4} \dot{x}^2 + \frac{1}{2} p_i' \cdot \dot{x} - e \dot{x} \cdot A^{\text{pot}} - e p_i' \cdot A^{\text{pot}} \right)\,,   \\
W_i^{\text{rad}}  &=  - \frac{i}{\hbar^{\frac{3}{2}}} e \int_{-\infty}^{+\infty} \d t (\dot{x} + p_i')_{\mu} \sum_{\eta=\pm} \int \d \Phi(k) \,\bigl[a_{\eta}(k) \varepsilon^{\mu*}_{\eta}(k) e^{-\frac{i}{\hbar} k \cdot x} + h.c. \bigr] .
\]
If we restore the power counting in $\hbar$ for the coupling $e \to e / \sqrt{\hbar}$ and we define a (perturbative) \emph{trajectory-dependent} classical current
\[
    J_i^{\mu}(t, \v{x}) := i e (\dot{x}+ p_i')^{\mu} \,,
\]
we can write 
\[
W_i^{\text{rad}}  = -\int_{-\infty}^{+\infty} \d t \sum_{\eta=\pm} \int \d \Phi(\bar{k})\,\bigl[a_{\eta}(k) (J_i(t, \v{x}) \cdot \varepsilon^*_{\eta}(\bar{k})) e^{-i \bar{k}\cdot x} - h.c.\bigr].
\]
At this point we can express the current in frequency modes $J_i^{\mu}(\omega, x)$, 
\[
    J_i^{\mu}(t, \v{x}) = \int \frac{\d \bar{\omega}}{2 \pi} \, e^{-i \bar{\omega} t} \tilde{J}_i^{\mu}(\bar{\omega}, \v{x})\,,
\]
which gives
\[
    W_i^{\text{rad}} = \sum_{\eta=\pm} \int \d \Phi(\bar{k}) \, \left(\tilde{J}_i^*(-\omega_k, \v{x}) \cdot \varepsilon_{\eta}(\bar{k}) a^{\dagger}_{\eta}(k) - h.c.\right).
\]
This is indeed a coherent state, in the form of a displacement operator. We wish to remind the reader that at this stage we have not yet expanded the trajectory in perturbation theory, since \eqref{eq:unitaryevol_rad} is an exact result.
Effectively, at every order in perturbation theory, there will be an effective on-shell current which is the classical ``source" of coherent radiation for each trajectory. What remains to be established, perturbatively, is how these classical currents interact to produce the final conjectured coherent state.

\section{Classical Spin \& Colour Dynamics from the Eikonal}
\label{sec:spin}
So far, we have only discussed the classical dynamics of spinless charges in electrodynamics or point masses in gravity. 
However, the application of scattering amplitudes to classical systems involving colour and spin is also an important topic \cite{Guevara:2018wpp,Guevara:2019fsj,Chung:2018kqs,Maybee:2019jus,Chung:2019duq,Arkani-Hamed:2019ymq,Moynihan:2019bor,Burger:2019wkq,Bern:2020buy,Aoude:2020onz,Aoude:2020mlg,Emond:2020lwi,Aoude:2021oqj,Haddad:2021znf}, motivated in large part by the dynamics of inspiraling black holes with spin angular momentum. 
In this section, we expand our discussion to include minimum uncertainty spin and colour states. For simplicity, however, we revert to the purely conservative case throughout this section, leaving radiation of spin to future work.

\subsection{Final state including spin}
In section \ref{sec:eikonal-review}, we derived a formula for the final state involving an eikonal operator acting on the two-particle state. In general, states are labelled by an assortment of quantum numbers which can also evolve during a scattering event, and so in this section we will generalise the eikonal $S$-matrix operator to include Lie-algebra valued quantum numbers, focussing on spin for clarity. 

We begin with a generalisation of equation~\eqref{eq:psi} to include spin, resulting in a state of the form
\[\label{initspinstate}
\ket{\psi} &\equiv \int \sd\Phi(p_1,p_2)\phi_b(p_1,p_2)\xi_1^{a_1}\xi_2^{a_2}a^\dagger(p_1)_{a_1}a^\dagger(p_2)_{a_2}\ket{0}\\
&\equiv \int \sd\Phi(p_1,p_2)\phi_b(p_1,p_2)\xi_1^{a_1}\xi_2^{a_2}\ket{p_1,a_1;p_2,a_2} \\
&\equiv \int \sd\Phi(p_1,p_2)\phi_b(p_1,p_2)\ket{p_1,\xi_1;p_2,\xi_2} \,,
\]
where $a_1$ and $a_2$ are the spins (and the indices are summed over).
When spin is included, the expectation value of the $T$ matrix is given by
\begin{equation}
	\braket{p_1',a_1';p_2',a_2'|T|p_1,a_1;p_2,a_2} = \mathcal{A}_4(p_1,p_2\rightarrow p_1',p_2')_{a_{1}a_{2}}^{a'_{1} a'_{2}}\delta^{(4)}(p_1+p_2-p_1'-p_2')\,.
\end{equation}
Notice that the scattering amplitudes are now matrices in spin space. In view of this matrix structure, we follow~\cite{Aoude:2020mlg,AccettulliHuber:2020oou,Chen:2021huj} and assume that eikonal
exponentiation takes a matrix form\footnote{We continue to ignore the quantum remainder in this discussion.}
\begin{equation}\label{eikonal_spin}
\exp\bigg({i\chi(x_\perp)/\hbar}\bigg)^{a_1',a_2'}_{a_1,a_2} = \delta^{a_1'}_{a_1}\delta^{a_2'}_{a_2}+ i \int \hat{\sd}^4q \, \del(2p_1\cdot q)\del(2p_2\cdot q)\,e^{-iq\cdot x/\hbar}\, \amp_4{}^{a_1',a_2'}_{a_1,a_2} \,.
\end{equation}
Note that the four-point amplitude with spin depends on more variables than simply Mandelstam $s$ and $q^2$. Similarly the eikonal $\chi$ depends on more variables; we explicitly indicated the dependence on $x_\perp$ above because it is particularly important.

Acting with the $S$-matrix on the initial state eq. \eqref{initspinstate} and following steps similar those in section \ref{sec:KMOC_eikonal}, we find the out state to be
\[
S\ket \psi  = \ket{\psi} +  \int & \sd \Phi\left(p_1',p_2'\right) \dd^4 q \, \phi_b\left(p_{1}'-q, p_{2}'+q\right) \ket{p_{1}', a_{1}' ; p_{2}', a_{2}'}\\& \times \hat{\delta}\left(2p_1'\cdot q - q^2\right)\hat{\delta}\left(2p_2'\cdot q + q^2\right) i\mathcal{A}_4{}_{a_{1} a_{2}}^{a'_{1} a'_{2}} \, \xi_1^{a_{1}} \xi_2^{a_{2}} \,,
\]
which clearly recovers the spinless case of equation~\eqref{eq:stepToFinalState} if we take the spin group to be trivial. Again following section~\ref{sec:KMOC_eikonal}, we invert the Fourier transform in eq. \eqref{eikonal_spin} to find
\[
i\del(2\tilde{p}_1\cdot q)\del(2\tilde{p}_2\cdot q) \amp_4{}^{a_1' a_2'}_{a_1,a_2} = \frac{1}{\hbar^4}\int \d^4 x \, e^{iq\cdot x/\hbar}\left\{\exp\bigg({i\chi(x_\perp)/\hbar}\bigg)^{a_1',a_2'}_{a_1,a_2} - \delta^{a_1'}_{a_1}\delta^{a_2'}_{a_2}\right\} \,.
\]
As a result, we can express the final state in terms of the eikonal operator as
\[
\label{eq:spinEikonalState}
S\ket \psi  = \frac{1}{\hbar^4}\int \dPhi(p_1',p_2')\,\ket{p_{1}', a_{1}' ; p_{2}', a_{2}'} \int &\dd^4  q \d^4 x \, {\phi}_{b} ( p_1'-q,p_2'+q)   e^{iq\cdot x/\hbar}\\
&\times\exp\bigg({i\chi(x_\perp)/\hbar}\bigg)^{a_1',a_2'}_{a_1,a_2} \, \xi_1^{a_{1}} \xi_2^{a_{2}} \,.
\]

While we have constructed this with spin in mind, it is applicable to any operator with quantum numbers $a_i$, and so we will now consider the expectation value of a generic Lie-algebra valued operator $\cl{O}$.
The change in an observable due to a scattering event is defined as
\begin{equation}
	\Delta\cl{O} = \braket{\Psi|S^\dagger \hat{\cl{O}}S|\Psi} - \braket{\Psi|\hat{\cl{O}}|\Psi} = \braket{\Psi|S^\dagger [\hat{\cl{O}},S]|\Psi}.
\end{equation}
For concreteness, we restrict ourselves to an important class of operators with sensible classical limits, namely the momentum operator $\Pop^\mu$, the Pauli-Lubanski operator $\Wop^{\mu}$ defined as
\begin{equation}
	\Wop^{\mu} \equiv \frac{1}{2} \epsilon^{\mu \nu \rho \sigma} \mathbb{P}_{\nu} \mathbb{S}_{\rho \sigma} \,,
\end{equation}
and (in the Yang-Mills case) the colour charge operator $\mathbb{C}^a$. Such operators act on one particle states, are at most linear in the momentum and obey momentum conservation, meaning they can be written
\begin{equation}
	\braket{p,a|\cl{O}|k,b} = \Del(p-k)\cl{O}^a_{~b}(p).
\end{equation}
Examining an operator that acts only on the spin space of particle one, we find
\[
\cl{O}_1S\ket{\psi} =  \frac{1}{\hbar^4}\int \dPhi(p_1',p_2')\, &\ket{p_1',b_1;p_2',a_2'} \int \dd^4  q \, \d^4 x\, \phi_b( p_1'-q,p_2'+q) e^{iq\cdot x/\hbar}\\
&\qquad\times\cl{O}_1(p_1')_{~a_1'}^{b_1}\exp\bigg({i\chi(x_\perp)/\hbar}\bigg)^{a_1',a_2'}_{a_1,a_2}  \, \xi_1^{a_{1}} \xi_2^{a_{2}} \,,
\]
where, in order to evaluate the operator, we have inserted a complete set of spin states
\begin{equation}
	\mathds{1} = \sum_{b_1,b_2}\int d\Phi(k_1,k_2)~ \ket{k_1,b_1;k_2,b_2}\bra{k_1,b_1;k_2,b_2}.
\end{equation}
Similarly, acting with $S\cl{O}_1\ket{\psi}$ gives
\[
S\cl{O}_1\ket{\psi} =  \frac{1}{\hbar^4}\int \dPhi(p_1',p_2')\, &\ket{p_1',b_1;p_2',a_2'} \int \dd^4  q \, \d^4 x \, \phi_b( p_1'-q,p_2'+q) e^{iq\cdot x/\hbar}\\
&\qquad\times\exp\bigg({i\chi(x_\perp)/\hbar}\bigg)^{b_1,a_2'}_{a_1',a_2}\cl{O}_1(p_1'-q)_{~a_1}^{a_1'} \, \xi_1^{a_{1}} \xi_2^{a_{2}}\, .
\]
Notice that we encounter the operator $\cl{O}_1(p_1'-q)$ at a shifted value of the momentum. 
Since the class of operators we consider is at most linear in the momentum, we proceed by writing
\[
\cl{O}_1(p_1'-q) = \cl{O}_1(p_1') - \cl{O}_1(q) \,.
\]
Notice that our notation $\cl{O}_1(q)$ indicates that momentum factors in the operator are evaluated at momentum $q$.
Using this notation, we can eventually write
\[\label{o1commutator}
&[\cl{O}_1(p_1'),S]\ket{\psi} = \frac{1}{\hbar^4}\int \dPhi(p_1',p_2')\, \ket{p_1',b_1;p_2',a_2'} \int \dd^4  q \, \d^4 x \, \phi_b( p_1'-q,p_2'+q) e^{iq\cdot x/\hbar} \\
&~~~~~\times\left\{\left[\cl{O}_1(p_1'),\exp \left(i\chi(x_\perp)/\hbar\right)\right]^{b_1,a_2'}_{a_1,a_2} + \exp\bigg({i\chi(x_\perp)/\hbar}\bigg)^{b_1,a_2'}_{a_1',a_2} \cl{O}_1(q)_{~a_1}^{a_1'}\right\}  \, \xi_1^{a_{1}} \xi_2^{a_{2}} \, .
\]
Two matrix structures have appeared under the integral: one is a commutator, while the second is a more simple matrix product.
This structure is a generalisation of the structure discussed in~\cite{Maybee:2019jus}, 
where the second term was called the ``direct'' term in contrast to the ``commutator'' term.

We will shortly perform the $q$ and $x$ integrals in equation~\eqref{o1commutator} using stationary phase. 
In preparation for doing so, it is convenient to rewrite the equation such that the matrix 
$
\exp({i\chi(x_\perp)/\hbar})
$
stands to the left of any other matrices. 
Then in the evaluation of the overlap $\bra{\psi} S^\dagger [\cl{O}_1(p_1'),S]\ket{\psi}$, the matrix and (on the solution of the
stationary phase conditions) its inverse will simplify.
We only have to move the eikonal operator though the commutator term in equation~\eqref{o1commutator}; we can do so using
the Baker-Hausdorff lemma in the form \cite{sakurai_napolitano_2017}
\[
	[\cl{O},e^{i\chi/\hbar}] =e^{i\chi/\hbar} \left( -\frac{i}{\hbar}[\chi,\cl{O}] - \frac{1}{2\hbar^2}[\chi,[\chi,\cl{O}]] + \cdots + \frac{(-i)^n}{n!\hbar^n}[\chi,[\chi,[\chi,...[\chi,\cl{O}]]]...]\right) \,,
\label{eq:BHlemma}
\]
where the last term contains $n$ nested commutators involving $\chi$.

So far, we have not yet taken advantage of any simplifications available for large, classical spins. 
Classical spin representations must be large, in the sense that the spin quantum number $S$ times $\hbar$ is a classical angular momentum $S\hbar$. 
It can be useful to think of this as the limit $S \rightarrow \infty$, $\hbar \rightarrow 0$ with $S\hbar$ fixed. 
Large spin representations have the property that
\[
\label{eq:expectExample}
\langle S_{\mu\nu} S_{\rho\sigma} \rangle = \langle S_{\mu\nu} \rangle \langle S_{\rho\sigma} \rangle + \textrm{small} \,,
\]
where $S_{\mu\nu}$ is a spin Lorentz generator and $\langle S_{\mu\nu} \rangle$ is its expectation value on the spin state. 
The key point here is that the small correction term is of order $S \hbar^2$ compared to the explicit term on the right-hand
side, which is of order $S^2 \hbar^2$ (see appendix A of~\cite{delaCruz:2020bbn} for a recent review). We would therefore like
to take advantage of this kind of simplification in the context of equation~\eqref{o1commutator}.

In fact, we can easily take advantage of this simplification provided we first use the Baker-Hausdorff lemma~\eqref{eq:BHlemma}
to move the eikonal operator to the left. 
The reason is that this exposes the infinite series on the right-hand-side of equation~\eqref{eq:BHlemma} which is a series
in inverse powers of $\hbar$. 
These inverse powers are compensated by powers in commutators of operators; for example in 
the case of spin we have
\[
\label{eq:Wcommutator}
[\Wop_{\mu}, \Wop_{\nu} ] = -i \hbar \epsilon_{\mu\nu\rho\sigma} \Wop^\rho \Pop^\sigma \,.
\]
Had we not simplified the matrix structure and first tried to replace operators by expectation values we would make an error because
the small correction in equation~\eqref{eq:expectExample}, which is intimately related to the commutator~\eqref{eq:Wcommutator},
would be omitted.

Having taken advantage of the Baker-Hausdorff lemma, then, we may replace all operators by commutators. 
In doing so, we must retain the infinite set of non-vanishing commutators.
This is easily done: we simply introduce the Poisson bracket notation defined by
\[
\{ \langle \Wop_{\mu} \rangle, \langle \Wop_{\nu} \rangle \} = - \epsilon_{\mu\nu\rho\sigma} \langle \Wop^\rho \rangle \langle \Pop^\sigma \rangle \,.
\]
Notice that the Poisson brackets are directly inherited from the underlying algebraic structure of the quantum field theory. We have scaled out appropriate factors of $i\hbar$.

We therefore pass from equation~\eqref{o1commutator} to
\[
[\cl{O}_1(p_1'),S]\ket{\psi} &= \frac{1}{\hbar^4}\int \dPhi(p_1',p_2')\int \dd^4  q \, \d^4 x \, \phi_b( p_1'-q,p_2'+q)\, e^{iq\cdot x/\hbar}\\
&\hspace{-10pt}\times \left(\left\{\cl{O}_1(p_1'),e^{i\chi(x_\perp,s; \langle w \rangle)/\hbar}\right\}_\textrm{B.H.} + \cl{O}_1(q) e^{i\chi(x_\perp,s; \langle w \rangle)/\hbar}\right)\ket{p_1',\xi_1;p_2',\xi_2}, 
\]
where all operators have been replaced with scalar functions, at the expense of introducing Poisson brackets. The notation is
\[
\label{eq:bhlemma2}
\left\{\cl{O}_1(p_1'),e^{i\chi(x_\perp,s; \langle w \rangle)/\hbar}\right\}_\textrm{B.H.}
\equiv 
e^{i\chi/\hbar} \left( -\{\cl{O},\chi\} - \frac{1}{2}\{\chi,\{\cl{O},\chi\}\} + \cdots \right).
\]
We emphasise that $\{ \cdot, \cdot \}_\textrm{B.H.}$ is \emph{not} a Poisson bracket: it is simply convenient notation for the commutator
structure in equation~\eqref{eq:BHlemma}. 

In this form, the eikonal has also become a scalar function $\chi(x_\perp,s; \langle w \rangle, \ldots)$ which now depends on the expectation value of operators, eg Pauli-Lubanski $\langle w \rangle$ (and/or in the Yang-Mills case, the colour).
We can therefore perform the $x$ and $q$ integrals by stationary phase, following precisely the methods of section~\ref{sec:conservativeImpulse}.
The stationary phase conditions are
\[
\label{eq:stationaryImpulse2}
q_{\mu}  &= -\frac{\partial}{\partial x^{\mu}} \chi (x_{\perp}) \,.
\]
Note again that the scalar eikonal depends on variables other than $x_\perp$ which we have suppressed. 
Because of this additional dependence, $q^\mu_*$ need not be proportional to $x_\perp^\mu$, and will in general point along some other direction in the plane perpendicular to $\tilde{p}_1$ and $\tilde{p}_2$. The final result for the out state is then
\[
[\cl{O}_1(p_1'),S]\ket{\psi} &= \int \dPhi(p_1',p_2')\, \phi_b( p_1'-q_*,p_2'+q_*)e^{iq_*\cdot (x_*)/\hbar}\\
&\times \left(\left\{\cl{O}_1(p_1'),e^{i\chi(x_{\perp*})/\hbar}\right\}_\textrm{B.H.} + \cl{O}_1(q_*) e^{i\chi(x_{\perp*})/\hbar}\right)\ket{p_1',\xi_1;p_2',\xi_2}. 
\]
We can evaluate $\bra{\psi}S^\dagger$ on the stationary phase too, which results in
\begin{equation}\label{key1}
	\bra{\psi}S^\dagger = \int d\Phi(p_1,p_2)\phi_b^\dagger(p_1-q_*,p_2+q_*)e^{-iq_*\cdot (x_*)/\hbar}e^{-i\chi^\dagger(x_{\perp*})}\bra{p_1,\xi_1;p_2,\xi_2},
\end{equation}
such that
\[\label{DeltaO1}
\Delta\cl{O}_1 &= \braket{\psi|S^\dagger[\cl{O}_1,S]|\psi}\\
&= \int d\Phi(p_1,p_2)|\phi(p_1-q_*,p_2+q_*)|^2
\left( e^{-i\chi^\dagger(x_{\perp*})/\hbar} \left\{\cl{O}_1(p_1),e^{i\chi(x_{\perp*})/\hbar}\right\}_\textrm{B.H.} + \cl{O}_1(q_*)\right)\\
&= \Lexp  e^{-i\chi^\dagger(x_{\perp*})/\hbar} \left\{\cl{O}_1(p_1),e^{i\chi(x_{\perp*})/\hbar}\right\}_\textrm{B.H.} + \cl{O}_1(q_*)\Rexp \,.
\]
The Baker-Hausdorff lemma in the form of equation~\eqref{eq:bhlemma2} immediately tells us how to expand the Poisson bracket, leading to a neat expression for the change in $\cl{O}_1$
\[\label{DeltaO12}
\Delta\cl{O}_1 = \cl{O}_1(q_*) -\{\cl{O}_1(p_1),\chi\} - \frac{1}{2}\{\chi,\{\cl{O}_1(p_1),\chi\}\} - \frac{1}{6}\{\chi,\{\chi,\{\cl{O}_1(p_1),\chi\}\}\} + \cdots.
\]
where we recall that $\chi$ depends on $x_{\perp*}^\mu$ along with spin, colour etc, and we have dropped the brackets since we are in the fully classical regime. 
It is interesting to compare this expression to equation 7.20 in reference~\cite{Bern:2020buy}.
Evidently our equation~\eqref{DeltaO12} captures a similar exponentiation, though details of the structure are different.
It would be interesting to compare and contrast these two approaches in more detail in future.

As a first check, we consider the momentum operator of particle one, which acts trivially in spin-space meaning we can choose
\begin{equation}\label{key2}
	\cl{O}_1 = \mathbb{P}_1^\mu,~~~~~\rightarrow~~~~~ {\cl{O}_1}^{a}_{~b}(p) = p_1^\mu\delta^{a}_{~b}.
\end{equation}
Since the identity commutes with everything in spin space, only the $\cl{O}(q_*)$ term contributes, giving
\begin{equation}\label{key3}
	\Delta p_1^\mu = \KMOCav{q_*^\mu} = -\KMOCav{\pd^\mu\chi(x_\perp)},
\end{equation}
which matches the expression given in eq. \eqref{eq:impulsePerp}. Note, however, that in general the $x_\perp^\mu$ dependence of $\chi$ will be different in the spinning case, which will lead to additional spin contributions to the impulse.

Let us emphasise that we expect equation~\eqref{DeltaO12} to hold also for the change in the spin and the colour since the associated operators are at most linear in the momentum operator. 
We will examine these two cases in more detail below. 
It is useful to note that for operators without momentum dependence, the term involving $q_*$ vanishes, since it was induced by a shift in momentum and the linearity of the operator.

\subsection{Zero variance in spin space}

It is possible to prove, similarly to what we have done for the radiation case in section~\ref{sec:Rad_State}, that the generic spin quantum state for the massive particle is necessarily a coherent spin state once we impose the uncertainty principle. Let's first define a coherent spin state according to the covariant $SU(2)$ Schwinger construction, following \cite{Aoude:2021oqj}. 
A generic spin vector can be expressed using a set of $2s + 1$ harmonic oscillators
\[
[a^a,a^{\dagger}_b] = \delta^a_b \qquad \v{S} = \frac{\hbar}{2} a^{\dagger}_a \v{\sigma}^a_b a^b ,
\label{eq:spinSU2_oscillators}
\]
where $\v{\sigma}^a_b$ are the standard Pauli matrices. Using \eqref{eq:spinSU2_oscillators}, it is easy to see that the spin vector $ \v{S}$ obeys the expected algebra $[S^i,S^j] = i \hbar \epsilon^{i j k} S^k$. Spin coherent states are then defined as eigenstates of $a_a$:
\[
\ket{\alpha^S} := e^{-\frac{1}{2}\tilde{\alpha}_a \alpha^a} e^{\alpha^a a^{\dagger}_a} \ket{0} \rightarrow a_a \ket{\alpha^S} = \alpha_a \ket{\alpha^S} \, .
\]
We can now represent the most generic quantum spin state of the initial and the final massive particles in our two-body scattering problem in terms of a density matrix.  Such a construction has  been developed in the quantum optics literature \cite{giraud2008classicality}, and for our setup it corresponds to dressing the external incoming and outgoing massive momentum states with the following spin density matrix 
\[
\hat{\rho}^S_{\text{in}} := \int \prod_{i=1}^2  \, \d^2 \alpha^{S}_i \mathcal{P}^S_{\text{in}}(\alpha_1^S,\alpha_2^S) \ket{\alpha_1^S} \ket{\alpha_2^S} \bra{\alpha_2^S} \bra{\alpha_1^S} \nonumber \\
\hat{\rho}^S_{\text{out}} := \int \prod_{i=1}^2  \, \d^2 \alpha^{S}_i \mathcal{P}^S_{\text{out}}(\alpha_1^S,\alpha_2^S) \ket{\alpha_1^S} \ket{\alpha_2^S} \bra{\alpha_2^S} \bra{\alpha_1^S} ,
\]
where in the classical limit the P-representation is separable \cite{giraud2008classicality}
\[
\mathcal{P}^S_{\text{in}}(\alpha_1^S,\alpha_2^S) &=  \mathcal{P}^S_{\text{in}}(\alpha_1^S)  \mathcal{P}^S_{\text{in}}(\alpha_2^S) \nonumber \\
\mathcal{P}^S_{\text{out}}(\alpha_1^S,\alpha_2^S) &=  \mathcal{P}^S_{\text{out}}(\alpha_1^S)  \mathcal{P}^S_{\text{out}}(\alpha_2^S) .
\]
We now impose the uncertainty principle for the generic expectation value of the spin operator, 
\[
\text{Tr}_{\rho^S_{\text{in}}}(\v{S}_i \v{S}_j) &\stackrel{\hbar \to 0}{=} \text{Tr}_{\rho^S_{\text{in}}}(\v{S}_i) \text{Tr}_{\rho^S_{\text{in}}}(\v{S}_j) \qquad \text{for} \,\,\,i=1,2 \nonumber \\
\text{Tr}_{\rho^S_{\text{out}}}(\v{S}_i \v{S}_j) &\stackrel{\hbar \to 0}{=} \text{Tr}_{\rho^S_{\text{out}}}(\v{S}_i) \text{Tr}_{\rho^S_{\text{out}}}(\v{S}_j) \qquad \text{for} \,\,\,i=1,2 \, .
\]
Following some simple steps, it is easy to see that this is equivalent to demanding the zero-variance property for both $\mathcal{P}^S_{\text{in}}(\alpha_i^S)$ and $\mathcal{P}^S_{\text{out}}(\alpha_i^S)$ ($i=1,2$). Therefore, we conclude that 
\[
\mathcal{P}^S_{\text{in}}(\alpha_1^S,\alpha_2^S) &= \delta(\alpha_1^S - \alpha_{1,*,\text{in}}^S) \delta(\alpha_2^S - \alpha_{2,*,\text{in}}^S) \nonumber \\
\mathcal{P}^S_{\text{out}}(\alpha_1^S,\alpha_2^S) &= \delta(\alpha_1^S - \alpha_{1,*,\text{out}}^S) \delta(\alpha_2^S - \alpha_{2,*,\text{out}}^S) .
\]
This implies that spin coherent states are suitable to represent the quantum-mechanical state of classical spin particles, and more generally an exponentiation of the spin degrees of freedom is required in the classical limit. This is also consistent with the fact that we should expect to not have any entanglement in the spin sector \cite{Aoude:2020mlg}.

\subsection{Classical Colour from the Eikonal} 
We now turn to explicit examples. Before discussing spin, it is useful to explore the simpler example of classical colour observables. In this case, the eikonal has colour indices and the operator we consider is the colour operator $\mathbb{C}$. The change in the classical colour is then given by eq. \eqref{DeltaO1}
\begin{equation}
	\Delta c^a_1 = \Lexp e^{-i\chi^\dagger(x_{\perp*},c^i)/\hbar} \left\{c_1^a,e^{i\chi(x_{\perp*},c^i)/\hbar}\right\}_\textrm{B.H.}\Rexp,
\end{equation}
where $c^a = \braket{\psi|\mathbb{C}|\psi}$, and the $\cl{O}(q_*)$ term has dropped out since the colour does not depend on momentum. The first few terms in the Baker-Hausdorff expansion are given by
\begin{equation}\label{key5}
	\Delta c^a_1 = -\{c_1^a,\chi\} -\frac12\{\chi,\{c_1^a,\chi\}\} + \cdots,
\end{equation}
Using the properties of Poisson brackets, namely that $\{A,f(B)\} = \{A,B\}\frac{\pd f}{\pd B}$, along with the relation $\{c^a,c^b\} = f^{abc}c^c$, we can write the first bracket as
\begin{equation}\label{key6}
	\{c_1^a,\chi\} = \{c_1^a,c_1^b\}\frac{\pd\chi}{\pd c_1^b} = f^{abc}\frac{\pd\chi}{\pd c_1^b}c_1^c. 
\end{equation}

At order $\cl{O}(g^4)$, we expect contributions from the tree-level and one-loop in the first bracket, as well as the tree-level terms in the second bracket. The tree-level and one-loop eikonals are given by \cite{delaCruz:2020bbn,delaCruz:2021gjp}
\[
	\chi_1(x_{\perp},c^i) &= g^2(c_1\cdot c_2) \frac{\gamma}{4\pi\sqrt{\gamma^2-1}}\log\left|\frac{x_\perp^2}{L^2}\right|,\\
	\chi_2(x_{\perp},c^i) &= \frac{g^4(c_1\cdot c_2)^2(m_1+m_2)}{32\pi m_1m_2\sqrt{\gamma^2-1}|x_\perp|},
\]
where $L$ is an infrared cutoff. We can plug the tree-level eikonal into the bracket to find, at first order,
\begin{align}
	\Delta c^{a,0}_1 &= -\{c_1^a,\chi_1\} = \frac{\gamma g^2 f^{abc}c_1^bc_2^c}{4\pi\sqrt{\gamma^2-1}}\log\left|\frac{x_\perp^2}{L^2}\right|.
\end{align}
The second bracket is given by
\begin{align}\label{key7}
	\Delta c^{a,1}_1 &= -\frac12\{\chi,\{c_1^a,\chi\}\}\\
	&= -\frac12\frac{g^4\gamma^2}{(4\pi)^2(\gamma^2-1)}\log^2\left|\frac{x_\perp^2}{L^2}\right|f^{abc}\left[\{\chi,c_2^b\}c_1^c + c_2^b\{\chi,c_1^c\}\right]\\
	&= -\frac12\frac{g^4\gamma^2}{(4\pi)^2(\gamma^2-1)}\log^2\left|\frac{x_\perp^2}{L^2}\right|\left[f^{abc}f^{bde}c_1^cc_1^dc_2^e - f^{abc}f^{bde}c_2^cc_2^dc_1^e\right].
\end{align}
The contribution from the one-loop eikonal is evaluated in the first bracket to give
\[
	\{c_1^a,\chi_2\} &= \frac{g^4(m_1+m_2)}{32\pi m_1m_2\sqrt{\gamma^2-1}|x_\perp|}\left\{c_1^a,(c_1\cdot c_2)^2\right\}\\
					  &= -\frac{g^4(c_1\cdot c_2)f^{abc}c_1^bc_2^c}{16\pi\sqrt{\gamma^2-1}|x_\perp|}\left(\frac{1}{m_1} + \frac{1}{m_2}\right).
\]
Putting this together gives the full $\cl{O}(g^4)$ colour impulse
\[
	\Delta c_1^a &= \frac{\gamma g^2 f^{abc}c_1^bc_2^c}{4\pi\sqrt{\gamma^2-1}}\log\left|\frac{x_\perp^2}{L^2}\right| + \frac{g^4(c_1\cdot c_2)f^{abc}c_1^bc_2^c}{16\pi\sqrt{\gamma^2-1}|x_\perp|}\left(\frac{1}{m_1} + \frac{1}{m_2}\right) \\&- \frac12\frac{g^4\gamma^2}{(4\pi)^2(\gamma^2-1)}\log^2\left|\frac{x_\perp^2}{L^2}\right|\left[f^{abc}f^{bde}c_1^cc_1^dc_2^e - f^{abc}f^{bde}c_2^cc_2^dc_1^e\right].
\]
This matches the NLO colour impulse in the literature \cite{delaCruz:2021gjp,delaCruz:2020bbn}.
\subsection{Classical Spin from the Eikonal}
The angular impulse for particle one is given by
\begin{align}\label{deltaW}
	\Delta s_1^\mu &= \frac{1}{m_{1}}\braket{\Psi|S^{\dagger}[\Wop_{1}^{\mu},S]| \Psi}\\
	&= \Lexp e^{-i\chi^\dagger(x_{\perp*},s_i)/\hbar} \left\{s_1^\mu(p_1),e^{i\chi(x_{\perp*},s_i)/\hbar} \right\}_\textrm{B.H.} + s_1^\mu(q_*)\Rexp,
\end{align}
where the expection of the Pauli-Lubanski pseudovector is
\begin{equation}
\braket{p,a_i|\Wop^{\mu}|p',a_j} \equiv ms_{ij}^\mu(p)\hat{\delta}(p-p').
\end{equation}
The spin is more complicated to consider than colour, since the Pauli-Lubanski operator is a product of both the linear and angular momentum operators. This means that the $q_*$-dependent piece contributes to the spin and we find that the expansion is therefore given by 
\begin{equation}
	\Delta s_1^\mu = s_1^\mu(q_*) -\{s_1^\mu,\chi\} -\frac12 \{\chi,\{s_1^\mu,\chi\}\} + \cdots
\end{equation}
The first term is the spin vector of particle one evaluated at $q_*$, and so at leading order we can write it as
\begin{equation}\label{key8}
	s_1^\mu(q_*) = \frac{1}{2m_1}\epsilon^{\mu\nu\rho\sigma}q_{*\nu}S_{\rho\sigma}(p_1),
\end{equation}
where the spin tensor is defined as $S_{\rho\sigma}(p_1) = \frac{1}{m_1}\epsilon_{\rho\sigma\mu\nu}p_{1}^\mu s_{1}^\nu(p_1)$. We can input this definition to find that the leading order contribution can be expressed
\[
	s_1^\mu(q_*) &= \frac{1}{2m_1^2}\epsilon^{\mu\nu\rho\sigma}q_{*\nu}\epsilon_{\rho\sigma\lambda\tau}p_{1}^\lambda s_{1}^\tau(p_1) = \frac{1}{m_1^2}\left((q_*\cdot p_1)s_1^\mu(p_1) - (s_1\cdot q_*)p_1^\mu\right)\\
	&= \frac{1}{m_1^2}\bigg((\Delta p_1\cdot p_1)s_1^\mu(p_1) - (s_1\cdot \Delta p_1)p_1^\mu\bigg).
\]
We can compute the angular impulse in GR at the spin-$1/2\times$spin-$0$ order by considering the lowest order eikonal phase (see e.g.~\cite{Liu:2021zxr})
\begin{equation}\label{key9}
	\chi = -\frac{2Gm_1m_2}{\sqrt{\gamma^2-1}}\left[(2\gamma^2-1)\log\bigg|\frac{x_\perp}{L}\bigg| - \frac{2\gamma}{m_1}\frac{\epsilon^{\mu\nu\rho\sigma}u_{1\mu}x_{\perp\nu}u_{2\rho}s_{1\sigma}}{|x_\perp|^2} \right].
\end{equation}
The angular impulse is given by
\[
\Delta s_1^\mu &= s_1^\mu(q_*) -\{s_1^\mu,s_1^\nu\}\frac{\pd\chi}{\pd s_1^\nu}\\
&= -\frac{1}{m_1^2}p_1^\mu(s_1(p_1)\cdot\Delta p_1)-\frac{1}{m_1}\epsilon^{\mu\nu\rho\sigma}\frac{\pd\chi}{\pd s_1^\nu}p_{1\rho}s_{1\sigma},
\]
where we have used $\{s_1^\mu, s_1^\nu\} = \frac{1}{m_1}\epsilon^{\mu\nu\rho\sigma}p_{1\rho}s_{1\sigma}$.
The derivative with respect to $s_1^\mu$ is given by
\begin{equation}\label{key10}
	\frac{\pd\chi}{\pd s_{1\alpha}} =  \frac{4Gm_2\gamma}{\sqrt{\gamma^2-1}}\frac{\epsilon^{\alpha\mu\nu\rho}u_{1\mu}x_{\perp\nu}u_{2\rho}}{|x_\perp|^2},
\end{equation}
such that the angular impulse is finally
\begin{equation}\label{key11}
	\Delta s_1^\mu = -\frac{1}{m_1^2}(s_1\cdot \Delta p_1)p_1^\mu - \frac{4Gm_2\gamma}{\sqrt{\gamma^2-1}}\frac{\epsilon^{\mu \nu \rho \sigma} s_{1 \nu} u_{1 \rho} \epsilon_{\sigma \alpha \beta \gamma} u_{1}^{\alpha} u_{2}^{\beta} x_\perp^{\gamma}}{|x_\perp|^2},
\end{equation}
which matches the existing literature \cite{Maybee:2019jus} and where we have dropped the higher order $\Delta p_1\cdot p_1$ term.
\section{Discussion}
\label{sec:discussion}

The central theme of this paper has been that the classical limit emerges from scattering amplitudes via an infinite set of relationships satisfied by
multiloop, multileg amplitudes in a transfer expansion. These relationships arise from requiring negligible uncertainty in the measurement of 
observables computed from amplitudes in the correspondence limit where the classical approximation is valid. One can write scattering amplitudes
in an exponential form as a result of these relationships, as has long been recognised in the conservative sector through the eikonal approximation.
We have argued that the same holds for radiative physics. 

While we focused on the eikonal approach to classical dynamics, it is worth emphasising that we could have phrased our discussion in terms of other,
closely related, quantities. Recently the radial action has received particular attention in the literature~\cite{Bern:2021dqo}; this radial action is a 
classical limit of the usual quantum phase shifts~\cite{Kol:2021jjc}, and is a close cousin of the eikonal function~\cite{Bautista:2021wfy}.

The eikonal and the radial action arise directly in other approaches to gravitational dynamics. While this paper started with quantum field theory,
undergraduate classes start more simply with worldline actions --- and indeed pragmatic approaches to gravitational phenomena start with
analogous worldline theories~\cite{Goldberger:2004jt,Mogull:2020sak,Mougiakakos:2021ckm,Jakobsen:2021smu,Jakobsen:2021lvp,Jakobsen:2021zvh}. Nevertheless there is one inescapable fact about the dynamics that quantum field theory makes blatantly
clear: radiation is \emph{not} suppressed in fully relativistic physics. Thus in our approach radiation is encoded in $\alpha$, fully as important
as $\chi$. 
Similarly the physics of radiation reaction is
clear from the outset and is a consequence of a basic principle: conservation of momentum.

One of our central observations was that six-point tree amplitudes are suppressed in the classical region. We showed explicitly that this suppression hold in scalar QED, but our more general arguments indicate that the suppression should hold in general relativity~\cite{Britto:2021pud} and in perturbative/classical applications of Yang-Mills theory.
This suppression of the six-point tree amplitude is consistent with classical intuition. Classical electrodynamics, for example, involves two main ingredients: knowledge of the electromagnetic field, and knowledge of the particle motion. As our work has focussed on determining the asymptotic properties of particles and waves, the relevant part of the classical field is the radiation field: this is the part which determines the energy and momentum in the field, for example. Similar remarks hold in gravity.
In a quantum approach, asymptotic particle motion is captured by the eikonal function (or the radial action.) The radiation field is described at the leading quantum level by the five-point amplitude~\cite{Luna:2017dtq,KMOC,Cristofoli:2021vyo}. Our work generalises this last statement: we have argued that the all-order radiation field is described by the waveshape parameter $\alpha$, itself determined by the all-order five-point amplitude together with the eikonal function.
As the four- and five-point amplitudes are enough to describe the classical dynamics, it makes sense that higher-point amplitudes are classically suppressed.
Incidentally, our work does not mean that six (or higher) point amplitudes can never be of use in classical computations: they may appear for example inside cuts as building-blocks of other observables. 

From a classical perspective, the waveshape $\alpha$ is essentially~\cite{Cristofoli:2021vyo} a Newman-Penrose scalar: either $\phi_2$ in gauge theory or $\psi_4$ in gravity. Our contention is then that a complete description of the classical radiative dynamics is given by knowledge of one of these scalars, and knowledge of the eikonal function $\chi$. It is important for us in this context that the \emph{full} impulse can be deduced by differentiation of $\chi$, not just the scattering angle. Indeed in a radiative context, knowledge of the angle needs to be supplemented by knowledge of the outgoing energy to fully determine the outgoing particle trajectory.

Spin angular momentum is also a key aspect of classical dynamics, in particular in the context of black hole dynamics. In section~\ref{sec:spin} we built on our eikonal description of classical motion to make contact with spin (and colour) at the conservative level. This is particularly natural for us since it is well-known by now that classical spinning particles have a natural interpretation as coherent states of large occupation number in spin space. In Yang-Mills theory, the Wong dynamics of colour arises in a very analogous manner~\cite{delaCruz:2020bbn} through coherent states in colour space~\cite{delaCruz:2021gjp}.

We emphasise that we made no attempt to prove that the final semi-classical state really is of the precise form given in equation~\eqref{eq:finalStateProposal1}. We provided arguments in support of the exponentiation based on path-integral techniques, and indeed at leading non-trivial order it is clear that the radiation is described by a coherent state. Beyond this order much less is known. It certainly seems possible
that the waveshape $\alpha$ could depend, in general, on more variables. The structure of the final state we suggested does allow for radiation reaction,
so that the $\chi$ and $\alpha$ are not independent of one another. However it could certainly be that tail effects induce a closer link between the two
functions. Since progress in this field is currently very rapid we can look forward to exploring the exponentiation of radiation in far greater detail before long.

It is remarkable how much our community has learned in recent years by studying the interface between classical and quantum field theory (see references~\cite{Carrasco:2021bmu,Carrasco:2020ywq,Cristofoli:2020hnk,Brandhuber:2019qpg,Emond:2019crr,Cristofoli:2019neg,Bjerrum-Bohr:2019nws,Bjerrum-Bohr:2019kec,Cristofoli:2020uzm,AccettulliHuber:2020dal,Bjerrum-Bohr:2021vuf,Brandhuber:2021kpo,Gomez:2021shh,Damgaard:2021rnk,AccettulliHuber:2020dal,AccettulliHuber:2020oou,Brandhuber:2021bsf,Chacon:2021wbr,White:2020sfn,Chacon:2021lox,Guevara:2021yud} for some recent highlights). 
The expansion of scattering amplitudes in momentum transfer was an important topic in this paper.
Perhaps in the future the structure of this expansion can be used to simplify
multiloop collider computations by inputting known lower-loop information. 
We hope that an exploration of amplitudes in heavy particle effective theories~\cite{Damgaard:2019lfh,Aoude:2020onz,Haddad:2020tvs} could be another productive future direction for the community.
This class of effective theory is intimately related to the expansion in powers of momentum transfer, and in view of the phenomenological
importance of heavy quark effective theory there could be a lot to explore.

Looking to the future, our work leaves a great many open questions. We explicitly discussed the suppression of the six-point tree amplitude, but consideration of higher-point correlators indicates that all $n$-point trees are suppressed for $n>5$ in Yang-Mills theory and in gravity. It would be interesting to check this. Further, if the semi-classical final state is determined by exponentiating only four- and five-point amplitudes, it must follow that \emph{every} higher-point amplitude, in the classical regime, is determined by four- and five-point amplitudes.  

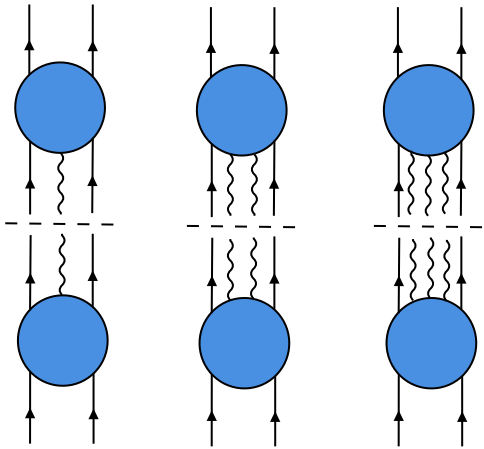
\begin{figure}[h]
\begin{center}

\tikzset{every picture/.style={line width=0.75pt}} %set default line width to 0.75pt        

\begin{tikzpicture}[x=0.75pt,y=0.75pt,yscale=-1,xscale=1]
%uncomment if require: \path (0,8388); %set diagram left start at 0, and has height of 8388

%Shape: Ellipse [id:dp26309117276292016] 
\draw  [fill={rgb, 255:red, 74; green, 144; blue, 226 }  ,fill opacity=1 ] (79.64,8304.71) .. controls (71.15,8295.57) and (71.56,8281.18) .. (80.56,8272.56) .. controls (89.56,8263.93) and (103.73,8264.35) .. (112.21,8273.49) .. controls (120.7,8282.63) and (120.29,8297.03) .. (111.29,8305.65) .. controls (102.29,8314.27) and (88.12,8313.85) .. (79.64,8304.71) -- cycle ;
%Straight Lines [id:da1053522991510536] 
\draw    (79.6,8274) -- (79.89,8236.06) ;
\draw [shift={(79.74,8255.03)}, rotate = 90.44] [fill={rgb, 255:red, 0; green, 0; blue, 0 }  ][line width=0.08]  [draw opacity=0] (5.36,-2.57) -- (0,0) -- (5.36,2.57) -- cycle    ;
%Straight Lines [id:da8381122007915189] 
\draw    (79.6,8340.97) -- (79.64,8304.71) ;
\draw [shift={(79.62,8322.84)}, rotate = 90.06] [fill={rgb, 255:red, 0; green, 0; blue, 0 }  ][line width=0.08]  [draw opacity=0] (5.36,-2.57) -- (0,0) -- (5.36,2.57) -- cycle    ;
%Straight Lines [id:da7775380664861671] 
\draw    (111.3,8341) -- (111.29,8305.65) ;
\draw [shift={(111.3,8323.32)}, rotate = 89.99] [fill={rgb, 255:red, 0; green, 0; blue, 0 }  ][line width=0.08]  [draw opacity=0] (5.36,-2.57) -- (0,0) -- (5.36,2.57) -- cycle    ;
%Straight Lines [id:da5594386601395533] 
\draw    (95.64,8266.21) .. controls (93.97,8264.55) and (93.97,8262.88) .. (95.63,8261.21) .. controls (97.29,8259.54) and (97.28,8257.87) .. (95.61,8256.21) .. controls (93.94,8254.54) and (93.94,8252.88) .. (95.6,8251.21) .. controls (97.26,8249.54) and (97.26,8247.88) .. (95.59,8246.21) .. controls (93.92,8244.55) and (93.91,8242.88) .. (95.57,8241.21) .. controls (97.23,8239.54) and (97.23,8237.88) .. (95.56,8236.21) -- (95.56,8235.39) -- (95.56,8235.39) ;
%Straight Lines [id:da4846019256837213] 
\draw    (110.88,8272.16) -- (110.89,8235.39) ;
\draw [shift={(110.88,8253.77)}, rotate = 90.01] [fill={rgb, 255:red, 0; green, 0; blue, 0 }  ][line width=0.08]  [draw opacity=0] (5.36,-2.57) -- (0,0) -- (5.36,2.57) -- cycle    ;
%Shape: Ellipse [id:dp8146783818286132] 
\draw  [fill={rgb, 255:red, 74; green, 144; blue, 226 }  ,fill opacity=1 ] (110.88,8156.34) .. controls (119.37,8165.48) and (118.95,8179.88) .. (109.96,8188.5) .. controls (100.96,8197.12) and (86.79,8196.7) .. (78.3,8187.56) .. controls (69.82,8178.42) and (70.23,8164.03) .. (79.23,8155.41) .. controls (88.22,8146.79) and (102.39,8147.21) .. (110.88,8156.34) -- cycle ;
%Straight Lines [id:da6806225730035012] 
\draw    (110.92,8187.06) -- (110.63,8225) ;
\draw [shift={(110.77,8206.03)}, rotate = 90.44] [fill={rgb, 255:red, 0; green, 0; blue, 0 }  ][line width=0.08]  [draw opacity=0] (5.36,-2.57) -- (0,0) -- (5.36,2.57) -- cycle    ;
%Straight Lines [id:da11657802919891536] 
\draw    (110.92,8120.08) -- (110.88,8156.34) ;
\draw [shift={(110.9,8138.21)}, rotate = 90.06] [fill={rgb, 255:red, 0; green, 0; blue, 0 }  ][line width=0.08]  [draw opacity=0] (5.36,-2.57) -- (0,0) -- (5.36,2.57) -- cycle    ;
%Straight Lines [id:da26543854926210897] 
\draw    (79.22,8120.06) -- (79.23,8155.41) ;
\draw [shift={(79.22,8137.73)}, rotate = 89.99] [fill={rgb, 255:red, 0; green, 0; blue, 0 }  ][line width=0.08]  [draw opacity=0] (5.36,-2.57) -- (0,0) -- (5.36,2.57) -- cycle    ;
%Straight Lines [id:da42158057809742133] 
\draw    (94.87,8194.84) .. controls (96.54,8196.5) and (96.55,8198.17) .. (94.89,8199.84) .. controls (93.23,8201.51) and (93.23,8203.17) .. (94.9,8204.84) .. controls (96.57,8206.5) and (96.58,8208.17) .. (94.92,8209.84) .. controls (93.26,8211.51) and (93.26,8213.17) .. (94.93,8214.84) .. controls (96.6,8216.5) and (96.61,8218.17) .. (94.95,8219.84) .. controls (93.29,8221.51) and (93.29,8223.17) .. (94.96,8224.84) -- (94.96,8225.67) -- (94.96,8225.67) ;
%Straight Lines [id:da7692387015335118] 
\draw    (79.64,8188.9) -- (79.63,8225.67) ;
\draw [shift={(79.63,8207.28)}, rotate = 90.01] [fill={rgb, 255:red, 0; green, 0; blue, 0 }  ][line width=0.08]  [draw opacity=0] (5.36,-2.57) -- (0,0) -- (5.36,2.57) -- cycle    ;
%Straight Lines [id:da2797362963968857] 
\draw  [dash pattern={on 4.5pt off 4.5pt}]  (67.22,8230.11) -- (122.56,8230.78) ;
%Shape: Ellipse [id:dp9472440219031224] 
\draw  [fill={rgb, 255:red, 74; green, 144; blue, 226 }  ,fill opacity=1 ] (170.3,8306.04) .. controls (161.82,8296.91) and (162.23,8282.51) .. (171.23,8273.89) .. controls (180.22,8265.27) and (194.39,8265.69) .. (202.88,8274.83) .. controls (211.37,8283.96) and (210.95,8298.36) .. (201.96,8306.98) .. controls (192.96,8315.6) and (178.79,8315.18) .. (170.3,8306.04) -- cycle ;
%Straight Lines [id:da5364856570837855] 
\draw    (170.27,8275.33) -- (170.56,8237.39) ;
\draw [shift={(170.41,8256.36)}, rotate = 90.44] [fill={rgb, 255:red, 0; green, 0; blue, 0 }  ][line width=0.08]  [draw opacity=0] (5.36,-2.57) -- (0,0) -- (5.36,2.57) -- cycle    ;
%Straight Lines [id:da008949574729196552] 
\draw    (170.27,8342.31) -- (170.3,8306.05) ;
\draw [shift={(170.29,8324.18)}, rotate = 90.06] [fill={rgb, 255:red, 0; green, 0; blue, 0 }  ][line width=0.08]  [draw opacity=0] (5.36,-2.57) -- (0,0) -- (5.36,2.57) -- cycle    ;
%Straight Lines [id:da5785231734767462] 
\draw    (201.97,8342.33) -- (201.96,8306.98) ;
\draw [shift={(201.96,8324.66)}, rotate = 89.99] [fill={rgb, 255:red, 0; green, 0; blue, 0 }  ][line width=0.08]  [draw opacity=0] (5.36,-2.57) -- (0,0) -- (5.36,2.57) -- cycle    ;
%Straight Lines [id:da020478317249902567] 
\draw    (179.64,8268.88) .. controls (177.97,8267.22) and (177.97,8265.55) .. (179.63,8263.88) .. controls (181.29,8262.21) and (181.28,8260.54) .. (179.61,8258.88) .. controls (177.94,8257.21) and (177.94,8255.55) .. (179.6,8253.88) .. controls (181.26,8252.21) and (181.26,8250.55) .. (179.59,8248.88) .. controls (177.92,8247.22) and (177.91,8245.55) .. (179.57,8243.88) .. controls (181.23,8242.21) and (181.23,8240.55) .. (179.56,8238.88) -- (179.56,8238.06) -- (179.56,8238.06) ;
%Straight Lines [id:da234876713258251] 
\draw    (201.55,8273.49) -- (201.56,8236.72) ;
\draw [shift={(201.55,8255.11)}, rotate = 90.01] [fill={rgb, 255:red, 0; green, 0; blue, 0 }  ][line width=0.08]  [draw opacity=0] (5.36,-2.57) -- (0,0) -- (5.36,2.57) -- cycle    ;
%Shape: Ellipse [id:dp8627458688606935] 
\draw  [fill={rgb, 255:red, 74; green, 144; blue, 226 }  ,fill opacity=1 ] (201.55,8157.68) .. controls (210.03,8166.82) and (209.62,8181.21) .. (200.62,8189.83) .. controls (191.63,8198.46) and (177.46,8198.04) .. (168.97,8188.9) .. controls (160.48,8179.76) and (160.9,8165.36) .. (169.89,8156.74) .. controls (178.89,8148.12) and (193.06,8148.54) .. (201.55,8157.68) -- cycle ;
%Straight Lines [id:da9513238759055793] 
\draw    (201.58,8188.39) -- (201.3,8226.33) ;
\draw [shift={(201.44,8207.36)}, rotate = 90.44] [fill={rgb, 255:red, 0; green, 0; blue, 0 }  ][line width=0.08]  [draw opacity=0] (5.36,-2.57) -- (0,0) -- (5.36,2.57) -- cycle    ;
%Straight Lines [id:da957826835280086] 
\draw    (201.58,8121.42) -- (201.55,8157.68) ;
\draw [shift={(201.57,8139.55)}, rotate = 90.06] [fill={rgb, 255:red, 0; green, 0; blue, 0 }  ][line width=0.08]  [draw opacity=0] (5.36,-2.57) -- (0,0) -- (5.36,2.57) -- cycle    ;
%Straight Lines [id:da9219062229383121] 
\draw    (169.88,8121.39) -- (169.89,8156.74) ;
\draw [shift={(169.89,8139.06)}, rotate = 89.99] [fill={rgb, 255:red, 0; green, 0; blue, 0 }  ][line width=0.08]  [draw opacity=0] (5.36,-2.57) -- (0,0) -- (5.36,2.57) -- cycle    ;
%Straight Lines [id:da5187131902135318] 
\draw    (179.54,8195.51) .. controls (181.21,8197.18) and (181.22,8198.84) .. (179.55,8200.51) .. controls (177.89,8202.18) and (177.9,8203.85) .. (179.57,8205.51) .. controls (181.24,8207.18) and (181.24,8208.84) .. (179.58,8210.51) .. controls (177.92,8212.18) and (177.93,8213.85) .. (179.6,8215.51) .. controls (181.27,8217.18) and (181.27,8218.84) .. (179.61,8220.51) .. controls (177.95,8222.18) and (177.96,8223.85) .. (179.63,8225.51) -- (179.63,8226.33) -- (179.63,8226.33) ;
%Straight Lines [id:da07261091208599413] 
\draw    (170.3,8190.23) -- (170.3,8227) ;
\draw [shift={(170.3,8208.61)}, rotate = 90.01] [fill={rgb, 255:red, 0; green, 0; blue, 0 }  ][line width=0.08]  [draw opacity=0] (5.36,-2.57) -- (0,0) -- (5.36,2.57) -- cycle    ;
%Straight Lines [id:da5543269048009081] 
\draw  [dash pattern={on 4.5pt off 4.5pt}]  (157.89,8231.44) -- (213.22,8232.11) ;
%Shape: Ellipse [id:dp03440259519803579] 
\draw  [fill={rgb, 255:red, 74; green, 144; blue, 226 }  ,fill opacity=1 ] (263.64,8306.04) .. controls (255.15,8296.91) and (255.56,8282.51) .. (264.56,8273.89) .. controls (273.56,8265.27) and (287.73,8265.69) .. (296.21,8274.83) .. controls (304.7,8283.96) and (304.29,8298.36) .. (295.29,8306.98) .. controls (286.29,8315.6) and (272.12,8315.18) .. (263.64,8306.04) -- cycle ;
%Straight Lines [id:da16862605523753982] 
\draw    (263.6,8275.33) -- (263.89,8237.39) ;
\draw [shift={(263.74,8256.36)}, rotate = 90.44] [fill={rgb, 255:red, 0; green, 0; blue, 0 }  ][line width=0.08]  [draw opacity=0] (5.36,-2.57) -- (0,0) -- (5.36,2.57) -- cycle    ;
%Straight Lines [id:da004717247009009773] 
\draw    (263.6,8342.31) -- (263.64,8306.05) ;
\draw [shift={(263.62,8324.18)}, rotate = 90.06] [fill={rgb, 255:red, 0; green, 0; blue, 0 }  ][line width=0.08]  [draw opacity=0] (5.36,-2.57) -- (0,0) -- (5.36,2.57) -- cycle    ;
%Straight Lines [id:da5677919067765058] 
\draw    (295.3,8342.33) -- (295.29,8306.98) ;
\draw [shift={(295.3,8324.66)}, rotate = 89.99] [fill={rgb, 255:red, 0; green, 0; blue, 0 }  ][line width=0.08]  [draw opacity=0] (5.36,-2.57) -- (0,0) -- (5.36,2.57) -- cycle    ;
%Straight Lines [id:da09691754419840293] 
\draw    (294.88,8273.49) -- (294.89,8236.72) ;
\draw [shift={(294.88,8255.11)}, rotate = 90.01] [fill={rgb, 255:red, 0; green, 0; blue, 0 }  ][line width=0.08]  [draw opacity=0] (5.36,-2.57) -- (0,0) -- (5.36,2.57) -- cycle    ;
%Shape: Ellipse [id:dp28790599931459626] 
\draw  [fill={rgb, 255:red, 74; green, 144; blue, 226 }  ,fill opacity=1 ] (294.88,8157.68) .. controls (303.37,8166.82) and (302.95,8181.21) .. (293.96,8189.83) .. controls (284.96,8198.46) and (270.79,8198.04) .. (262.3,8188.9) .. controls (253.82,8179.76) and (254.23,8165.36) .. (263.23,8156.74) .. controls (272.22,8148.12) and (286.39,8148.54) .. (294.88,8157.68) -- cycle ;
%Straight Lines [id:da08782890765087048] 
\draw    (294.92,8188.39) -- (294.63,8226.33) ;
\draw [shift={(294.77,8207.36)}, rotate = 90.44] [fill={rgb, 255:red, 0; green, 0; blue, 0 }  ][line width=0.08]  [draw opacity=0] (5.36,-2.57) -- (0,0) -- (5.36,2.57) -- cycle    ;
%Straight Lines [id:da5778057333754831] 
\draw    (294.92,8121.42) -- (294.88,8157.68) ;
\draw [shift={(294.9,8139.55)}, rotate = 90.06] [fill={rgb, 255:red, 0; green, 0; blue, 0 }  ][line width=0.08]  [draw opacity=0] (5.36,-2.57) -- (0,0) -- (5.36,2.57) -- cycle    ;
%Straight Lines [id:da2455098309021546] 
\draw    (263.22,8121.39) -- (263.23,8156.74) ;
\draw [shift={(263.22,8139.06)}, rotate = 89.99] [fill={rgb, 255:red, 0; green, 0; blue, 0 }  ][line width=0.08]  [draw opacity=0] (5.36,-2.57) -- (0,0) -- (5.36,2.57) -- cycle    ;
%Straight Lines [id:da2601562691157633] 
\draw    (263.64,8190.23) -- (263.63,8227) ;
\draw [shift={(263.63,8208.61)}, rotate = 90.01] [fill={rgb, 255:red, 0; green, 0; blue, 0 }  ][line width=0.08]  [draw opacity=0] (5.36,-2.57) -- (0,0) -- (5.36,2.57) -- cycle    ;
%Straight Lines [id:da7025366392212558] 
\draw  [dash pattern={on 4.5pt off 4.5pt}]  (251.22,8231.44) -- (306.56,8232.11) ;
%Straight Lines [id:da026812009019757888] 
\draw    (191.54,8195.51) .. controls (193.21,8197.18) and (193.22,8198.84) .. (191.55,8200.51) .. controls (189.89,8202.18) and (189.9,8203.85) .. (191.57,8205.51) .. controls (193.24,8207.18) and (193.24,8208.84) .. (191.58,8210.51) .. controls (189.92,8212.18) and (189.93,8213.85) .. (191.6,8215.51) .. controls (193.27,8217.18) and (193.27,8218.84) .. (191.61,8220.51) .. controls (189.95,8222.18) and (189.96,8223.85) .. (191.63,8225.51) -- (191.63,8226.33) -- (191.63,8226.33) ;
%Straight Lines [id:da48282846234471744] 
\draw    (191.11,8267.94) .. controls (189.46,8266.27) and (189.47,8264.6) .. (191.14,8262.94) .. controls (192.81,8261.28) and (192.82,8259.61) .. (191.16,8257.94) .. controls (189.51,8256.27) and (189.52,8254.6) .. (191.19,8252.94) .. controls (192.86,8251.28) and (192.87,8249.61) .. (191.21,8247.94) .. controls (189.56,8246.27) and (189.57,8244.6) .. (191.24,8242.94) .. controls (192.91,8241.28) and (192.92,8239.61) .. (191.26,8237.94) -- (191.27,8237.27) -- (191.27,8237.27) ;
%Straight Lines [id:da766622009103787] 
\draw    (270.84,8268.88) .. controls (269.17,8267.22) and (269.17,8265.55) .. (270.83,8263.88) .. controls (272.49,8262.21) and (272.48,8260.54) .. (270.81,8258.88) .. controls (269.14,8257.21) and (269.14,8255.55) .. (270.8,8253.88) .. controls (272.46,8252.21) and (272.46,8250.55) .. (270.79,8248.88) .. controls (269.12,8247.22) and (269.11,8245.55) .. (270.77,8243.88) .. controls (272.43,8242.21) and (272.43,8240.55) .. (270.76,8238.88) -- (270.76,8238.06) -- (270.76,8238.06) ;
%Straight Lines [id:da20520844386785275] 
\draw    (287.51,8269.14) .. controls (285.86,8267.47) and (285.87,8265.8) .. (287.54,8264.14) .. controls (289.21,8262.48) and (289.22,8260.81) .. (287.56,8259.14) .. controls (285.91,8257.47) and (285.92,8255.8) .. (287.59,8254.14) .. controls (289.26,8252.48) and (289.27,8250.81) .. (287.61,8249.14) .. controls (285.96,8247.47) and (285.97,8245.8) .. (287.64,8244.14) .. controls (289.31,8242.48) and (289.32,8240.81) .. (287.66,8239.14) -- (287.67,8238.47) -- (287.67,8238.47) ;
%Straight Lines [id:da9225451615943663] 
\draw    (279.51,8267.94) .. controls (277.86,8266.27) and (277.87,8264.6) .. (279.54,8262.94) .. controls (281.21,8261.28) and (281.22,8259.61) .. (279.56,8257.94) .. controls (277.91,8256.27) and (277.92,8254.6) .. (279.59,8252.94) .. controls (281.26,8251.28) and (281.27,8249.61) .. (279.61,8247.94) .. controls (277.96,8246.27) and (277.97,8244.6) .. (279.64,8242.94) .. controls (281.31,8241.28) and (281.32,8239.61) .. (279.66,8237.94) -- (279.67,8237.27) -- (279.67,8237.27) ;
%Straight Lines [id:da4125717487291387] 
\draw    (269.67,8225.67) .. controls (268.02,8223.99) and (268.03,8222.32) .. (269.71,8220.67) .. controls (271.39,8219.02) and (271.41,8217.35) .. (269.76,8215.67) .. controls (268.11,8213.99) and (268.12,8212.32) .. (269.8,8210.67) .. controls (271.48,8209.02) and (271.5,8207.35) .. (269.85,8205.67) .. controls (268.2,8203.99) and (268.21,8202.32) .. (269.89,8200.67) .. controls (271.57,8199.02) and (271.59,8197.35) .. (269.94,8195.67) -- (269.96,8193.66) -- (269.96,8193.66) ;
%Straight Lines [id:da3021745505561111] 
\draw    (286.87,8225.27) .. controls (285.2,8223.6) and (285.2,8221.94) .. (286.87,8220.27) .. controls (288.54,8218.6) and (288.54,8216.94) .. (286.87,8215.27) .. controls (285.2,8213.6) and (285.2,8211.94) .. (286.87,8210.27) .. controls (288.54,8208.6) and (288.54,8206.94) .. (286.87,8205.27) .. controls (285.2,8203.6) and (285.2,8201.94) .. (286.87,8200.27) .. controls (288.54,8198.6) and (288.54,8196.94) .. (286.87,8195.27) -- (286.87,8194.07) -- (286.87,8194.07) ;
%Straight Lines [id:da5455532890854453] 
\draw    (278.31,8226.34) .. controls (276.66,8224.67) and (276.67,8223) .. (278.34,8221.34) .. controls (280.01,8219.68) and (280.02,8218.01) .. (278.36,8216.34) .. controls (276.71,8214.67) and (276.72,8213) .. (278.39,8211.34) .. controls (280.06,8209.68) and (280.07,8208.01) .. (278.41,8206.34) .. controls (276.76,8204.67) and (276.77,8203) .. (278.44,8201.34) .. controls (280.11,8199.68) and (280.12,8198.01) .. (278.46,8196.34) -- (278.47,8195.67) -- (278.47,8195.67) ;

\end{tikzpicture}

\end{center}
\caption{Cuts contributing to the four-point amplitude. The two-loop amplitude built from five-point trees is classically relevant, while power counting shows that the three-loop contribution from six-point trees does not contribute to $\chi$ at this order. It will be interesting to explore, for example, the four-loop contribution from seven-points trees.}
\label{fig:loopcuts}
\end{figure}

Another interesting question following this work relates to the structure of the eikonal function $\chi$ at multiloop order. We have seen that the six-point tree amplitude must be suppressed in gauge theory and gravity. As shown in figure~\ref{fig:loopcuts}, the square of this tree contributes to the four-point three-loop amplitude, as determined for example through unitarity. Power-counting of coupings and $\hbar$'s indicates that this cut does not contribute to the eikonal function at this order. Since our work implies further suppression of higher-point trees, it will be interesting to determine what class of cuts are relevant for a unitarity-based construction of $\chi$. 

Finally, let us reiterate the philosophy underlying our article. We are taking scattering amplitudes as the \emph{definition} of our theory. While this definition
has its attractions, allowing us to bootstrap the theory from very basic first principles~\cite{Benincasa:2007xk,Arkani-Hamed:2017jhn}, there is a price: we must understand the many successes of
more standard definitions of quantum field theory, particularly in terms of path integrals and actions, without actually having an action. We focussed
on the uncertainty principle and the emergence of the classical limit, encountering two pieces of functional data, $\chi$ and $\alpha$, which are defined in the classical regime. Now a shortcoming of our quantum-first approach is that bound state dynamics is currently poorly understood. However, classical
objects are typically solutions of differential equations. The solutions of these equations can often be analytically continued~\cite{Kalin:2019rwq,Kalin:2019inp,Saketh:2021sri,Bini:2020hmy,Cho:2021arx} in a simple manner 
from the scattering case to the bound case. It would be very exciting if both $\chi$ and $\alpha$ can be continued in a similar way: this would
lead to a direct, quantum-first connection between amplitudes and binary black hole physics. Already the literature contains encouraging
results pointing in this direction~\cite{Goldberger:2017vcg,Bautista:2021inx}.

\acknowledgments
Our work benefited from discussions and correspondence with Paolo Di Vecchia, Carlo Heissenberg, Franz Herzog, David Kosower, Rodolfo Russo, Ofri Telem, Gabriele Veneziano, Justin Vines and Mao Zeng. We also thank Carlo, Rodolfo and Gabriele for sharing slides~\cite{carlosTalk, rodolfosTalk, gabrielesTalk} from their recent Saclay talks with us. Finally, we thank the Galileo Galilei Institute for Theoretical Physics (GGI) for hosting a workshop, conference and training week on ``Gravitational scattering, inspiral, and radiation'' which informed and enriched our work.  This project has received funding from the European Union's Horizon 2020 research and innovation program under the Marie Sklodowska-Curie grant agreement No.764850 ``SAGEX''. AC is supported by the Leverhulme Trust (RPG-2020-386). NM is supported by STFC grant ST/P0000630/1 and the Royal Society of Edinburgh Saltire Early Career Fellowship.
For the purpose of open access, the author has applied a Creative Commons Attribution (CC BY) licence to any Author Accepted Manuscript version arising from this submission.

\appendix

\section{Further details at six points}
In this appendix we provide the full details of the gauge fixed calculation which were not included in \ref{sec:6_point_tree}. 

For $A_{(2,2)}$ we have diagrams in the same classes as in the main text. There is only one location for the cubic emission on particle 1 with our gauge choice, and we will label the diagrams by 1 (2) for emission on the outgoing (incoming) particle 2 line. We find
\[
D_6(Q_1^2Q_2^2)|_{cubic,1}  &= \frac{1}{\hbar^5 }  \frac{\varepsilon_1\cdot(2\barq_1)\varepsilon_2\cdot (2p_2)(2p_2+\hbar \barq_2 +\hbar \bar k_2)\cdot (2p_1+\hbar \barq_1 -\hbar \bar k_1)}{(\barq_2 - \bar k_2)^2 (2p_2\cdot \bar k_2) (2p_1\cdot (\barq_1-\bar k_1) + \hbar (\bar q_1 -\bar k_1)^2)}
\\
&\qquad\qquad\qquad\qquad + 2 \left(p_1\cdot (\bar k_2 + \barq_2) + p_2 \cdot (\barq_1 - \bar k _1)\right)   \Bigg]
\\
&=- \frac{(\varepsilon_1\cdot \barq_1)(\varepsilon_2\cdot p_2)}{\hbar^4 (\barq_2 - \bar k_2)^2 (p_2\cdot \bar k_2) (p_1\cdot \bar k_1)} \Bigg[  \frac{4p_1\cdot p_2}{\hbar} - \frac{4p_1\cdot p_2\bar q_1 \cdot\bar k_1}{p_1\cdot \bar k_1}
\\
&\qquad\qquad\qquad\qquad +  \left(4p_1\cdot \bar k_2 + 2p_1 \cdot\bar k_1 +2 p_2 \cdot \bar k_2 \right)   \Bigg].
\label{D_6(Q_1^2Q_2^2)|_{cubic,1}}
\]
To get the last line we used momentum conservation to replace, for example, $k_1-q_1 \to q_2 -k_2$ and then dropped the $p_i\cdot \bar q_i$ as this gets an extra $\hbar$ from the on shell conditions. The second is
\[
D_6&(Q_1^2Q_2^2)|_{cubic,2}  = \frac{1}{\hbar^5 }  \frac{[\varepsilon_1\cdot(2\barq_1)\varepsilon_2\cdot (2p_2 +\hbar \barq_2)][(2p_2+\hbar \barq_2 -\hbar \bar k_2)\cdot (2p_1+\hbar \barq_1 -\hbar \bar k_1)]}{(\barq_2 - \bar k_2)^2 (2p_1\cdot (\barq_1-\bar k_1) + \hbar (\bar q_1 -\bar k_1)^2(2p_2\cdot (\barq_2-\bar k_2) + \hbar (\bar q_2 -\bar k_2)^2}
\\
&= \frac{1}{\hbar^4 (\barq_2 - \bar k_2)^2 (p_2\cdot \bar k_2) (p_1\cdot \bar k_1)}
\Bigg[\frac{ 4p_1\cdot p_2(\varepsilon_1\cdot \barq_1)(\varepsilon_2\cdot p_2)}{\hbar} + 4p_1\cdot p_2 (\varepsilon_1\cdot\barq_1)( \varepsilon_2 \cdot \barq_2 ) 
\\
& + 2(\varepsilon_1\cdot \barq_1)(\varepsilon_2\cdot p_2)(p_1\cdot\bar k_1  + p_2 \cdot \bar k_2) - 4p_1\cdot p_2(\varepsilon_1\cdot \barq_1)(\varepsilon_2\cdot p_2) \left( \frac{\bar q_1\cdot \bar k_1 }{p_1\cdot \bar k_1}+\frac{\bar q_2\cdot \bar k_2 }{p_2\cdot \bar k_2}\right) \Bigg]  \,.
\label{D_6(Q_1^2Q_2^2)|_{cubic,2}}
\]
Now for the diagrams with cubic and quartic vertices. They both have the quartic vertex on particle 1 and two locations for cubic emission on particle 2. These, again labelled by 1 (2), are, 
\[
D_6(Q_1^2Q_2^2)_{cubic/quartic,1} &= -\frac{2}{\hbar^5 } \frac{\varepsilon_1 \cdot (2p_2)\varepsilon_2\cdot (2p_2+\hbar \barq_2)}{(\barq_2 -\bar k_1)^2 (2p_2\cdot \bar k_2)} 
\\
&= -\frac{1}{\hbar^4 (\barq_2 -\bar k_1)^2 (p_2\cdot \bar k_2)} \left[\frac{ 4\varepsilon_1 \cdot (p_2)\varepsilon_2\cdot p_2)}{\hbar} + 2(\varepsilon_1\cdot \barq_2) (\varepsilon_2\cdot p_2)\right]
\] 
and the second 
\[
D_6&(Q_1^2Q_2^2)_{cubic/quartic,2} = -\frac{2}{\hbar^5 } \frac{\varepsilon_1 \cdot (2p_2+\hbar \barq_2-\hbar \bar k_2)\varepsilon_2\cdot (2p_2+2\hbar \barq_2)}{(\barq_1 -\bar k_1)^2 (2p_2\cdot(\barq_2- \bar k_2) + \hbar (\barq_2-\bar k_2)^2} 
\\
&= +\frac{1}{\hbar^4 (\barq_1 -\bar k_1)^2 (p_2\cdot \bar k_2)} \Bigg[\frac{ 4(\varepsilon_1 \cdot p_2)(\varepsilon_2\cdot p_2)}{\hbar} + 2 (\varepsilon_2\cdot p_2)(\varepsilon_1\cdot(\barq_2-\bar k_2))
\\
 &\qquad\qquad\qquad\qquad\qquad \qquad 4(\varepsilon_1\cdot p_2)( \varepsilon_2\cdot \barq _2) - \frac{4(\varepsilon_1 \cdot p_2)(\varepsilon_2\cdot p_2)(\barq_2 \cdot \bar k_2)}{p_2\cdot \bar k_2} \Bigg].
\]
Where we used the on-shell condition $2p_2\cdot \barq _2 = - \hbar \barq_2 ^2$, and then Taylor expanded the propagator denominator.
Note how the two most singular terms cancel when we combine these diagrams. 

The final diagram doesn't require the gauge fixing to simplify anything as it is already the correct order in a general gauge. This is
\[
D_6(Q_1^2Q_2^2)_{quartic} = \frac{4 \varepsilon_1\cdot \varepsilon_2}{\hbar^4 (\barq_2-\bar k_2)^2}.
\]
Combining these gives the result \eqref{eq:ampq12q22final} in the main text.

\section{Projection in the plane of scattering}
\label{app:projector}
The relation between the eikonal impact parameter $x^{\mu}_{\perp}$ and $b^{\mu}$ can be stated clearly by first introducing some notation. Let us define the following four-vectors in momentum space
\[
\begin{cases}\label{eq:projectors-definition}
e_{0}^{\mu} \equiv N_{0} (\tilde{p}^{\mu}_{1}+\tilde{p}^{\mu}_{2}) \\
e_{q}^{\mu} \equiv N_{q} (\tilde{p}^{\mu}_{1}-\tilde{p}^{\mu}_{2})-N_{0q}(\tilde{p}^{\mu}_{1}+\tilde{p}^{\mu}_{2}) \ ,
\end{cases}
\]
where the normalization factors $N_{0}$, $N_{q}$ and $N_{0q}$ are fixed by requiring $e^{2}_{0}=1$, $e^{2}_{q}=-1$ and $e_{0} \cdot e_{q}=0$. By definition of $x^{\mu}_{\perp}$, the following identities hold: 
\[
e_{0} \cdot x_{\perp}=0 \quad , \quad e_{q} \cdot x_{\perp}=0 \ .
\]
As a consequence, we can write the projection of $x^{\mu}$ on the plane orthogonal to $\tilde{p}^{\mu}_{1}$ and $\tilde{p}^{\mu}_{2}$ as
\[ \label{eq:projection-x}
x^{\mu}_{\perp}=x^{\mu}-(x \cdot e_{0} ) e_{0}^{\mu} +(x \cdot e_{q}) e^{\mu}_{q} \ ,
\]
where the different signs in the the last two terms are a consequence of $e_{0}^{\mu}$ being time-like while $e_{q}^{\mu}$ space-like.  Using \eqref{eq:projection-x} we can easily compute some of the derivatives involved in the evaluation of the stationary phase on $x^{\mu}$ such as
\[
\frac{\partial x_{\perp}^{2}}{\partial x_{\mu}}=2x_{\perp, \nu}(\eta^{\mu \nu}-e_{0}^{\mu}e_{0}^{\nu}+e_{q}^{\mu}e_{q}^{\nu})=2x^{\mu}_{\perp} \ .
\]
Another example where the use of \eqref{eq:projection-x} is useful is when we apply the stationary phase for the integral over $q^{\mu}$. In this case, the stationary condition for $q^{\mu}$ can be expressed as
\[
x^{\mu}=b^{\mu}+ \left. q_{\nu, *}\frac{\partial x_{\perp}^{\nu}}{\partial q_{\mu}} \right|_{q=q_{*}} \ ,
\]
where $q_{*}^{\mu}$ satisfies the stationary phase condition on $x^{\mu}$ given by $q^{\mu}_{*}=-2\chi'(x_{\perp})x^{\mu}_{\perp}$.  One of the advantages in the definition \eqref{eq:projectors-definition} of $e_{0}^{\mu}$ is that it is $q^{\mu}$ independent so that the previous stationary condition can be expressed as
\[
x^{\mu}=b^{\mu}+ \left. q_{\nu, *}\frac{\partial}{\partial q_{\mu}}\bigg[(x \cdot e_{q})e_{q}^{\nu} \bigg] \right|_{q=q_{*}} \, .
\label{eq:toBeikonalStep}
\]
Since $q_{*}^{\mu}$ is parallel to $x^{\mu}_{\perp}$, we know that $q_* \cdot e_q \propto x_\perp \cdot e_q = 0$, and so equation~\eqref{eq:toBeikonalStep} simplifies to
\[
x^{\mu}=b^{\mu}+\left. q_{\nu, *}(x \cdot e_{q})\frac{\partial e_{q}^{\nu}}{\partial q_{\mu}} \right|_{q=q_{*}} \, .
\]
The remaining derivative can be easily performed using the definition of $e_{q}^{\mu}$. The result is 
\[
\left.\left. q_{\nu}\frac{\partial e_{q}^{\nu}}{\partial q_{\mu}}\right|_{q=q_{*}}=-(N_{q} \: q^{\mu}) \right|_{q=q_{*}} \, .
\]
We can then write the eikonal impact parameter as
\[
x^{\mu}_{\perp}=b^{\mu}-(x \cdot e_{0})e_{0}^{\mu}-\left.\big[ (x \cdot e_{q})N_{q} q^{\mu}\big] \right|_{q=q_{*}}+\left.\big[(x \cdot e_{q}) e_{q}^{\mu}\big]\right|_{q=q_{*}} \ .
\]
The scalar products $x \cdot e_{0}$ and $x \cdot e_{q}$, which can be viewed as Lagrange multipliers for the phase which we are minimizing, are fixed by requiring the eikonal impact parameter to be orthogonal to $e_{0}^{\mu}$ and $e_{q}^{\mu}$. A straightforward calculation gives
\[
x^{\mu}_{\perp}=b^{\mu}-\left.\big[(e_{q} \cdot b) (N_q q^{\mu}-e^{\mu}_{q}) \big]\right|_{q=q_{*}} \ .
\]
Expressing $e^{\mu}_{q}$ in terms of $p^{\mu}_{1} $, $p^{\mu}_{2}$ and $q^{\mu}$ we obtain
\[
x^{\mu}_{\perp}=b^{\mu}-\left.\bigg[(e_{q}\cdot b) \big[N_{0q}(p^{\mu}_{1}+p^{\mu}_{2})-N_{q}(p^{\mu}_{1}-p^{\mu}_{2}) \big] \bigg]\right|_{q=q_{*}} \, ,
\]
which agrees --- when evaluated in the center of mass frame --- with the expression for the eikonal impact parameter in \eqref{eq:x-eikonal}, where $\tilde{N}_{q}=-(e_{q} \cdot b) N_{q}$ and $\tilde{N}_{0q}=(e_{q} \cdot b) N_{0q}$.

\section{Unitarity of the conservative final state}
\label{app:unitarity}
After stationary phase in $q^{\mu}$ and $x^{\mu}$, the final state in the conservative sector (that is, no radiation and hence no radiation reaction) is
\[
S|\psi\rangle=\int \mathrm{d} \Phi(p_{1}^{\prime}, p_{2}^{\prime})\left|p_{1}^{\prime}, p_{2}^{\prime}\right\rangle  e^{i q_{*}(s) \cdot x_{*}(s) / \hbar} e^{i \chi\left(x_{*,\perp}(s) ; s\right) / \hbar} \phi_{b}\left(p_{1}^{\prime}-q_{*}(s), p_{2}^{\prime}+q_{*}(s)\right) \ ,
\]
where $q_{*}(s)$ and $x_{*}(s)$ are solutions of the stationary phase condition depending on the Mandelstam variable $s=(p_1'+p_2')^2$.  To check unitarity of the final state we proceed by computing  
\[
\langle \psi| S^{\dag} S |\psi \rangle=\int \mathrm{d} \Phi(p'_{1}, p'_{2}) \, | \phi\left(p'_{1}-q_{*}(s), p'_{2}+q_{*}(s)\right)|^{2} \, ,
\]
taking advantage of the absolute value to simplify the wavepackets using equation~\eqref{eq:twophi}.
It is useful to think of $q_*=\Delta p$ as the total impulse on each particle, recognising $p'_1 - \Delta p$ and $p'_2 + \Delta p$ as the \emph{initial} momenta of the particles:
\[
\langle \psi| S^{\dag} S |\psi \rangle=\int \mathrm{d} \Phi(p'_{1}, p'_{2}) \, | \phi\left(p_{1}, p_{2}\right)|^{2} \, ,
\]
In the CM frame, these initial momenta are rotations of the final momenta through the scattering angle $\Psi$ so that we may also write
\[
\langle \psi| S^{\dag} S |\psi \rangle=\int \mathrm{d} \Phi(p'_{1}, p'_{2}) \, | \phi\left(R_1(\Psi) p'_{1}, R_2(\Psi) p'_{2}\right)|^{2} \, ,
\]
where $R_i(\Psi)$ perform the appropriate rotation from initial to final momenta for each particle. 
In a general frame, these operators are Lorentz transformations.

The basic idea now is to change variables from the final momenta $p_i'$ to the initial momenta $R_i(\Psi) p_i'$, exploiting Lorentz invariance of the phase-space integration measure, aiming to show that
\[
\langle \psi| S^{\dag} S |\psi \rangle=\int \mathrm{d} \Phi(p'_{1}, p'_{2}) \, | \phi\left(p'_{1},  p'_{2}\right)|^{2}  = 
\langle \psi |\psi \rangle = 1 \,.
\]
However, the rotations depend on the momenta of each particle, and hence the change of variables is not immediate.

To examine this issue, consider a helpful choice of wavepacket defined by
\[
\phi(p_1, p_2) = \sqrt{2 U \cdot p_1 \, 2 U \cdot p_2} \, \phi_{\textrm{CM}} ((p_1 + p_2)/2) \phi_{\textrm{rel}} (p_1)  \,,
\]
with $U$ being the time-like unit vector defining the CM frame. 
The factor $\phi_{\textrm{CM}} ((p_1 + p_2)/2)$ in this wavefunction controls the CM motion, so it could be chosen to be a product of four Gaussians (of appropriately narrow width) which define the CM energy and constrain the total spatial momenta in the CM frame to be zero (to a sufficiently good approximation).
The factor $\phi_{\textrm{rel}} (p_1)$ controls the relative motion in the CM frame. 
It is helpful to think of this factor as defining the direction of the spatial momentum $\v{n} \equiv \v{p}_1 / | \v{p}_1|$ in the CM frame, in particular choosing $\phi_{\textrm{rel}} (p_1) = \phi_{\textrm{rel}} (\v{n})$.
Then the wavepacket for the initial momenta can be written in terms of final momenta as
\[
\phi(p_1, p_2) = \sqrt{2 U \cdot p'_1 \, 2 U \cdot p'_2} \, \phi_{\textrm{CM}} ((p'_1 + p'_2)/2) \phi_{\textrm{rel}} (R_1(\Psi) \v{n}' )  \,,
\]
using the fact that $U\cdot \Delta p = 0$. 
Here $\v{n}' \equiv \v{p}'_1 / | \v{p}_1|$ in the unit vector in the direction of the final momentum of particle 1 in the CM frame.

Rewriting the integral as an integral over the CM momentum $P' = (p'_1 + p'_2)/2$ and the relative momentum $p'_\mathrm{rel} = p'_1 - p'_2$ we have
\[
\langle \psi| S^{\dag} S |\psi \rangle=\int \mathrm{d}^3 P' \, |\phi_{\textrm{CM}} (P')|^2  \int \mathrm{d}^3 p'_\textrm{rel} \, |\phi_{\textrm{rel}} (R_1(\Psi)  \v{n}' )|^2 \,.
\]
Now the key point is whether we can change variable in the relative motion to remove the rotation. The answer is yes: although the rotation depends on the magnitude of the relative momentum, it does not depend on the angle of the relative momentum.
We conclude that
\[
\langle \psi| S^{\dag} S |\psi \rangle=\int \mathrm{d}^3 P' \, |\phi_{\textrm{CM}} (P')|^2  \int \mathrm{d}^3 p'_\textrm{rel} \, |\phi_{\textrm{rel}} (\v{n}' )|^2 = \langle \psi |\psi \rangle = 1 \, ,
\]
as desired.

\section{Review of Schwinger proper time method}
\label{app:Schwinger}

In this appendix, we present a short review of the Schwinger proper
time method, which allows propagators in quantum
field theory can be rewritten in terms of path integrals in ordinary
quantum mechanics. We set $\hbar = 1$ only for this appendix, but we have restored for the application of this formalism in section \ref{sec:Schwinger}. Consider first the Feynman propagator for a free scalar field of mass $m$:
\[
D_{F}(x-y)=\int \frac{d^{d} k}{(2 \pi)^{d}} \frac{i}{k^{2}-m^{2}} e^{i k \cdot(x-y)}
\]
which satisfies the position space equation
\[
i(\hat{H}-i \varepsilon) D_{F}(x-y)=\delta^{(d)}(x-y), \quad \hat{H}=\square+m^{2}
\label{eq:KG_operator}
\]
where we have introduced the Klein-Gordon operator $\hat{H}$, which appears in the quadratic terms in the scalar field Lagrangian. Equation \eqref{eq:KG_operator} corresponds to the well-known fact that the propagator is associated with the inverse of the operator $\hat{H}$, and we may formally write
\[
-i(\hat{H}-i \varepsilon)^{-1}=\int_{0}^{\infty} d T e^{-i T(\hat{H}-i \varepsilon)},
\label{eq:Schwinger_time}
\]
where the $i \varepsilon$ prescription guarantees convergence of the integral, and $T$ is conventionally called a Schwinger parameter. The integral in \eqref{eq:Schwinger_time} contains the operator
\[
\hat{U}(T)=e^{-i \hat{H} T},
\]
and if we interpret $T$ as a time variable. this has the known form of the evolution operator in quantum mechanics where $\hat{H}$ is the Hamiltonian, given by
\[
\hat{H}=-\hat{p}^{2}+m^{2}.
\label{eq:H_scalar}
\]
We can introduce a Hilbert space of states on which this Hamiltonian acts. Complete sets are provided by the position or momentum states $\{|x\rangle\}$ or $\{|p\rangle\}$ (eigenstates of $\hat{x}$ and $\hat{p}$ respectively). Consider the evolution operator sandwiched between a state of given initial position and final momentum. For a small time separation $\delta T$, one finds
\[
\bra{p} e^{-i \hat{H} \delta T}\ket{x}=e^{-i H \delta T+\mathcal{O}\left(\delta T^{2}\right)} \langle p|x \rangle
\]
where $H$ denotes the replacement of the position and momentum operators in $\hat{H}$ with their corresponding eigenvalues. Note that, strictly speaking, we must ensure that the Hamiltonian is first
written in Weyl-ordered form, such that all momentum operators appear to the left of all position ones. This is not an issue for the Hamiltonian of eq.\eqref{eq:H_scalar}, but will be relevant when a gauge field is included. The above simple result applies only for small time separations. For a large time separation $T$, one may divide the time interval into $N$ steps
\[
\delta T=\frac{T}{N}
\]
where the limit $N \rightarrow \infty$ will eventually be taken. Then we may introduce a complete set of intermediate position and momentum states at each time-step, so that the matrix element of the evolution operator between a final momentum state $\left|p_{f}\right\rangle$ and initial position state $\left|x_{i}\right\rangle$ becomes
\[
\begin{aligned}
\bra{p_{f}} e^{-i \hat{H} T}\ket{x_{i}} &=\int d x_{1} \ldots d x_{N} \int d p_{0} \ldots d p_{N-1} \bra{p_{f}} e^{-i \hat{H} \delta t} \ket{x_{N}} \\
& \times  \langle x_{N}|p_{N-1} \rangle \bra{p_{N-1}} e^{-i \hat{H} \delta t} \ket{x_{N-1}} \langle x_{N-1}|p_{N-2} \rangle \bra{p_{N-2}} e^{-i \hat{H} \delta t} \ket{x_{N-2}} \ldots \\
& \times\langle  x_{1}|p_{0}  \rangle \bra{p_{0}}e^{-i \hat{H} \delta t}\ket{x_{i}}.
\end{aligned}
\]
Introducing the notation $x_{0} \equiv x_{i}, p_{N} \equiv p_{f}$  for the fixed boundary conditions, we can write the previous equation more succinctly as
\[
\begin{aligned}
\bra{p_{f}} e^{-i \hat{H} T}\ket{x_{i}} &=\int d x_{1} \ldots \int d x_{N} \int d p_{0} \ldots d p_{N-1} \exp \left[-i \sum_{k=0}^{N} H\left(p_{k}, x_{k}\right) \delta T\right] \\
& \times \prod_{k=0}^{N} \langle p_{k}|x_{k} \rangle \prod_{k=0}^{N-1} \langle x_{k+1}|p_{k} \rangle \\
&=\int d x_{1} \ldots \int d x_{N} \int d p_{0} \ldots d p_{N-1} \exp \left[-i \sum_{k=0}^{N} H\left(p_{k}, x_{k}\right) \delta T\right] \\
& \times\left[\prod_{k=0}^{N-1} \exp \left(i \frac{p_{k} \cdot\left(x_{k+1}-x_{k}\right)}{\delta T} \delta T\right)\right] e^{-i p_{f} \cdot x_{N}},
\end{aligned}
\label{eq:Schwinger_pathintegral}
\]
where in the second line we have used that the inner product of position and momentum states gives
\[
\langle x|p \rangle=e^{-i p \cdot x}.
\]
We have written \eqref{eq:Schwinger_pathintegral} in a form such that
its continuum limit $N \rightarrow \infty$ --
\eqref{eq:Schwinger_pathintegral2} -- may be straightforwardly
recognised.

This has the form of a double path integral in position and momentum,
subject to the boundary conditions we imposed above. By carrying out
further manipulations, \eqref{eq:Schwinger_pathintegral2} can be
related to the usual expression for the evolution operator sandwiched
between two position states:
\[
\bra{x_{f}} e^{-i \hat{H} T} \ket{x_{i}}=\int_{x(0)=x_{i}}^{x(T)=x_{f}} \mathcal{D} x \exp \left[i \int_{0}^{T} d t L(x, \dot{x})\right],
\]
where $L(x, \dot{x})$ is the Lagrangian\footnote{To do this, one can
  add an arbitrary potential to the Hamiltonian, and complete the
  Gaussian path integral over $p$.}.However,
\eqref{eq:Schwinger_pathintegral2} will be more convenient for our
purposes. We have here considered a free particle, but the extension
to a particle in a background gauge field is straightforward: one
simply replaces the Hamiltonian of \eqref{eq:H_scalar} with the
corresponding quadratic operator when a gauge field is present, as in
\eqref{HA}.
\bibliographystyle{JHEP}
%\bibliography{bibliography_coherence}

\providecommand{\href}[2]{#2}\begingroup\raggedright\endgroup

\end{document}